\documentclass[11pt]{article}
\usepackage{graphicx,amsmath,amsthm,amssymb,amscd}
\font\Bbb=msbm10                
\newcommand\R{\ifmmode{\mbox{\Bbb R}}\else{{\Bbb R}}\fi}
\newcommand\I{\ifmmode{\mbox{\Bbb I}}\else{{\Bbb I}}\fi}
\newcommand\N{\ifmmode{\mbox{\Bbb N}}\else{{\Bbb N}}\fi}
\newcommand\Z{\ifmmode{\mbox{\Bbb Z}}\else{{\Bbb Z}}\fi}
\newcommand\Q{\ifmmode{\mbox{\Bbb Q}}\else{{\Bbb Q}}\fi}

\newtheorem{theorem}{Theorem}[section]
\newtheorem{lemma}[theorem]{Lemma}
\newtheorem{prop}[theorem]{Proposition}

\newtheorem{corollary}[theorem]{Corollary}
\newtheorem{remark}[theorem]{Remark}

\newtheorem{definition}[theorem]{Definition}

\newcommand{\rar}{\rightarrow}
\newcommand{\hrar}{\hookrightarrow}
\newcommand{\lrar}{\longrightarrow}
\newcommand{\sumApq}{\mathop{{\sum}}_{A=1}^{p+q}}
\newcommand{\sumAApq}{\mathop{{\sum}}_{A_{1}..A_{k}=1}^{p+q}}
\newcommand{\sumAonepq}{\mathop{{\sum}}_{A_{1}=1}^{p+q}}
\newcommand{\sumAtwopq}{\mathop{{\sum}}_{A_{2}=1}^{p+q}}
\newcommand{\sumAkpq}{\mathop{{\sum}}_{A_{k}=1}^{p+q}}

\newcommand{\sumizk}{\mathop{{\sum}}_{i=0}^{k}}

\newcommand{\sumkof}{\mathop{{\sum}}_{k=1}^{\infty}}
\newcommand{\sumkzf}{\mathop{{\sum}}_{k=0}^{\infty}}
\newcommand{\sumip}{\mathop{{\sum}}_{i=1}^{p}}
\newcommand{\sumaq}{\mathop{{\sum}}_{\alpha=1}^{q}}

\newcommand{\summop}{\mathop{{\sum}}_{m=1}^{p}}
\newcommand{\sumaoq}{\mathop{{\sum}}_{\au=1}^{q}}
\newcommand{\summor}{\mathop{{\sum}}_{m=1}^{r}}
\newcommand{\sumaos}{\mathop{{\sum}}_{\au=1}^{s}}

\newcommand{\sumijpq}{\mathop{{\sum}}_{i=1}^{p}\mathop{{\sum}}_{j=1}^{q}}
\newcommand{\summn}{\mathop{{\sum}}_{m,n}}

\newcommand{\Par}{\partial}
\newcommand{\ddt}{\frac{d}{dt}}
\newcommand{\ddl}{\frac{d}{d \lambda}}

\newcommand{\au}{{\alpha}}
\newcommand{\eps}{{\epsilon}}
\newcommand{\epsp}{{\epsilon}'}

\newcommand{\Cn}{\mathbb{C}}
\newcommand{\Rn}{\mathbb{R}}
\newcommand{\Kn}{\mathbb{K}}
\newcommand{\Zn}{\mathbb{Z}}
\newcommand{\Kpq}{\mathbb{K}^{p|q}}
\newcommand{\Krs}{\mathbb{K}^{r|s}}
\newcommand{\zz}{{\{(0,0)\}}}
\newcommand{\Lm}{\Lambda}
\newcommand{\zLm}{{}^{0}\Lambda}
\newcommand{\oLm}{{}^{1}\Lambda}

\newcommand{\Ginf}{G^{\infty}}

\newcommand{\eGinf}{{}^{\epsilon}G^{\infty}}
\newcommand{\GinfU}{G^{\infty}(U)}

\newcommand{\eGinfU}{{}^{\epsilon}G^{\infty}(U)}

\newcommand{\Ginfgx}{G^{\infty}_{g(x)}}
\newcommand{\Cinf}{C^{\infty}}

\newcommand{\Fcal}{\mathcal{F}}
\newcommand{\rcal}{r}
\newcommand{\Lcal}{\mathcal{L}}
\newcommand{\Mcal}{\mathcal{M}}
\newcommand{\Mcale}{{\mathcal{M}}_{e}}
\newcommand{\Mcalee}{{\mathcal{M}}_{e}^{\eps}}
\newcommand{\Dcale}{{\mathcal{D}}_{\eps}}
\newcommand{\Dcalep}{{\mathcal{D}}_{\eps '}}
\newcommand{\Ncal}{\mathcal{N}}
\newcommand{\Scal}{\mathcal{S}}
\newcommand{\Gcal}{\mathcal{G}}
\newcommand{\Hcal}{\mathcal{H}}
\newcommand{\Vcal}{\mathcal{V}}
\newcommand{\Wcal}{\mathcal{W}}
\newcommand{\Pcal}{\mathcal{P}}

\newcommand{\gfrak}{\frak{g}}
\newcommand{\vfrak}{\frak{v}}
\newcommand{\mfrak}{\frak{m}}
\newcommand{\gk}{{\frak{g}}^{k}}
\newcommand{\hfrak}{\frak{h}}
\newcommand{\gz}{{\frak{g}}^{0}}
\newcommand{\hz}{{\frak{h}}^{0}}

\newcommand{\gLie}{{\frak{g}}_{Lie}}
\newcommand{\gLiehat}{\widehat{ {\frak{g}}_{Lie} } }

\newcommand{\Em}{E_{m}}
\newcommand{\Ea}{\tilde{E}_{\au}}
\newcommand{\EM}{E_{M}}
\newcommand{\EN}{E_{N}}
\newcommand{\EK}{E_{K}}

\newcommand{\BM}{\mathcal{BM}}
\newcommand{\BN}{\mathcal{BN}}
\newcommand{\BS}{\mathcal{BS}}
\newcommand{\BG}{\mathcal{BG}}
\newcommand{\BH}{\mathcal{BH}}
\newcommand{\BB}{\mathcal{B}}

\newcommand{\GG}{\mathcal{G} \times \mathcal{G} }

\newcommand{\GmS}{\Gcal / \Scal }

\newcommand{\muS}{{\mu}_{\Scal}}
\newcommand{\muG}{{\mu}_{\Gcal}}

\newcommand{\phiv}{{\phi}_{v}}
\newcommand{\psigI}{{\psi}_{g}^{-1}}
\newcommand{\psihI}{{\psi}_{h}^{-1}}
\newcommand{\psih}{{\psi}_{h}}

\newcommand{\psiS}{{\psi}_{\Scal}}
\newcommand{\psiSI}{{\psi}_{\Scal}^{-1}}
\newcommand{\phiS}{{\phi}_{\Scal}}

\newcommand{\Am}{\mathcal{A}_{\mathcal{M}}}
\newcommand{\An}{\mathcal{A}_{\mathcal{N}}}
\newcommand{\As}{\mathcal{A}_{\mathcal{S}}}
\newcommand{\Abm}{\mathcal{A}_{\mathcal{BM}}}

\newcommand{\Xx}{X_{x}}
\newcommand{\Xl}{X_{\lambda}}
\newcommand{\Xv}{X^{v}}

\newcommand{\TG}{T \Gcal}

\newcommand{\TxzG}{{T}_{x}^{0} \Gcal}

\newcommand{\TezG}{{T}_{e}^{0} \Gcal}
\newcommand{\TezS}{{T}_{e}^{0} \Scal}

\newcommand{\TxBG}{{T}_{x} \BG}

\newcommand{\TxM}{T_{x}\Mcal}
\newcommand{\TxBM}{T_{x}\BM}
\newcommand{\TfxN}{{T}_{f(x)}\Ncal}
\newcommand{\TxzM}{{T}_{x}^{0}\Mcal}
\newcommand{\TxoM}{{T}_{x}^{1}\Mcal}
\newcommand{\TxeM}{{T}_{x}^{\epsilon}\Mcal}

\newcommand{\TyeN}{{T}_{y}^{\epsilon}\Ncal}

\newcommand{\TeG}{{T}_{e} \Gcal}
\newcommand{\TeS}{{T}_{e} \Scal}

\newcommand{\TezDe}{{T}_{e}^{0} \Dcale}

\newcommand{\TyH}{{T}_{y} \Hcal}
\newcommand{\TxyH}{{T}_{xy} \Hcal}
\newcommand{\TezH}{{T}_{e}^{0} \Hcal}
\newcommand{\TxzH}{{T}_{x}^{0} \Hcal}
\newcommand{\TyzH}{{T}_{y}^{0} \Hcal}
\newcommand{\TxyzH}{{T}_{xy}^{0} \Hcal}

\newcommand{\dxg}{d_{x}g}

\newcommand{\delx}{d_{e}l_{x}}
\newcommand{\delexpX}{d_{e}l_{exp(X)}}
\newcommand{\dylx}{d_{y}l_{x}}
\newcommand{\dMfH}{d_{M}f(H)}

\newcommand{\pzA}{\frac{\partial}{\partial z^{A}}}
\newcommand{\pfzA}{\frac{\partial f}{\partial z^{A}}}

\newcommand{\PfJrep}{ \biggl( \frac{ \partial ( {\phi}^{J} \circ f \circ {\psi}^{-1}) } {\partial u^{A_{2}} } \biggr) }
\newcommand{\psiB}{\bar{\psi}}
\newcommand{\psiBI}{{\bar{\psi}}^{-1}}

\newcommand{\phiB}{\bar{\phi}}
\newcommand{\phiBJ}{{\bar{\phi}}^{J}}

\newcommand{\phiJ}{{\phi}^{J}}
\newcommand{\psiI}{{\psi}^{-1}}
\newcommand{\phiI}{{\phi}^{-1}}
\newcommand{\phih}{{\phi}_{h}}
\newcommand{\phihI}{{\phi}_{h}^{-1}}
\newcommand{\psig}{{\psi}_{g}}
\newcommand{\phigI}{{\phi}^{-1}_{g}}

\newcommand{\UBx}{{\bar{U}}_x}
\newcommand{\VBx}{{\bar{V}}_x}

\newcommand{\Mij}{{M}_{ij}}
\newcommand{\Hta}{\tilde{H}^{\alpha}}
\newcommand{\ei}{e_{i}}
\newcommand{\ea}{\tilde{e}_{\alpha}}
\newcommand{\Oa}{{\theta}^{\alpha}}
\newcommand{\gm}{\gamma}
\newcommand{\lk}{{\lambda}^{k}}
\newcommand{\Rs}{{R}_{s}}

\newcommand{\Yl}{{Y}_{\lambda}}
\newcommand{\fl}{{f}_{\lambda}}
\newcommand{\liszero}{{|}_{\lambda =0}}
\newcommand{\delf}{{\delta}_{f}}
\newcommand{\dhat}{\hat{{\delta}}}
\newcommand{\gez}{{\frak{g}}^{0}_{e}}

\begin{document}

\begin{centering}

{\huge INFINITE DIMENSIONAL SUPER LIE GROUPS  }

\vspace{1.4cm}
{\large James Cook and Ronald Fulp}\\

\vspace{.5cm}

 Department of Mathematics, North Carolina State University, Raleigh,
NC 27695-8205.\\

E-mail: jscook3@ncsu.edu  and  fulp@math.ncsu.edu\\
\rm
\vspace{.5cm}

\rm
\vspace{.5cm}
\vspace{.5cm}



\begin{abstract}A super Lie group is a group whose operations are $G^{\infty}$ mappings in the sense of Rogers. Thus the underlying supermanifold possesses an atlas whose transition functions are $G^{\infty}$ functions. Moreover the images of our charts are open subsets of a graded infinite-dimensional Banach space since our space of supernumbers is a Banach Grassmann algebra with a countably infinite set of generators. 

In this context, we prove that if $\hfrak$ is a closed, split sub-super Lie algebra of the super Lie algebra of a super Lie group $\Gcal,$ then $\hfrak$ is the super Lie algebra of a sub-super Lie group of $\Gcal.$ Additionally, we show that if $\gfrak$ is a Banach super Lie algebra satisfying certain natural conditions, then there is a super Lie group $\Gcal$  such that the even part of $\gfrak$ is the even part of the super Lie algebra of $\Gcal.$ In general, the module structure on $\gfrak$ is required to obtain  $\Gcal,$ but the ``structure constants" involving the odd part of $\gfrak$ can not be recovered without further restrictions. We also show that if $\Hcal$ is a closed sub-super Lie group of a super Lie group $\Gcal,$ then $\Gcal \rar \Gcal/\Hcal$ is a principal fiber bundle.

Finally, we show that if $\gfrak$ is a graded Lie algebra over $\Cn,$ then there is a super Lie group whose super Lie algebra is the Grassmann shell of $\gfrak.$ We also briefly relate our theory to techniques used in the physics literature.

We emphasize that some of these theorems are known when the space of supernumbers is finitely generated in which case one can use finite-dimensional techniques. The issues dealt with here are that our supermanifolds are modeled on graded Banach spaces and that all mappings must be morphisms in the $\Ginf$ category.

\end{abstract}
\end{centering}
\vspace{1cm}
{\bf M.S. classification:} 58B25, 17B65, 81R10, 57P99

\newpage

\section{Introduction}
Although mathematicians and physicists have been developing the theory of super Lie groups for over a quarter of a century, there remains a gap in one of the formulations of this theory. From almost the beginning, there have been at least two distinct approaches to the foundations of the superanalysis underlying the theory. Chronologically, the first of these is based on techniques reminiscent of ideas from algebraic geometry. We think of this approach as the sheaf theoretic development of supermathematics even when the theory of sheaves may not explicitly appear in some specific treatments of the subject. Certainly, Berezin, Leites, and Kostant $\cite{Leites},\cite{Kostant}$ were forerunners of this method and for that matter of the entire theory.

A second approach to the formulation  of superanalysis and supermanifolds was initiated separately and differently by Rogers $\cite{Rsuperman}$ and DeWitt $\cite{Dewitt}$. Their work is more closely related to traditional ideas in manifold theory. Much work has been done describing both the sheaf theoretic and manifold theoretic descriptions of supermanifolds and how they are related, but we mention only a few whose work has directly impacted our work here, namely Rogers $\cite{Rsuperman},\cite{Rexamples},\cite{RLie}$, Batchelor $\cite{Batchelor}$, and Bruzzo $\cite{Bruzzo}$.

The gap which we perceive to exist has to do with the treatment of super Lie groups due to Rogers $\cite{RLie}$. She, in fact, has laid out the basic theory of supermanifolds based on a space $\Lambda$ of supernumbers which is in fact a Banach algebra generated by either a finite or a countably infinite number of Grassmann generators. Her supermanifolds are locally modeled on Banach spaces $\Kpq=(\Lambda^0)^p\times (\Lambda^1)^q$  where either $\Lambda=\Lambda_N$ has $N$ generators or $\Lambda=\Lambda_{\infty} $ has an infinite number of such generators. In her paper on super Lie groups $\cite{RLie}$ she derives basic theorems about super Lie groups, but the deeper results are obtained only when $\Lambda=\Lambda_N.$ In this case it turns out that, with considerable effort, one can reduce the deeper theorems to corresponding theorems for ordinary finite dimensional Lie groups. It is asserted that it would be interesting to develop these ideas in case $\Lambda=\Lambda_{\infty}$ and that there are explicit areas of quantum field theory where such results would be useful. This same conclusion is asserted in the book by Freund $\cite{Freund}$.

It is our purpose to fill this gap in the Rogers approach to super Lie groups. Infinitely generated Grassmann algebras are both more and less complicated than in the finitely generated case. Since there is no generator of maximal order, there are no ambiguities in the top dimension.  In the finitely generated case, the highest order derivatives of a function are not unique; this ambiguity sporadically surfaces and can be a source of difficulty which continually requires consideration. On the other hand, in the infinitely generated case, we are not able to appeal to corresponding theory of finite dimensional Lie groups. We {\it are} able to utilize the theory of Banach Lie groups at various points of our development, but even when we are able to do so, we often must develop the machinery needed to assure that we remain in the ``supersmooth category".

Our notation throughout the paper is an amalgam of that of Rogers $\cite{Rsuperman}$ and Buchbinder and Kuzenko $\cite{BB}$. We strive for completeness, thus Section 1 provides the basic definitions and results from Rogers $\cite{Rsuperman},\cite{RLie}$ essential to our development. Section 2 provides the connections needed between smooth Banach manifolds and supersmooth super manifolds. Section 3 examines what is required in order that a subset of a supermanifold be a sub-supermanifold. 

Section 4 contains our main results. This includes a determination of when a sub-super Lie algebra $\hfrak$ of the super Lie algebra $\Lcal(\Gcal)$ of a super Lie group $\Gcal$ is in fact the super Lie algebra of a sub-super Lie group of $\Gcal.$ We also find conditions under which the even part of an abstract Banach super Lie algebra is the even part of the super Lie algebra of some super Lie group $\Gcal.$  Given a super Lie algebra $\gfrak$ we show that there exists a super Lie group whose $\Ginf$ structure is determined by the even part of $\gfrak,$  but the odd part participates only through its module structure. Without other conditions, which we do not obtain here, the super Lie structure on $\gfrak$ is not recovered from the super Lie group $\Gcal.$  In fact different super Lie structures on the odd part do not change the super group $\Gcal,$ although different module structures on $\gfrak$ will result in different super group structures on $\Gcal.$
Along the way we also show that if $\Hcal$ is a closed sub-super Lie group of a super Lie group $\Gcal,$ then $\Gcal \rar \Gcal/\Hcal$ is a principal fiber bundle. We emphasize that all of this work requires an infinite number of Grassmann generators of our space of supernumbers. 

Finally, in the last section of the paper, we show how to apply our results to those types of super Lie groups prevalent in the physics literature. In that context super Lie groups often arise by beginning with a super Lie algebra which is used to construct a super Lie group using the exponential mapping and the Campbell-Hausdorff formula. This is an effective procedure but does not address the issue of finding a super smooth atlas for the group. In particular, one also has no way of determining the topology of the super Lie group. Our theory settles these issues when the underlying module structures utilize infinitely generated supernumbers as scalars; we emphasize that the finitely generated case was dealt with by Rogers $\cite{RLie}$. In this last section we show how our results relate to procedures utilized in the physics literature especially for super Lie groups and super Lie algebras of matrices with supernumbers as entries.  Additionally, we show that for every graded Lie algebra $\gfrak$ over $\Cn,$ there exists a super Lie group $\Gcal$ whose super Lie algebra is the Grassmann shell $\gLiehat$ of the Lie algebra $\gfrak.$  

The authors are grateful to T. Ratiu who provided us with information and references regarding the theory of Banach Lie groups. He is, of course, in no way responsible for any  misunderstanding or misuse of these ideas in this paper.

\section{Preliminary Definitions and Results}
We first briefly describe our conventions regarding supernumbers. Let $\Kn$ denote either the field $\Rn$ of real numbers  or the field $\Cn$ of complex numbers. For each positive integer $p$ let $\mathcal{I}_{p}$ denote the set of all mult-indices $I=(i_1,i_2,\cdots,i_p)$ of positive integers such that
$1\leq i_1 <  i_2 < \cdots < i_p.$ Let $\{\zeta^1,\zeta^2,\zeta^3,\cdots \} $ denote a countably infinite set of distinct Grassmann generators of the Grassmann algebra of formal power series 
\begin{equation} \label{E:one}
z=\sum_{p=0}^{\infty}\sum_{I \in \mathcal{I}_{p}}^{}z_{I}{\zeta}^{I}
\end{equation}
where $\zeta^I$ is defined to be the formal product $\zeta^{i_1}\zeta^{i_2}\cdots \zeta^{i_p} $
for each multi-index $I.$ Our supernumbers are by definition the set $\Lambda$ of all such formal series $z$ for which the norm of $z$ defined by 
\begin{equation} \label{E:two}
||z|| = \sum_{p=0}^{\infty} \sum_{I \in \mathcal{I}_{p}} |z_{I}|. 
\end{equation}
is finite. Although $\mathcal{I}_{p}$ is infinite, it is countable and the series 
$\sum_{I \in \mathcal{I}_{p}} |z_{I}|$ is therefore absolutely convergent. See $\cite{Rsuperman}$ for full details. 
Let $\zLm$ denote the set of all even supernumbers, i.e., $z$ is in $ \zLm $ if and only if 
\begin{equation} \label{E:three}
z=\sum_{p=0}^{\infty}\sum_{I \in \mathcal{I}_{2p}}^{}z_{I}{\zeta}^{I}
\end{equation}
Such supernumbers are called commuting supernumbers where the $p=0$ term is understood to be an element of $\Kn $. Similarly, odd supernumbers are by definition those $z \in \oLm $ for which 
\begin{equation} \label{E:four}
z=\sum_{p=0}^{\infty}\sum_{I \in \mathcal{I}_{2p+1}}^{}z_{I}{\zeta}^{I}.
\end{equation}
Odd supernumbers are anticommuting supernumbers. The supernumbers in $\zLm \cup \oLm $ are called pure. Pure supernumbers have parity $\epsilon = \pm 1 $ where $\epsilon (\zLm ) = 0 $ and $\epsilon (\oLm ) = 1 $. The soul of a supernumber is defined to be
\begin{equation} \label{E:five}
z_{S}=\sum_{p=1}^{\infty}\sum_{I \in \mathcal{I}_{p}}^{}z_{I}{\zeta}^{I}.
\end{equation}
The body of a supernumber is denoted $z_{B} \in \Kn $ and is defined for each $z \in \Lm $ by $z = z_{B} + z_{S} $. 

\noindent
Our space $\Lambda$ of supernumbers  is denoted $B_{\infty} $ in $\cite{Rsuperman}$ where $\mathbb{K} = \Rn $. We have not included a notion of super conjugation as we do not assume that $\Kn $ has a conjugation. When $\Kn =\Cn $ we could give $\Lm $ a conjugation as is done in $\cite{BB}$. In other treatments supernumbers are often formed from a finite set of $L$-Grassmann generators, for example in $\cite{Rsuperman}$ this is denoted $B_{L} $. We consistently work in the infinite case throughout this work. 

\begin{prop} \label{P:twoPone}
The set $\Lambda$ of supernumbers is a Grassmann algebra and is a Banach space relative to its norm. Moreover, for $z,w\in \Lambda, \quad ||zw|| \leq ||z|| ||w|| $ and consequently  $\Lm $  is a Banach algebra over $\Kn $.
\end{prop}

\noindent
The proof can be found in $\cite{Rsuperman}$.

\begin{definition} 
Let $\Kpq$ denote the set of all $(p+q)$-tuples $z=(x^1,\dots,x^p,{\theta}^{1},\dots , {\theta}^{q}) $ where
$x^{m} \in \zLm $ for $m=1,2,\dots , p$ and $\Oa \in \oLm $ for $\au = 1,2,\dots , q $. In a more compact notation we also write $z = (z^M)$ for $ M=1,2,\dots , p+q $. The norm on $\Kpq $ is induced from the norm on $\Lm $,
\begin{equation} \label{E:six}
|| \ z \ || = \summop ||x^m|| + \sumaoq || \Oa || = \sum_{M=1}^{p+q} || z^M ||.
\end{equation}
We define the dimension of $\Kpq$ to be $(p|q)$.
\end{definition}

\noindent
Since $\Lm $ is complete it is straightforward to prove that $\Kpq $ is complete and consequently $\Kpq$ is a Banach space.

\begin{definition} 
A $\mathbb{Z}_{2}$-graded vector space $U$ over a field $\Kn $ is a vector space over $\Kn $ with subspaces $U_0$ and $U_{1}$ such that $U = U_{0}\oplus U_{1}$. Vectors in $U_{0} $ are termed even and have parity $\epsilon (U_{0} ) = 0$ whereas vectors in $U_{1}$ are termed odd and have parity $\epsilon (U_{1} ) = 1 $. Suppose $dim(U_{0})=p $ and
$dim(U_{1})=q $ then we say U has graded dimension $(p,q)$. 
\end{definition}

\noindent
Algebraists often refer to such spaces as called superspaces, however we will reserve that term for spaces built over $\Lm $. Graded will always refer to $\mathbb{Z}_{2}$-grading in this paper.

\begin{definition} 
A $\mathbb{Z}_{2}$-graded algebra $V=V_0 \oplus V_1 $ is a graded vector space with a multiplication such that 
$1 \in V_0 $ and $V_{r}V_{s} \subset V_{r+s} $ mod 2. 
\end{definition}

\begin{definition} 
A graded-commutative algebra W is a $\mathbb{Z}_{2}$-graded algebra such that for all $v \in W_{r}$ and $w \in W_{s} $  we have $vw = {(-1)}^{rs}wv $ for $ r,s=0,1 $. 
\end{definition}

\begin{definition} 
A graded Lie algebra is a graded vector space $ U=U_{0} \oplus U_{1} $ over $\Kn $ with a bilinear bracket $[ \ , \ ]: U \times U \rar U $ which is graded $[U_{r},U_{s}] \subset U_{r+s} $ for $r,s = 0,1 $, and for all $a,b,c \in U_{0} \cup U_{1} $ with parities ${\eps}_{a},{\eps}_{b},{\eps}_{c} $ satifies the graded Jacobi indentity
\[ {(-1)}^{{\eps}_{a}{\eps}_{c}} [a,[b,c]] +
   {(-1)}^{{\eps}_{b}{\eps}_{a}} [b,[c,a]] +
   {(-1)}^{{\eps}_{c}{\eps}_{b}} [c,[a,b]] = 0 \]
and the graded skewsymmetry condition,
\[ [a,b] = -{(-1)}^{{\eps}_{a}{\eps}_{b}}[b,a]. \] 
\end{definition}

\noindent
Algebraists often refer to such graded Lie algebras as superalgebras, however we will reserve that term for algebras built over $\Lm $. An associative graded algebra can be given the structure of a graded Lie algebra by defining the bracket to be
\[ [a,b] = ab -(-1)^{{\eps}_{a}{\eps}_{b}}ba. \]
This is the supercommutator which functions both as a commutator and an anticommutator depending on the inputs. A graded commutative algebra has a trivial supercommutator and will be called a graded-Abelian Lie algebra. Many things are known as graded Lie algebras over $\Cn $, see $\cite{KAC}$ for the classification of all finite dimensional graded Lie algebras and a rehashing of much of classical Lie theory in the graded case.

\begin{definition} 
A graded left $\Lm $-module is a graded vector space $U=U_0 \oplus U_1 $ which is a left module which respects the parity structures of $U$ and $\Lm $; that is $\zLm U_{r} \subset U_{r} $ and $\oLm U_{r} \subset U_{r+1} $ for $r \in \{0,1\} = \mathbb{Z}_{2}$. Likewise a graded right $\Lm $-module is a graded vector space $U=U_0 \oplus U_1 $ which is a right module which respects the parity structures of $U$ and $\Lm $; that is $U_{r}\zLm  \subset U_{r} $ and $U_{r} \oLm \subset U_{r+1} $ for $r \in \{0,1\} = \mathbb{Z}_{2}$.
A $\Lm $ bimodule is a left-right $\Lm $-module U that satifies an intertwining relations $(\au v) \beta)=\au (v \beta)$  for all $\au,\beta \in \Lm $, $v \in U $, and $\gamma w = {(-1)}^{{\eps}_{\gamma}{\eps}_{w}}w \gamma $ for all $w \in U_0 \cup U_1 $ and $ \gamma \in \zLm \cup \oLm $.   
\end{definition}

\noindent
If $\Kn = \Cn $ and we add conjugation to a $\Lm $-bimodule we would have the structure of a supervector space similar to that defined by DeWitt. We will focus on left module actions, but this is not a necessary restriction, many constructions allow a right multiplication of $\Lm $ and a conjugation if we have $\Kn =\Cn $. We will avoid the issue of conjugation in this paper.

\begin{definition} 
A graded Lie left $\Lm $-module is a graded Lie algebra W over $\Kn $ which is a left $\Lm $-module such that
\[ [\au X , Y ] = \au [X,Y] \]
for all $\au \in \Lm $ and $X,Y \in W $.
\end{definition}

\begin{definition} 
A graded Lie right $\Lm $-module is a graded Lie algebra W over $\Kn $ which is a right $\Lm $-module such that
\[ [ X , Y \au] =  [X,Y] \au \]
for all $\au \in \Lm $ and $X,Y \in W $.
\end{definition}

\begin{prop} \label{P:twoPten}
Given a left $\Lm$-module $V$ we can construct a right $\Lm$-module according to the rule
\begin{equation} \label{E:seven}
X \au \equiv (-1)^{\eps(X) \eps (\au )}\au X.
\end{equation}
for all $X \in V_{0} \cup V_{1}$ and $\au \in \zLm \cup \oLm$. Likewise, a graded Lie left $\Lm $-module $W$ is given a natural graded right Lie $\Lm $-module under the same rule.
\end{prop}

\noindent
It is trivial to verify that $V$ has a right $\Lm $-module structure as defined in the proposition.Consider the following to see that W is a right Lie $\Lm$-module,
\begin{equation} \label{E:eight}
\begin{array}{ll}
[X,Y \au] &= (-1)^{\eps (X) \eps (Y \au)+1}[Y \au,X ] \\
          &= (-1)^{\eps (X)(\eps (Y)+\eps (\au))+1}[(-1)^{\eps (Y) \eps ( \au)}\au Y,X] \\
	  &= (-1)^{\eps (X)\eps (Y)+\eps (\au) (\eps (X) +\eps (Y)) +1}\au [Y,X] \\
	  &= (-1)^{\eps (X)\eps (Y)+\eps (\au) (\eps (X) +\eps (Y)) +1}
	     (-1)^{\eps (\au) \eps ([Y,X])} [Y,X]\au \\
	  &= (-1)^{\eps (X)\eps (Y)+\eps (\au) (\eps (X) +\eps (Y)) +1}
	     (-1)^{\eps (\au) (\eps (Y)+\eps (X))}
	     (-1)^{\eps (X)\eps(Y) +1}[X,Y]\au \\
	  &= [X,Y]\au
\end{array}
\end{equation}
we have employed the useful relations $\eps (Y \au)=\eps (Y)+\eps (\au)$ and $\eps ([Y,X])=\eps (Y)+\eps (X)$ for all pure $X,Y \in W$ and pure $\au \in \Lm$ to make the needed cancellations. This calculation shows that we can always induce a right Lie-$\Lm $-module structure on W given that W is a left Lie-$\Lm$-module. \\
Notice that if $V$ is a left $\Lm $ module then it is also a right $\Lm $ module, this is almost a {\it supervector space}. When $\Kn = \Cn$ and there is a conjugation on $V$ which is an anti-involution respecting the $\Lm$ module structure then we say that $V$ is a supervector space.

\begin{definition}
A super Lie algebra is a left Lie $\Lm$-module. 
\end{definition}

\noindent
As we have discussed this automatically induces a right Lie $\Lm$-module structure. Other authors include a conjugation in the definition of a {\it super Lie algebra} but we allow the base field $\Kn$ to be $\Rn$ or $\Cn$ only in the later case is it natural to introduce a super conjugation on $\Lm $.

\begin{definition} 
Let $V$ be a graded left $\Lm $ module and let $m=1,2,\dots p $, $\au = 1,2,\dots q $ and $\Em \in V_{0}$, $\Ea \in V_{1}$ then we call $ \{\Em, \Ea \} $ a pure basis of graded dimension $(p,q)$ if there exist $v^{m},\tilde{v}^{\au} \in \Lm $ for each $v\in V $ such that
\[ v = \summop v^{m} \Em + \sumaoq \tilde{v}^{\au} \Ea = \sum_{M=1}^{p+q}v^{M} \EM  \]
where we also denote $ \{\Em, \Ea \}  = \{ \EM \} $ with $\EM = \Em $ for $M=m=1,2,\dots p $ and $E_{M} = \Ea $ for $M=p+\au = p+1, p+2,\dots p+q $. For convenience denote $\eps ( \EM )$ by ${\eps}_{M}$.  
\end{definition}

\noindent
One should notice that $\Kpq $ is not a left $\Lm $-module, multiplication by $\oLm $ distorts the structure of $\Kpq $. It will be important to distinguish the difference between graded dimension $(p,q)$ and supermanifold dimension $(p|q)$. It is fairly obvious that we can obtain left $\Lm $ modules from graded vector spaces by simply tensoring with $\Lm $, however not all left $\Lm $-modules have such structure (see Example 4.2a in $\cite{Rsuperman}$). Hence the class of left $\Lm $ modules is larger than that of graded vector spaces.

\begin{definition} 
Given a Lie left $\Lm $-module V of graded dimension $(p,q)$ with pure basis $\{ \EM \} $, $M=1,2, \dots,p+q $ there exist structure constants ${f}^{K}_{MN} \in \Lm $ such that $ [ \EM,\EN ] = \sum_{K=1}^{p+q}{f}^{K}_{MN} \EK $ for all $ M,N=1,2,\dots ,p+q $. If V possesses a pure basis for which $ s({f}^{K}_{MN})=0 $ for all $ M,N,K $ then we say that V is a conventional Lie left $\Lm $-module, otherwise we say V is unconventional.
\end{definition}

\noindent
In fact, conventional Lie left $\Lm $-modules correspond to graded Lie algebras. Our treatment of super Lie groups will include unconventional Lie left $\Lm $-modules which to our knowledge are not fully classified at this time. 

\begin{definition} 
Let U be open in $\Kpq $ and let $f: U \rar \Lm $. Then \\
(1.) \ f is said to be $G^{0}$ on U if f is continuous on U. \\
(2.) \ f is said to be $G^{1}$ on U if there exist p+q functions $G_{M}f: U \rar \Lm $, $M=1,2,\dots , p+q $ and a
function $\eta : \Kpq \rar \Lm $ such that, if $(a,b),(a+h,b+k) \in U$
\[ f(a+h,b+k) = f(a,b) + \summop h^{m}(G_{m}f)(a,b) + \sumaoq k^{\au}(G_{p+\au}f)(a,b) + ||(h,k)||\eta (h,k) \]
where $ || \eta (h,k) || \rar 0 $ as $|| \eta (h,k) || \rar 0 $. \\
(3.) \ for a positive integer s, f is said to be $G^{s}$ on U if f is $G^{1}$ on U and it is possible to choose
$G_{M}f: U \rar \Lm $, $M=1,2,\dots , p+q $ which are $G^{s-1}$ on U. \\
(4.) \ f is said to be $\Ginf $ on U if f is $G^{s}$ for every postive integer s. \\
(5.) \ for a postive integer s, let $g: U \rar {\Lm}^{s} $ where ${\Lm}^{s}$ is the Cartesian product of s-copies
of $\Lm $, and let $P_{M}:{\Lm}^{s} \rar \Lm $ be the projection onto the M-th factor ($P_{M}(c^1,c^2,\dots , c^{p+q}) = c^{M}$). Then g is said to be $G^{s} $ or $\Ginf $ if each component function $g^{M} = P_{M} \circ g $, $1\leq M \leq p+q $ is likewise $G^{s} $ or $\Ginf $.
\end{definition}

\noindent
We comment that there are ambiguities that arise in choosing the functions $G_{M}f$ in the case that the underlying Grassmann algebra has only finitely many generators. That ambiguity has been dealt with in various ways by different authors. To deal with this difficulty,  Rogers introduced the "z-mapping" in $\cite{Rsuperman}$ and later the $GH^{\infty}$ functions $\cite{Rlater};$ additionally Rothstein $\cite{Roth}$ suggested another solution and  Bruzzo $\cite{Bruzzo}$ introduced the notion of a G-function. All of these are similar in spirit to Roger's original definition which is of course inspired by classical analysis. We avoid the controversy by focusing on the case of infinitely many Grassmann generators, in this case the ambiguity is not present. As a consequence we are forced to use infinite dimensional Banach manifolds in our treatment, as opposed to Roger's who was able to capitalize on the fact that super Lie groups over $B_{L}$ ( for $L < \infty $ ) are also finite dimensional analytic Lie groups. 

\begin{prop} \label{P:twoPfifteen}
If U is open in $\Kpq $ and $f \in \Ginf (U)$, then $ f \in C^{\infty} (U,\Lm ) $ the space of all $C^{\infty}$ maps of U into $\Lm $. In particular, regarding the s-th total derivative of f as a multi-linear transformation from $({\Kpq})^{s}$ to $\Lm $,
\[  d^{s}_cf(H_{1},\dots,H_{s})=  [ d^{s}f(c) ][H_{1},\dots,H_{s} ] = {\sum}_{M_{1},\dots , M_{s} = 1}^{p+q}{H}_{1}^{M_{1}} \cdots {H}_{s}^{M_{s}}
                                       (G_{M_{s}} \cdots G_{M_{1}}f)(c) \]
for all $c \in U $ and $[H_{1},\dots,H_{s} ] \in ({\Kpq})^{s} $				       
\end{prop}

\noindent
This is Proposition 2.8 of $\cite{Rsuperman}$. The $d^{s}$ is an iterated Frechet derivative and is explained in $\cite{Lang}$ for the infinite dimensional case. 

 
\begin{prop} \label{P:twoPsixteen}
Let U be open in $\Kpq $, $f,g \in \Ginf (U) $, $a \in \zLm \cup \oLm $, and $\lambda \in \Kn $. Then \\
(1.) \ $f+g \in \Ginf (U)$ and $G_{M}(f+g) = G_{M}f + G_{M}g $ for $1 \leq M \leq p+q$ \\
(2.) \ $\lambda f \in \Ginf (U)$ and $G_{M}(\lambda f) = \lambda G_{M}f $ for $1 \leq M \leq p+q$ \\
(3.) \ If E and Q represent projection maps of $\Lm $ onto $\zLm $ and $\oLm $, respectively, then $E \circ f$ and $ Q \circ f $ are in $\Ginf (U)$. Moreover $\Ginf (U) $ is a graded vector space with
\[ \Ginf (U)_{0} = \{ f \in \Ginf (U) \ | \ E \circ f = f \} \qquad \Ginf (U)_{1} = \{ f \in \Ginf (U) \ | \ Q \circ f = f \} \] 
We define $ \eps(\Ginf (U)_{r})=r $ for $r=0,1$ as usual. \\
(4.) \ $f \in \Ginf (U)_{0} \cup \Ginf (U)_{1} $ then $af \in \Ginf (U) $ with 
       $G_{M}(af) = {(-1)}^{\eps (a) \eps_{M}}aG_{M}f $ \\
(5.) \ $f,g \in \Ginf (U)_{0} \cup \Ginf (U)_{1}$ then $fg \in \Ginf (U) $ with
       $G_{M}(fg) = (G_{M}f)g + {(-1)}^{\eps (f) \eps_{M}}f G_{M}g $. \\
(6.) \ V open in $\Krs $ and $h \in \Ginf (V,\Kpq ) $ then $f \circ h \in \Ginf [h^{-1}(U) \cap V ] $ with
\[ G_{M} (f \circ h )(a) = {\sum}_{N=1}^{p+q} (G_{M}H^{K})(a)(G_{K}f)[h(a)] \]
for $H^{M} = P_{M} \circ H $ for $1 \leq M \leq p+q $ and for all $a \in h^{-1}(U) \cap V$, $K=1,2,\dots , r+s $. \\
(7.) \ If the interval I is open in $\Rn $ and $\tilde {h} \in \Cinf (I, \Kpq )$ then 
$f \circ \tilde{h} \in \Cinf [{\tilde{h}}^{-1}(U) \cap I , \Kpq ] $ and
\[ \frac{\Par}{\Par t} (f \circ {\tilde{h}} ) = {\sum}_{M=1}^{p+q} \frac{ \Par {\tilde{h}}^{M}(t)}{\Par t }
(G_{M}f)[\tilde{h} (t)] \]
for $t \in I $.
\end{prop}

\noindent
This is Proposition 2.12 of $\cite{Rsuperman}$. Parts (4.) and (5.) of the proposition above are easily extended by linearity to objects which are not pure. 

\begin{definition} 
Let $\Mcal $ be a Hausdorff topological space. \\
(1.) \ An $(p|q)$ open chart on $\Mcal $ over $\Lm $ is a pair $(U,\psi )$ with U open in $\Mcal $ and $\psi $ a homeomorphism of U onto an open subset of $\Kpq $. \\
(2.) \ An $(p|q)$ $G^{s}$ structure on $\Mcal $ over $\Lm $ is a collection $\{ (U_{\au},{\psi}_{\au}) \ | \ \au \in \mathcal{I} \} $ of open charts on $\Mcal $ such that (i) $\Mcal = {\cup}_{\au \in \mathcal{I}} U_{\au}$, (ii) for $U_{\au} \cap U_{\beta} \ne \emptyset $ the mapping ${\psi}_{\beta} \circ {\psi}_{\au}^{-1} $ is a $\Ginf $ mapping of 
${\psi}_{\au}(U_{\au} \cap U_{\beta}) $ onto ${\psi}_{\beta}(U_{\au} \cap U_{\beta})$, and (iii) the collection
$\{ (U_{\au},{\psi}_{\au}) \ | \ \au \in \mathcal{I} \} $ is a maximal collection of open charts for which (i) and (ii) hold. A collection for which (i) and (ii) hold but is not necessarily maximal is called a $(p|q)$ $G^{s}$ subatlas on $\Mcal $ over $\Lm $. \\
(3.) \ An $(p|q)$ dimensional $G^{s}$ supermanifold over $\Kpq $, is a Hausdorff topological space $\Mcal $ with an
$(p|q)$ $G^{s}$ structure on $\Mcal $ over $\Lm $. \\
(4.) \ Each $U_{\au} $ is called a coordinate neighborhood, and each ${\psi}_{\au}$ is a coordinate map. For each $\au \in \mathcal{I}$, $p+q$ local coordinate functions are defined by,
\[ u^{m} = P_{m} \circ {\psi}_{\au} \qquad v^{\beta} = P_{p+\beta} \circ {\psi}_{\au}  
                                    \qquad u^{M} = P_{M} \circ {\psi}_{\au} \]
where $m=1,2,\dots , p$, $\beta = 1,2,\dots, q$, and $M=1,2,\dots,p+q$. We use lower case Latin indices for the commuting coordinates, Greek indices for the anticommuting coordinates, and upper case Latin indices for both.\\
(5.) \ Setting $r=\infty$ defines the structure of a $\Ginf $ supermanifold.
\end{definition}

\noindent
There are other popular definitions used in the literature for supermanifold. For example, graded manifolds of Kostant $\cite{Kostant}$, or the DeWitt $\cite{Dewitt}$ or $H^{\infty}$-manifold, and the definition due to Berezin and Leites $\cite{Leites}$. All of these are included under the category of $\Ginf $-manifold as is discussed in $\cite{Rsuperman}$. The $\Ginf $ supermanifolds allow a richer class of topologies than the other definitions.

\begin{definition} 
Let $\Mcal $ be $\Ginf $ supermanifold and $\{ (U_{\au},{\psi}_{\au}) \ | \ \au \in \mathcal{I} \} $ a subatlas of $\Mcal $. If U is open in $\Mcal $ we define $\Ginf $ functions on U by 
\[ \Ginf (U) = \{ f \ | \ f: U \rar \Lm , \ with \ f \circ {\psi}_{\au}^{-1} 
                        \in \Ginf [ {\psi}_{\au} (U \cap U_{\au} )], \ \forall \au \in \mathcal{J} \}. \] 
Then  $\Ginf (p) $, the germ of $\Ginf $ functions at a point $p \in \Mcal $, is likewise defined by
\begin{equation} \notag
\Ginf (p) = \{ f \ | \ \exists \text{ an open neighborhood N of $p$ such that $f \in \Ginf (N)$} \}
\end{equation} 
We say two functions in $\Ginf (p)$ are equivalent iff they agree on some open set about $p$. Consequently it would be more rigorous to say that $\Ginf (p)$ is the set of equivalence classes of functions defined near $p$.
\end{definition}

\begin{prop} \label{P:twoPnineteen}
Given U open in $\Mcal $, then \\
(1.) \ $\Ginf (U)$ is a graded commutative algebra over $\Kn $ with,
\begin{equation} 
\begin{array}{l}
 \Ginf (U)_{0} = \{ f \in \Ginf (U) \ | \ f(U) \subset \zLm \} \\
 \Ginf (U)_{1} = \{ f \in \Ginf (U) \ | \ f(U) \subset \oLm \} 
\end{array}
\end{equation} 
(2.) \ $\Ginf (U)$ is a graded left $\Lm $ module with parity defined as in (1.).
\end{prop}

\noindent
In general we define the parity of functions according to the parity of their range. 

\begin{definition} 
Let V and W be graded $\Lm $-bimodules then we denote $L(V,W)$ to be the set of all ordinary $\Kn$-linear transformations from $V$ to $W$. A mapping $L \in L(V,W)$ is a left linear mapping if 
\[ L(v\au) = L(v)\au \]
for all $v \in V$ and $\au \in \Lm$. 
The space of all such left linear maps is denoted by $L^{+}(V,W) $. A mapping $L \in L(V,W)$ is a right linear mapping if
\[ L(\au v) = \au L(v) \]
for all $v \in V$ and $\au \in \Lm$. 
The space of all such right linear maps is denoted by $L^{-}(V,W) $.
In each of the above cases a linear map is deemed even if it preserves the parity of pure vectors and it is odd if it reverses the parity of pure vectors,
\[ \eps (L(v)) = \eps (L) +\eps (v) \]
for all $L \in L(V,W)$. When $V=W $ we use the usual notation $L(V,V)=End(V)$ with $End^+(V)$ and $End^{-}(V)$  for left and right endomorphisms, respectively.
\end{definition}

\begin{definition} 
Let $End^{+}[\Ginf (U)] $ denote the set of all left vector space endomorphisms of $\Ginf (U) $, i.e. $L \in End^{+}[\Ginf (U)] $ iff it is an endomorphism over $\Kn $ in the traditional sense and
\[ L(fa) = L(f)a \]
for all $a \in \Lm $ and all $f \in  \Ginf (U)$. 
\end{definition}

\noindent
We note that the super partial derivatives $G_{M} $ are in $End^{+}[\Ginf (U)] $. Other authors prefer to use right endomorphisms, for example $\cite{HT}$. Our notation is a synthesis of $\cite{BB}$ and $\cite{Rsuperman}$.

\begin{prop} \label{P:twoPtwentytwo}
Let U be open in $\Mcal $ then \\
(1.) \ $End^{+}[\Ginf (U)] $ is a graded commutative algebra over $\Kn $ with,
\begin{equation} 
\begin{array}{l} \notag
 End^{+}[\Ginf (U)]_{0} = \{ L \in End^{+}[\Ginf (U)] \ 
        | \ \eps (Lf) = \eps (f),  \ f\in  \Ginf (U)_{0} \cup \Ginf (U)_{1} \} \\
 End^{+}[\Ginf (U)]_{1} = \{ L \in End^{+}[\Ginf (U)] \ 
        | \ \eps (Lf) = \eps (f)+1,  \ f\in  \Ginf (U)_{0} \cup \Ginf (U)_{1} \}. \\
\end{array}
\end{equation} 
If $L \in End^{+}[\Ginf (U)]_{0} \cup End^{+}[\Ginf (U)]_{1}$ and $f \in \Ginf (U)_{0} \cup \Ginf (U)_{1}$ then
\[ \eps (Lf) = \eps (L) + \eps (f)  \]
(2.) \ $End^{+}[\Ginf (U)]$ is a graded left $\Lm $ module with parity defined as in (1.). \\
\end{prop}

\noindent
A similar proposition is true regarding $End^{-}[\Ginf (U)]$.

\begin{definition} 
Let U be open in $\Mcal $. A $\Ginf $ vector field on U is an element X of $End^{+}[\Ginf (U)]$ such that \\
(1.) \ $X(fg) = (Xf)g + {(-1)}^{\eps (f) \eps (X)}fXg$ for all $f,g \in \Ginf (U)_{0} \cup \Ginf (U)_{1}$ \\
(2.) \ $X(af) = {(-1)}^{\eps (a) \eps (X)}aXf$ for all $f \in \Ginf (U)_{0} \cup \Ginf (U)_{1}$ 
                                                              and $a \in \zLm \cup \oLm$ \\
The set of all $\Ginf $ vector fields is denoted $D^{1}(U)$.
\end{definition}

\noindent
We have affixed the qualifier $\Ginf $ to distinguish these vector fields from the ordinary $\Cinf $ vector fields which stem from the Banach space structure of $\Mcal $. Also although our definition is given for pure elements it should be clear how to extend linearly to impure functions and supernumbers.

\begin{prop} \label{P:twoPtwentyfour}
Let U be open in $\Mcal $ then $D^{1}(U)$ is a graded Lie left $\Lm $ module with bracket
\[ [X,Y] = XY - {(-1)}^{\eps (X) \eps (Y)}YX \] 
\end{prop}

\noindent
Since $\Ginf $ vector fields are in $End^{+}[\Ginf (U)]$ we already know how to grade them. This is Proposition 5.5 of $\cite{Rsuperman}$.

\begin{definition} 
Let $(U,\psi )$ be a chart on a $\Ginf $ supermanifold $\Mcal $ where $\psi = (u^{1},\dots , u^{p},v^{1},\dots , v^{q})$. For $m=1,2,\dots , p$, define
\[ \frac{\Par}{\Par u^{m}} : \Ginf (U) \rar \Ginf (U), \ \ where \ \ 
\frac{\Par f}{\Par u^{m}} \equiv [G_{m}(f \circ {\psi}^{-1}) ] \circ \psi 
\]
for all $f \in \Ginf (U) $. Also, for $\au = 1,2,\dots , q$ define
\[ \frac{\Par}{\Par v^{\au}} : \Ginf (U) \rar \Ginf (U), \ \ where \ \ 
\frac{\Par f}{\Par v^{\au}} \equiv [G_{p+\au}(f \circ {\psi}^{-1}) ] \circ \psi \] 
for all $f \in \Ginf (U) $. These are the coordinate derivatives.
\end{definition}

\begin{prop} \label{P:twoPtwentysix}
Let $(U,\psi )$ be a chart on a $\Ginf $ supermanifold $\Mcal $ of supermanifold dimension $(p|q)$. The coordinate derivatives are pure $\Ginf $ vector fields on U. In particular, for $m=1,2,\dots ,p$ $\Par {/} \Par u^m \in D^{1}(U)_{0} $, and for 
$\au = 1,2,\dots ,q$ $\Par {/} \Par v^{\au} \in D^{1}(U)_{1} $. In short, $\Par {/} \Par u^M \in D^{1}(U)_{{\eps}_{M}} $ for $M=1,2,\dots , p+q $. 
\end{prop}

\begin{prop} \label{P:twoPtwentyseven}
Let $(U,\psi )$ be a chart on a $\Ginf $ supermanifold $\Mcal $ where $\psi = (u^{1},\dots , u^{p},v^{1},\dots , v^{q})$, 
(1.) \ $D^{1}(U)$ is a graded left $\Ginf (U) $ module. \\
(2.) \ $D^{1}(U)$ is a free left $\Ginf (U) $ module with basis $\{ \Par {/} \Par u^{M} \}$ for $M=1,2,\dots p+q$. \\
The definition of a graded left $\Ginf (U) $ module is analogus to that of the graded left $\Lm $ module with supernumbers being replaced with $\Ginf $ functions.
\end{prop}

\section{Supermanifolds Viewed as Banach Manifolds}


Let $\Mcal$ be a supermanifold. Then one has a maximal $\Ginf$-atlas $\Am$ on $\Mcal$ such that for $\phi,\psi \in \Am,$  $\phi \circ {\psi}^{-1} : \psi (U \cap V ) \rar \phi (U \cap V)$ is a $\Ginf$ mapping from an open subset $\psi (U \cap V)$ of $\Kpq$ of $\Kpq$ to another open subset $\phi (U \cap V )$ . By Proposition 2.8 of  $\cite{Rsuperman}$  $\phi \circ {\psi}^{-1}$ is also a $\Cinf$ map.
 
\begin{prop}  \label{P:threePone}
If $\Mcal $ is a supermanifold with $\Ginf $-atlas $\Am $, then $\Mcal $ is also a Banach manifold relative to the unique maximal $\Cinf $-atlas, containing $\Am $. We denote this Banach manifold by $(\BM , \Abm)$ where, as sets $\BM=\Mcal $ and where $\Abm $ is the maximal $\Cinf $-atlas containing $\Am $.
\end{prop}

We will use $\BM $ when we wish to emphasize the Banach manifold structure of $\Mcal $. In practice we will work with the subatlas $\Am$ of the maximal atlas of the Banach manfold $\BM $ since it has the additional $\Ginf $ structure. \\

Recall that one definition of what it means to say $v$ is tangent to a Banach manifold is the which follows (see $\cite{Lang}$)

\begin{definition}
Let $M$ be a Banach manifold modeled on a Banach space $B$. We say that $v$ is tangent to $M$ at $x \in M$ and write $v \in T_{x}M$ iff $v$ is a mapping from the set of all $\Cinf$ charts of $M$ at $x$ into $B$ such that if $(U,\psi)$ and $(V,\phi)$ are $\Cinf$ charts of $M$ at $x$ then
\[ v( \psi ) = d_{\phi (x)}(\psi \circ \ \phiI ) (v( \phi ) ).\]
\end{definition}

\begin{remark} \label{E:tangent}
A tangent vector $v$ is uniquely determined by the latter transfomation law and its values on an atlas of $M$. So to define a tangent vector $v$ to $M$ at $x$ it suffices to define $v$ at all those charts of some atlas of $M$ which contain $x$ in their domain.
\end{remark}

We find the following slight modification of Roger's definition in $\cite{Rsuperman}$ to be useful in our context.

\begin{definition}
Let $\Mcal$ be a supermanifold and $x \in \Mcal$. We say that $v$ is a tangent to $\Mcal $ at $x$ and write $v \in T_{x}\Mcal $ iff $v$ is a mapping from $\Ginf (x) $ to $\Lm $ such that for some open set $U \subseteq \Mcal $ such that $x \in U$ and for some $\Ginf$ vector field $X \in  D^{1}(U)$,
\[ v(f) = X(f)(x) \]
for all $f \in \GinfU $. We say that $v$ is { \bf even} and write $v \in \TxzM$ iff $v( \Ginf (x)_{\eps}) \subseteq {}^{\eps}\Lm$ for $\eps = 0,1$. Likewise, $v$ is {\bf odd} and write $v \in \TxoM$ iff $v( \Ginf (x)_{\eps}) \subset {}^{\eps+1}\Lm$ for $\eps = 0,1$.
\end{definition}

Note that $\TxM $ is a graded vector space with $\TxM = \TxzM \oplus \TxoM $. Moreover $\TxM $ is a left $\Lm$-module which is called the {\bf tangent module} at $x \in \Mcal$.



\begin{definition}
Let $\Mcal$ and $\Ncal$ be supermanifolds and $g: \Mcal \rar \Ncal$ a class $G^{1}$ function we define $\dxg : \TxM \rar T_{g(x)}\Ncal$ by,
\begin{equation} \label{E:ten}
\begin{array}{l}
\dxg (\Xx)(f) \equiv \Xx(f \circ g)
\end{array}
\end{equation}
for all $f \in \Ginfgx$ and $\Xx \in \TxM$. 
\end{definition}
\begin{prop} \label{P:threePsix} 
Let $\Mcal$ and $\Ncal$ be supermanifolds and $g: \Mcal \rar \Ncal$ then $\dxg : \TxM \rar T_{g(x)}\Ncal$ is a parity preserving (even) right linear transformation, that is $\dxg \in L^{-}(\TxM,T_{g(x)}\Ncal)$.
\end{prop}
This follows from the fact that the parity of a composite function is determined as follows,
\begin{equation} \label{E:eleven}
\begin{array}{l}
f \circ g \in \eGinfU \iff f \in \eGinf.
\end{array}
\end{equation}
Thus, the parity of $g$ does not determine the parity of $f \circ g : \Mcal \rar \Lm$. This means that $\dxg$ is always parity preserving; for $y=g(x)$
\begin{equation} \label{E:twelve}
\begin{array}{l}
\dxg ( \TxeM ) \subseteq \TyeN .
\end{array}
\end{equation}
If $(U,\psi)$ is a chart at $x$ of $\Am$ with $\psi = (x^1,\dots,x^p,{\theta}^1,\dots ,{\theta}^q ) $ and $U \subset g^{-1}(V)$ for some chart $(V,\phi) \in \An $ with
$\phi = (y^1,\dots,y^r,{\beta}^1,\dots ,{\beta}^s)$ then the matrix of $\dxg $ is,
\[ [ \dxg ]_{\phi,\psi} = 
\begin{pmatrix}
(d_x (y^{j} \circ g )(\frac{\Par}{\Par x^{i}}) & 
(d_x (y^{j} \circ g )(\frac{\Par}{\Par \Oa })  \\
(d_x ({\beta}^{\gm} \circ g )(\frac{\Par}{\Par x^{i}}) & 
(d_x ({\beta}^{\gm} \circ g )(\frac{\Par}{\Par \Oa })  \\
\end{pmatrix} 
\]
where we note that the local coordinate representative of the Frechet derivative is a Grassmann valued matrix. Also notice that the matrix has the usual block decomposition 
\[
\begin{pmatrix}
A & B \\
C & D \\
\end{pmatrix} 
\]
where $A,D$ have entries from $\zLm$ and $B,C$ have entries from $\oLm$. 
\begin{remark}
If $\Mcal $ is a supermanifold and $g:\Mcal \rar \Lm$ is a class $G^{1}$ mapping the $T_{g(x)}\Lm $ is identified
with $\Lm $ for each $x \in \Mcal $ and $d_{x}g$ is regarded as the mapping from $T_{x}\Mcal $ to $\Lm$ defined by
\[ (d_{x}g)(\Xx ) = \Xx (g).\]
Notice that in this case $d_{x}g$ is not always even, ${\epsilon}(g) = {\epsilon}(d_{x}g)$ for all $x \in \Mcal $.
\end{remark}
\noindent
Obviously our definition of a tangent vector $v \in \TxM $ depends on the vector field $X$ used in the definition. We examine this dependence in more detail. Assume that $U,V$ are open in $\Mcal $, that $x \in U \cap V$, that $X$ is a vector field on $U$, that $Y$ is a vector field on $V$, and that
\[ v(f) = X(f)(x), \qquad v(g) = Y(g)(x) \]
for all $f \in \GinfU$, $g \in \Ginf (V)$. Then
\[ X(f)(x) = Y(f)(x) \]
for all $f \in \Ginf ( U \cap V )$. Moreover if $(\mathcal{O}, \psi )$ is a chart of $\Mcal $ at $x$ then, on $\mathcal{O} \cap U \cap V$,
\[ X = \sumApq {X}^{A}_{\psi} \pzA , \qquad Y = \sumApq {Y}^{A}_{\psi} \pzA \]
where $\psi = (z^{1},z^{2},\dots,z^{p+q})$ and where ${X}^{A}_{\psi},{Y}^{A}_{\psi}$ are $\Ginf $ maps from
$\mathcal{O} \cap U \cap V$ into $\Lm $. Moreover
\[ \sumApq {X}^{A}_{\psi}(x) \pfzA (x) = X(f)(x) = Y(f)(x) = \sumApq {Y}^{A}_{\psi}(x) \pfzA (x) \]
for all $ f \in \Ginf (\mathcal{O} \cap U \cap V)$. If we choose $f=z^{B}$, $1 \leq B \leq p+q$, we see that
\[ {X}^{B}_{\psi}(x) = {Y}^{B}_{\psi}(x) \]
for all $B$. \\

Notice that if $\Mcal $ is a supermanifold  then $T\Mcal = {\cup}_{p\in \Mcal} T_{p}\Mcal $ may be given a supermanifold structure just as in the case for ordinary manifolds. This follows using the $\Ginf$ transformation laws relating  two sets of components of tangent vectors to $\Mcal $. \\

Observe that there exists a well-defined mapping ${\beta}_{x} : \TxM \rar \TxBM $ defined by
\[ {\beta}_{x}(v)(\psi ) = ({X}^{1}_{\psi}(x),{X}^{2}_{\psi}(x),\cdots , {X}^{p+q}_{\psi}(x)) \]
for $v \in \TxM $ and $\psi $ a chart of $\BM $. Notice that we have defined ${\beta}_{x}(v)$ only on charts of $\BM $ at $x$ but if we show that the appropriate transformation law holds then ${\beta}_{x}(v)$ has a unique extension to all charts of $\BM $ at x ( see Remark $\ref{E:tangent}$) and thus uniquely defines an element of $\TxBM $. With this in mind let $(U,\psi ),(V,\phi )$ be charts of $\BM $ at $x$, and observe that
\begin{equation} \notag
\begin{array}{ll}
{\beta}_{x}(v)(\psi ) &= ({X}^{1}_{\psi}(x),{X}^{2}_{\psi}(x),\cdots , {X}^{p+q}_{\psi}(x)) \\
                      &= d_{\phi (x)} (\psi \circ \phiI )
		                     ({X}^{1}_{\phi}(x),{X}^{2}_{\phi}(x),\cdots , {X}^{p+q}_{\phi}(x)) \\
		      &= d_{\phi (x)} (\psi \circ \phiI ){\beta}_{x}(v)(\phi ).
\end{array}
\end{equation}
\begin{prop} \label{P:threePeight}
If $\Mcal $ is a supermanifold and $x \in \Mcal $ then ${\beta}_{x}$ is a $\zLm $-linear vector space isomorphism from $\TxzM $ onto $\TxBM$.
\end{prop}
\noindent
\begin{proof}
It is clear that ${\beta}_{x}$ is a $\zLm $-linear vector space homomorphism. We show that ${\beta}_{x}$ is injective. Assume that $v \in \TxzM$ such that ${\beta}_{x}(v)=0$. Then there is an open set $U \subseteq \Mcal $ and a vector field $X$ on $U$ such that $x \in U$, $v(f)=X(f)(x)$ for $f \in \GinfU$ and 
$0 = {\beta}_{x}(v)(\psi) = ({X}^{1}_{\psi}(x),{X}^{2}_{\psi}(x),\cdots , {X}^{p+q}_{\psi}(x))$ for all charts $\psi $ of $\Mcal $ at $x$. Thus $X=0$ and $v(f)=0$ for all $f \in \GinfU $. It follows that $v$ is zero on the germ $\Ginf (x)$ and ${\beta}_{x}$ is injective. \\

We now show that ${\beta}_{x}$ is surjective. Let $X_{x} \in \TxBM $ and recall that $X_{x}$ is a mapping from the set of all charts of $\BM $ into $B=\Kpq $. We want to find $v \in \TxM $ such that ${\beta}_{x}(v) = X_{x}$. First we need to find a vector field defined on an open subset of $\Mcal $ about $x$ which agrees with $X_{x}$ on charts of $\Mcal $. Choose any chart $(U,\psi )$ of $\Mcal $ at $x$. Then $X_{x}(\psi ) \in B=\Kpq $ and we can define a constant vector field $Y$ on $U$ by
\[ Y = \sumApq {X}^{A}_{x}(\psi) \pzA \]
where $\psi = ( z^{1},z^{2},\dots , z^{p+q} )$. Thus the functions ${Y}^{A}_{\psi } : U \rar \Lm $ are the constant functions ${Y}^{A}_{\psi }(u) \equiv {X}^{A}_{x}(\psi )$ for all $u \in U$. Notice that $Y \in {D}^{1}(U)_{0}$. Define $v : \Ginf (x) \rar \Lm $ by 
\[ v(f) = Y(f)(x) = \sumApq {X}^{A}_{x}(\psi) \pfzA (x). \]
Then for any chart $(V , \phi )$ of $\Mcal $ at $x$
\begin{equation} \notag
\begin{array}{ll}
{\beta}_{x}(v)(\phi ) &= ({Y}^{1}_{\phi}(x),{Y}^{2}_{\phi}(x),\cdots , {Y}^{p+q}_{\phi}(x)) \\
                      &= d_{\psi (x)} (\phi \circ \psiI )
		                     ({Y}^{1}_{\psi}(x),{Y}^{2}_{\psi}(x),\cdots , {Y}^{p+q}_{\psi}(x)) \\
		      &= d_{\psi (x)} (\phi \circ \psiI )
		                     ({X}^{1}_{x}(\psi ),{X}^{2}_{x}(\psi ),\cdots , {X}^{p+q}_{x}(\psi )) \\
		      &= d_{\psi (x)} (\phi \circ \psiI )(X_{x}(\psi ) ) \\
		      &= X_{x} (\phi ).
\end{array}
\end{equation}
Thus ${\beta}_{x}(v)(\phi ) = X_{x} (\phi )$ for all charts of $\Mcal,$ but since the charts of $\Mcal $ form a subatlas of the manifold structure of $\BM $, ${\beta}_{x}(v)$ can be uniquely extended to agree with $X_{x} $ at every chart of $\BM $. Thus ${\beta}_{x}$ is surjective. The proposition follows. 
\end{proof}

The mapping ${\beta}_{x}$ induces a mapping of vector fields as follows. Recall that a vector field on a Banach manifold $M$ is uniquely determined by defining a function $Y$ from charts $(U, \psi ) $ of $M$ into $\Cinf $-maps from $U$ into the Banach space $B$ on which $M$ is modeled. Of course if $(U,\psi )$ and $(V, \phi )$ are charts of $M$ such that $U \cap V \ne \emptyset $ the usual transformation holds,
\[ Y(\psi )(x) = d_{\phi (x)}(\psi \circ \phiI ) (Y (\phi )(x) )\]
for all $x \in U \cap V$. \\

Note that if $\mathcal{O} \subseteq \Mcal $ is open and $X \in {D}^{1} (\mathcal{O}),$ then for each $x \in \mathcal{O}$ we may define $X_{x} \in \TxM $ by
\[ X_{x} (f) = X(f)(x) \]
for all $f \in \Ginf (W)$ where $W$ is open and $x \in W \subseteq \mathcal{O}$. Thus if $(U, \psi )$ is a chart of $\Mcal $ at $x$,
\[ {\beta}_{x} (X_{x} )(\psi ) = ({X}^{1}_{\psi} (x),{X}^{2}_{\psi} (x),\cdots , {X}^{p+q}_{\psi} (x) ) \]
and the mapping $\beta (X)(\psi )$ given by $x \mapsto {\beta}_{x} (X_{x} )(\psi )$ is a $\Ginf $ function from $V$ into $B=\Kpq $. Since $\Ginf $ maps are necessarily $\Cinf $ maps we see that $\beta (X) $ is a {\bf vector field} on $\BM $ since as a maps of charts of $\Mcal $ it transforms correctly and thus can be extended to all charts of $\BM $.\newline

Thus we can write $v = {\sum}_{A=1}^{p+q}{X}_{\psi}^{A}(x) \Par / \Par z^{A}$ where 
$({X}_{x}^{1}(x),{X}_{x}^{2}(x),\dots ,{X}_{x}^{p+q}(x)) \in \Kpq $. If $\phi $ is another chart $\Ginf $ related to $\psi $ and $\phi = (w^{1},w^{2},\dots , w^{p+q} )$ then we can also write $v = {\sum}_{B=1}^{p+q}{X}_{\phi}^{B}(x) \Par / \Par w^{A}$. Moreover as in the classical case,
\[ ({X}_{\psi}^{1}(x),{X}_{\psi}^{2}(x),\dots ,{X}_{\psi}^{p+q}(x)) =d_{\phi (x)} ( \psi \circ {\phi}^{-1} )
({X}_{\phi}^{1}(x),{X}_{\phi}^{2}(x),\dots ,{X}_{\phi}^{p+q}(x)). \]

\noindent
Because the Banach space $B=\Kpq$ is a $\Lambda^0$ module, vector fields on $\BM $  have a $\zLm $-module structure.

\begin{corollary}
If $\mathcal{O} \subseteq \Mcal $ is an open subset of a supermanifold $\Mcal $ then $\beta $ is a $\zLm $-linear vector space injection of the $\zLm $-module of all vector fields $D^{1}(\mathcal{O} )$ on $\mathcal{O}$ into the $\zLm $-module of $\Cinf $-vector fields of the Banach manifold $\mathcal{O} \subset \BM $.
\end{corollary}

The mapping $\beta $ is not surjective since for $X \in D^{1}(\mathcal{O} )$ and for each chart $(U, \psi )$ of $\Mcal $, $\beta (X)(\psi ): U \rar \Kpq $ is a $\Ginf $-mapping and not every $\Cinf $-vector field on $\mathcal{ O} \subset \BM $ has this property. \\

Recall that if $U \subseteq \Kpq = \BB $ is open and $\tilde f: U \rar \Lm$ is a class $C^{k}$ mapping, then its p-fold Frechet derivative is a mapping from $U$ into symmetric multilinear maps from ${\BB}^{k} = \BB \times \BB \times \cdots \times \BB$ into $\Lm$. Thus for $x \in U$
\begin{equation} \notag
\begin{array}{l}
 {d}_{x}^{p}\tilde f:\BB \times \BB \times \cdots \times \BB \rar \Lm 
\end{array}
\end{equation}
is symmetric. It is obtained by iterating the Frechet derivatives, for example,
\begin{equation} \notag
\begin{array}{l}
{d}_{x}^{2}\tilde f(v,w) = d_{x}(y \rar (d_{y}\tilde f)(w))(v). 
\end{array}
\end{equation}

\begin{definition}
 Let $\Mcal$ and $\Ncal$ be supermanifolds and $f:\Mcal\rightarrow \Ncal$ a class $C^p$-mapping from $\BM$ into $\BN.$  Define a mapping $d_x^pf$ by
\begin{equation} \notag
d_{x}^pf : \TxBM \times \TxBM \times \cdots \times \TxBM \rar T_{f(x)}\Ncal
\end{equation}
where
\begin{equation} \notag
{d}_{x}^{p}f(v_1,v_2,\dots,v_p) = (d_{\phi(f(x))}{\phi}^{-1})\bigl({d}_{\psi(p)}^{p}(\phi \circ f \circ {\psi}^{-1})
({d}_{x}\psi(v_1),{d}_{x}\psi(v_1),\dots,{d}_{x}\psi(v_p))\bigr)
\end{equation}
and where  $(U,\psi)$ is any chart  $\Mcal$ and  $(V,\phi)$ is any  chart of $\Ncal$ such that $f^{-1}(V) \subseteq U.$ 
Since $\TxzM \subseteq \TxBM$ for each $x,$ notice that there is an induced mapping
\begin{equation} \notag
{d}_{x}^{p}f: \TxzM \times \TxzM \times \cdots \times \TxzM \rar T_{f(x)}\Ncal.
\end{equation} 
\end{definition}


\begin{theorem} \label{E:Btheorem}
Let $\Mcal$ and $\Ncal$ be supermanifolds
of dimension $(p|q)$ and $(r|s)$ respectively and let $f:\BM \rar \BN$ be a $\Cinf$ function. The function $f:\Mcal \rar \Ncal$ is a class $G^{l}$ function iff for every chart $(U,\psi)$ of $\Mcal$
and $(V,\phi)$ of $\Ncal $ such that $f^{-1}(V) \subset U$ there exist functions ${b}_{A_1 \dots A_{k}}^{\psi J}$ with $1\leq A_1 \dots A_{k} \leq p+q$, $1\leq J \leq r+s$ and $1<k\leq l$, such that 
\begin{equation} 
\begin{array}{ll} \notag
(1) &\text{ each function } {b}_{A_1 \dots A_{k}}^{\psi J} \text{ is in } G^{0}(U), and \\
(2) &\text{ for } x \in U \text{ and } X_1,X_2,\dots,X_k \in \TxzM, \\ 
    & {d}_{x}^{k}({\phiJ }\circ f )(X_1,\dots,X_k) = 
    \sumAonepq \cdots \sumAkpq {X}_{1}^{A_1} \cdots {X}_{k}^{A_k}
    {b}_{A_1 \dots A_{k}}^{\psi^{J}}(x). \ \ 
\end{array}
\end{equation}
\end{theorem}

\noindent
\begin{proof}
If $f$ is of class $G^{k}$ for $k \leq l$ then for charts $\psi,\phi$ of $\Mcal,\Ncal$ respectively
$\phi \circ f \circ {\psi}^{-1}$ is of class $G^{k}$. By Proposition 2.8 of $\cite{Rsuperman}$, where the partials are
of class $G^{0}$,
\begin{equation} \notag
{d}^k_{\psi(x)}({\phi}^{J}\circ f \circ {\psi}^{-1})(v_1,\dots,v_k) = \sumAApq {v}^{A_1} \cdots {v}^{A_k}
\frac{ {\Par}^{k} ( {\phi}^{J} \circ f \circ {\psi}^{-1} ) } { \Par u^{A_{k}} \cdots \Par u^{A_1} } (\psi (x))
\end{equation} 
for $1\leq J \leq r+s$ and $v_1,v_2,\dots,v_k \in \Kpq$. We identify $v_i$ with
$d_{x}\psi (X_{i})$ for arbitrary given $X_{1},X_{2},\dots,X_{k} \in \TxzM$ so that
\begin{equation} \notag
{d}^k_{x}({\phi}^{J}\circ f)(X_1,\dots,X_k) = \sumAApq {X}^{A_1} \cdots {X}^{A_k}
\frac{ {\Par}^{k} ( {\phi}^{J} \circ f ) } { \Par z^{A_{k}} \cdots \Par z^{A_1} } (x)
\end{equation} 
where $z^{A} \equiv {P}^{A} \circ \psi $ (recall that $P^A$ is the projection of $\Kpq$ onto its $A$-th factor). Thus we note that,
\begin{equation} \notag
{b}_{A_1 \dots A_{k}}^{\psi J}(x) = \frac{ {\Par}^{k} ( {\phi}^{J} \circ f ) } { \Par z^{A_{k}} \cdots \Par z^{A_1} } (x)
\qquad \text{for } x \in U.
\end{equation}
and (1) and (2) hold. 

Conversely, assume the existence of the functions ${b}_{A_1 \dots A_{k}}^{\psi^{J}}: U \rar \Lm$ with $k \leq l$  that satisfy conditions (1) and (2) above. We show $f$ is of class $G^{k}$ for all $k \leq l$. \\

Begin with the case $k=1$. Let $\psi,\phi$ of $\Mcal,\Ncal$ respectively and choose $U$ open in $\Mcal$ small enough so that $\phi \circ f \circ {\psi}^{-1}$ is defined on the open set $\psi(U)$. By hypothesis we have for $X \in \TxzM$ and $x\in U$,


\begin{equation} \label{E:starpg6}
d_{x}({\phiJ }\circ f)(X) = \sumApq X^{A} {b}_{A}^{{\psi J}}(x)
\end{equation} 
where ${b}_{A}^{{\psi} J}(x) \in G^{0}(U) $. Thus there are supernumbers ${b}_{A}^{{\psi J}}(x)$ that encode the Frechet derivative of $({\phi}^{J}\circ f)$ at $x$. Moreover, if we identify $H$ with $d_{x}\psi (X)$ we find from eq.($\ref{E:starpg6}$), 
\begin{equation} 
d_{\psi(x)}({\phi}^{J}\circ f \circ {\psi}^{-1})(H) = \sumApq H^{A} {b}_{A}^{{\psi J}}(\psi(x)).
\end{equation} 
This identity implies that ${\phi}^{J} \circ f \circ {\psi}^{-1}$ is of class $G^{1}$ on $\psi(U)$ for each $J$ and that $ {b}_{A}^{{\psi J}}(x)=G_A({\phi}^{J} \circ f \circ {\psi}^{-1})(\psi(x))$ in the notation of $\cite{Rsuperman}$. Hence ${\phi} \circ f \circ {\psi}^{-1}$ is of class $G^{1}$ on $\psi(U)$  and therefore, $f$ is of class $G^1$ on $U$. The case $k=1$ is proved. \\

Next we prove the case $k=2$. Consider the mapping $F$ from $\psi (U)$ to $\Lm$ defined by $F:y \mapsto d_{y}({\phi}^{J} \circ f \circ {\psi}^{-1})(V_2)$ where $V_2 \in \Kpq$ is given by $V_2 = d_x \psi(X_2 )$ for an arbitrary, but fixed, $X_2 \in \TxzM$. Then by construction, $V_2$ is an arbitrary element of the Banach space $B=\Kpq$ which does not change as $y$ changes. Consider the Frechet derivative of $F$ at $u=\psi(x)$. We Have for $X_1 \in \TxzM$ and $V_1 = d_x \psi(X_1)$ and, 
\begin{equation} \label{E:dFpg6}
\begin{array}{ll} 
d_u F (V_1) &= d_u (d ( {\phi}^{J} \circ f \circ {\psi}^{-1})(V_2))(V_1) \\
            &= {d}^{2}_{u}({\phi}^{J} \circ f \circ {\psi}^{-1})(V_1,V_2) \\
	    &= {d}^{2}_{\psi(x)}({\phi}^{J} \circ f \circ {\psi}^{-1})(d_x\psi(X_1),d_x\psi(X_2)) \\
	    &= {d}^{2}_{x}({\phi}^{J} \circ f)(X_1,X_2) \\
	    &= \sumAonepq \sumAtwopq {X}_{1}^{A_1}{X}_{2}^{A_2}{b}_{A_{1}A_{2}}^{{\psi J}}(x) \\
	    &= \sumAonepq \sumAtwopq {V}^{A_1}_{1}{V}^{A_2}_{2}{b}_{A_{1}A_{2}}^{{\psi J}}(x).
\end{array}
\end{equation} 

\noindent Since we have already shown that $f$ is of class $G^1$ on $U$ we have that 
$d_y({\phi}^{J} \circ f \circ {\psi}^{-1})(V_2)=\sumAtwopq {V}_{2}^{A_2} \PfJrep (y).$
From the definition of $F$ we note
\begin{equation} 
F(y)=\sumAtwopq {V}_{2}^{A_2} \PfJrep (y).
\end{equation} 
Thus for fixed $V_2$, we have
\begin{equation} \label{E:dFpg7}
d_u F(V_1)=\sumAtwopq (-1)^{{\epsilon}_{A_2}\epsilon(V_1)}{V}_{2}^{A_2} d_u \PfJrep (V_1).
\end{equation} 
And so, comparing eq.($\ref{E:dFpg6}$) and eq.($\ref{E:dFpg7}$) we find,
\begin{equation} 
\sumAtwopq (-1)^{{\epsilon}_{A_2}\epsilon(V_1)} {V}_{2}^{A_2} d_u \PfJrep (V_1) = \sumAonepq \sumAtwopq {V}^{A_1}_{1}{V}^{A_2}_{2}{b}_{A_{1}A_{2}}^{{\psi J}}(x).
\end{equation} 
Thus,
\begin{equation} 
\sumAtwopq  (-1)^{{\epsilon}_{A_2}\epsilon(V_1)}{V}_{2}^{A_2} d_u \PfJrep (V_1) = \sumAtwopq {V}^{A_2}_{2} \sumAonepq {(-1)}^{{\epsilon}_{A_1} {\epsilon}_{A_2}} {V}^{A_1}_{1}{b}_{A_{1}A_{2}}^{{\psi J}}(x).
\end{equation} 
This holds for all $V_{2}$ so,
\begin{equation} 
 d_u \PfJrep (V_1) = \sumAonepq {(-1)}^{{\epsilon}_{A_1} {\epsilon}_{A_2}+{\epsilon}_{A_2}\epsilon(V_1)} {V}^{A_1}_{1}{b}_{A_{1}A_{2}}^{{\psi J}}(x).
\end{equation} 
It follows that $\PfJrep$ is of class $G^1$ on $\psi(U) \subseteq \Kpq$, and 
for $u \in \psi (U)$. Since $\frac{ \Par ( {\phi}^{J} \circ f \circ \psiI )}{\Par u^{A_{2}}} $ is of class $G^{1}$ for each $A$, ${\phi}^{J} \circ f \circ \psiI $ is of class $G^2 $ on $\psi (U)$ for each $J$. Thus $f$ is of class $G^2 $ on $U$.  
An inductive argument using similar computations will show that $f$ is of class $G^k$ for all $k \leq l$ 
\end{proof} 


\noindent 
\begin{remark}
Given a $\Cinf $-mapping $f:\Mcal \rar \Ncal $ as in the theorem above we have conditions under which $f$ is of class $\Ginf $. One begins with maps
\begin{equation} 
{d}_{x}^{k}f : \TxBM \times \cdots \times \TxBM \rar \TfxN.
\end{equation} 
Then since $\TxzM \times \cdots \times \TxzM \subseteq \TxBM \times \cdots \times \TxBM $ one has a mapping on even vectors $X_1,X_2,\dots X_k \in \TxzM $. Moreover, one obtains the formula for even vectors
\begin{equation} 
{d}_{x}^{k}f(X_1,X_2,\dots,X_k) = \sumAApq ({X}^{A_1}_{1}{X}^{A_2}_{2}\cdots {X}^{A_k}_{k})
\biggl( \frac{ {\Par}^k f}{{\Par z}^{A_k} \cdots {\Par z}^{A_2}{\Par z}^{A_1}} \biggr)(x).
\end{equation} 
It now follows that this mapping can be extended to a mapping from $\TxM \times \TxM \times \cdots \times \TxM $ to $\TfxN$ where the components of pure tangent vectors $X_1,X_2,\dots,X_k$ may be in $\Kpq$ or possibly in $({\oLm})^{p} \times ({\zLm})^{q}$. If the vectors $X_1,X_2,\dots,X_k$ are not of definite parity then the components $({X}^{A_1}_{1},{X}^{A_2}_{2},\dots , {X}^{A_k}_{k})$ will reside in ${\Lm}^{k}$ in general. As an example, consider the case k=1. Observe that
\begin{equation} 
d_{x}f(X) = X_{x}(f) = \sumApq X^{A} \frac{ \Par f}{\Par z^A}(x)
\end{equation} 
makes sense for even and odd vectors $X_x \in \TxM$. It is interesting that the operation of $d_{x}f$ on $\TxzM$ defines its operation on the other half of $\TxM$ namely $\TxoM$. \\ 

In other words, a supermanifold ${\cal M}$ is modeled on $\Kpq $ but the $\Ginf $-tangent module "doubles the dimension". Even vectors are summed over {\bf ALL} of the even and odd coordinate vector fields (expanded against even and odd components in order that the vector field be even); so to have the derivative of some map defined for even vector fields actually means the map is defined on the coordinate vector field basis of the tangent module. This is why we can extend the derivative to act on both even and odd vector fields. Even vector fields have the $(p|q)$ data hidden in them, the tangent module at a point is the direct sum of the Banach space $\Kpq$ on which ${\cal M}$ is modeled and the Banach space  $({\oLm})^{p} \times ({\zLm})^{q}.$
\end{remark}


\begin{definition} Let $\Mcal$ be a supermanifold and $\vfrak$ a Banach supervector space. Provide $\vfrak^0$ with the supermanifold structure obtained by defining the obvious single global chart obtained from a basis of $\vfrak.$ Let $f$ denote a smooth function from $\Mcal$ into $\vfrak^0$ and let $\{f^B\}$ denote its components relative to a pure basis of $\vfrak$. We define the higher derivatives of $f^B$ at $w\in \Mcal$ inductively as follows. Define $d_wf^B:T_w\Mcal \rightarrow \Lambda$ by $d_wf^B(X)=X(f^B)$
for $X\in T_w\Mcal.$
Define $d_w^{k+1}f^B:T_w\Mcal \times T_w\Mcal \times \cdots T_w\Mcal \rightarrow \Lambda$ by 
$d_w^{k+1}f^B(X_1,X_2,\cdots,X_{k+1})=d_w[d^kf^B(X_2,X_3,\cdots ,X_{k+1})](X_1)$ for $X_1,X_2, \cdots,X_{k+1}\in T_w\Mcal.$ Here $d^kf^B(X_2,X_3,\cdots ,X_{k+1})$ denotes the function from $\Mcal$ into $\Lambda$ defined by $x\rightarrow d^k_xf^B(X_2,X_3,\cdots ,X_{k+1}).$
\end{definition}

We now consider an important special case of these ideas which we find useful in the last section of the paper. Let $\gfrak$ and $\vfrak$ denote  Banach super vector spaces. Consider $\gz$ as a supermanifold with a single global chart $\psi:\gz \rightarrow \Kpq$ whose components are defined by $\psi(x)=(u^1(x),u^2(x),\cdots,u^{p+q}(x)),$  $  x\in \gz.$ For each $x\in \gz, T_x\gz$ may be identified with $\gfrak$ by identifying the basis $\{u^B\}$ of $T_x\gz$ with a given fixed basis $\{e_B\}$ of $\gfrak.$ Similarly, choose a single coordinate chart on $\vfrak^0.$ Moreover if $f$ is a function from $\gz$ to $\vfrak^0,$ then denote its components relative to the chart on $\vfrak^0$ by the functions $f^B:\gz \rightarrow \Lambda.$ Recall that if $f$ is of class $C^{\infty},$ then it is also of class $G^{\infty}$ iff each  component function $f^B$ is of class $G^{\infty}.$ Notice that the components $f^1,f^2,\cdots f^p$ are all even while $f^{p+1},f^{p+2},\cdots,f^{p+q}$ are all odd. Also notice that the derivatives $d^k_wf^B$ of each component function are maps from $\gfrak^k=\gfrak \times \cdots \times \gfrak$ to $\Lambda$
at each $w\in \gz$ due to the identification of $\gfrak$ with $T_w\gz.$
 
\begin{definition} Let $\gfrak$ be a supervector space with basis $\{e_B\}$ and let $\beta :\gfrak^k\rightarrow \Lambda.$ We say that $\beta$ is multi-linear  over $\gz$ iff for some, pure basis $\{ e_{B} \}$ of $\gfrak$, 
\[ \beta(v_1,v_2,\cdots,v_k)=v_1^{A_1}v_2{^{A_2}}\cdots v_k^{A_k}\beta(e_{A_k},\cdots, e_{A_2},e_{A_1})\]
for $v_1,v_2,\cdots,v_k\in \gz.$
\end{definition}

\noindent
Notice that one must require that $\beta$ be defined on all of $\gfrak^k$ rather than $(\gz)^k,$ since it must be possible to evaluate $\beta$ at arbitrary elements of a basis of $\gfrak.$
This is also the case for higher derivatives such as $d^kf^B$ as defined above. This shows up explicitly in the proof of the following proposition.

\begin{prop} \label{P:threePfourteen}
Let $\gfrak$ and $\vfrak$ denote  Banach super vector spaces and $f: \gz \rightarrow \vfrak^0$ a $C^{\infty}$ function. Then $f$ is of class $G^{\infty}$ iff for each $x\in \gz$ and each positive integer $k,  d^k_xf^B : \gfrak^k\rightarrow \Lambda$ is multi-linear over $\gz$ for each component $f^B$ of $f.$ 
\end{prop} 

\begin{proof}
Assume first that $f:\gz \rightarrow \vfrak^0$ is of class $G^{\infty}$ and that $f^B$ is a component of $f.$ 
Choose a pure basis $\{ e_{B} \}$ of $\gfrak $ and define $u^{B}$ on $\gz $ by $u^{B}(\sum a^{K}e_{K}) = a^{B}$. Regard the $(u^{B})$ as coordinates on $\gz $. We first show for $x\in \gz$ and $v_1,v_2,\cdots v_k\in \gz,$  that
 $$d^k_xf^B(v_1,v_2,\cdots,v_k)=
v^{A_1}_1v^{A_2}_2\cdots v_k^{A_k}\frac {\partial^k f}{\partial u^{A_k}\cdots \partial u^{A_2}\partial u^{A_1}}(x).$$
The proof proceeds by induction. First observe that $d_xf^B(v)=v(f^B)=v^A\frac{\partial f^B}{\partial u^A}$ so the result is true for $k=1$ Now assume the result for arbitrary $k$ and we show that 
$$d^{k+1}_xf^B(v_1,v_2,\cdots,v_{k+1})=v^{A_1}_1v^{A_2}_2\cdots v_{k+1}^{A_{k}+1}\frac {\partial^{k+1} f}{\partial u^{A_{k+1}}\cdots \partial u^{A_2}\partial u^{A_1}}(x).$$ 
By definition 
$$d^{k+1}_xf^B(v_1,v_2,\cdots,v_{k+1})=d_x[d^{k}f^B(v_2,v_3,\cdots,v_{k+1})](v_1) =$$$$
d_x[v^{A_2}_2v^{A_3}_3\cdots v_{k+1}^{A_{k+1}}\frac {\partial^k f}{\partial u^{A_{k+1}}\cdots \partial u^{A_3}\partial u^{A_2}}](v_1)
=v_1^{A_1}\frac{\partial}{\partial u^{A_1}}[v^{A_2}_2v^{A_3}_3\cdots v_{k+1}^{A_{k+1}}\frac {\partial^k f}{\partial u^{A_{k+1}}\cdots \partial u^{A_3}\partial u^{A_2}}].$$
Now the partial derivative $\frac{\partial}{\partial u^{A_1}}$ can be pushed through the term
$v^{A_2}_2v^{A_3}_3\cdots v_{k+1}^{A_{k+1}}$ but in doing so it produces a sign change 
$\varepsilon=(-1)^{\varepsilon(A_1)\varepsilon(A_2)}(-1)^{\varepsilon(A_1)\varepsilon(A_3)}\cdots 
(-1)^{\varepsilon(A_1)\varepsilon(A_{k+1})}.$ Thus one obtains 
$$d^{k+1}_xf^B(v_1,v_2,\cdots,v_{k+1})=\varepsilon [v^{A_1}v^{A_2}_2v^{A_3}_3\cdots v_{k+1}^{A_{k+1}}\frac {\partial^k f}{\partial u^{A_1}\partial u^{A_{k+1}}\cdots \partial u^{A_3}\partial u^{A_2}}].$$ 
Now one must permute the order of the partials but one finds that 
$$\frac {\partial^k f}{\partial u^{A_1}\partial u^{A_{k+1}}\cdots \partial u^{A_3}\partial u^{A_2}}=
\varepsilon [\frac {\partial^k f}{\partial u^{A_{k+1}}\cdots \partial u^{A_3}\partial u^{A_2} \partial u^{A_1}}].$$
The two signs cancel to give the desired result
$$d^{k+1}_xf^B(v_1,v_2,\cdots,v_{k+1})=v^{A_1}_1v^{A_2}_2\cdots v_{k+1}^{A_{k}+1}\frac {\partial^{k+1} f}{\partial u^{A_{k+1}}\cdots \partial u^{A_2}\partial u^{A_1}}(x).$$ 
This finishes the first part of the proof. 

To complete the proof we must show that for each positive integer $k,$
$$d_x^{k}f^B(\frac{\partial}{\partial u^{A_1}},\frac{\partial}{\partial u^{A_2}}, \cdots \frac{\partial}{\partial u^{A_{k}}})=\frac {\partial^k f}{\partial u^{A_{1}} \partial u^{A_2}\cdots \partial u^{A_{k}}}(x).$$
This proof also proceeds by induction. The result is obvious when $k=1,$ since
$d_xf^B(\frac{\partial}{\partial u^A})=\frac{\partial f^B}{\partial u^A}.$   Assume, inductively, that for some positive $k,$ 
$$d_x^{k}f^B(\frac{\partial}{\partial u^{A_2}},\frac{\partial}{\partial u^{A_3}}, \cdots \frac{\partial}{\partial u^{A_{k+1}}})=\frac {\partial^k f}{\partial u^{A_{2}} \partial u^{A_3}\cdots \partial u^{A_{k+1}}}(x).$$
By the definition of $d_x^{k+1}f^B$ we have
$$d_x^{k+1}f^B(\frac{\partial}{\partial u^{A_1}},\frac{\partial}{\partial u^{A_2}}, \cdots \frac{\partial}{\partial u^{A_{k+1}}})=d_x[d^{k}f^B(\frac{\partial}{\partial u^{A_2}},\frac{\partial}{\partial u^{A_3}}, \cdots, \frac{\partial}{\partial u^{A_{k+1}}}))(\frac {\partial}{\partial u^{A_1}}]$$
$$=d_x(\frac {\partial^k f}{\partial u^{A_{2}} \partial u^{A_3}\cdots \partial u^{A_{k+1}}})(\frac{\partial}{\partial u^{A_1}})=\frac {\partial^{k+1} f}{\partial u^{A_{1}} \partial u^{A_2}\cdots \partial u^{A_{k+1}}}(x)$$ and the result follows.
From these two results, we have that for for all $k$ and for $v_1,v_2,\cdots v_{k} \in \gz$
$$d^{k}_xf^B(v_1,v_2,\cdots,v_{k})=v^{A_1}_1v^{A_2}_2\cdots v_{k}^{A_{k}}\frac {\partial^k f}{\partial u^{A_{k}}\cdots \partial u^{A_2}\partial u^{A_1}}(x)$$$$=(v^{A_1}_1v^{A_2}_2\cdots v_{k}^{A_{k+1}})d_x^{k}f^B(\frac{\partial}{\partial u^{A_{k}}},\cdots,\frac{\partial}{\partial u^{A_2}},  \frac{\partial}{\partial u^{A_{1}}}).$$ Thus $d_x^{k}f^B$ is $k$-multi-linear and consequently if $f$ is of class $\Ginf,$ then all the derivatives of the components of $f$  are multi-linear over $\gz.$ 

Conversely, assume that all the derivatives of the components of $f$ are multi-linear over $\gz.$ We show that $f$ is of class $\Ginf.$ In fact the result is an immediate consequence of Theorem $\ref{E:Btheorem}$ since we have that $$d^{k}_xf^B(v_1,v_2,\cdots,v_{k})=(v^{A_1}_1v^{A_2}_2\cdots v_{k}^{A_{k}})d_x^{k}f^B(\frac{\partial}{\partial u^{A_{k}}},\cdots,\frac{\partial}{\partial u^{A_2}},  \frac{\partial}{\partial u^{A_{1}}})$$$$=v^{A_1}_1v^{A_2}_2\cdots v_{k}^{A_{k}}\frac {\partial^k f}{\partial u^{A_{k}}\cdots \partial u^{A_2}\partial u^{A_1}}(x)$$ and the hypothesis of Theorem $\ref{E:Btheorem}$ holds with 
$b^J_{A_1A_2\cdots A_k}=\frac {\partial^k f}{\partial u^{A_{k}}\cdots \partial u^{A_2}\partial u^{A_1}}.$
The proposition follows. 
\end{proof}

\section{Submanifolds of Supermanifolds}

\begin{definition}
Let $\Mcal $ be a $(p|q) $ supermanifold and $\Scal \subseteq \Mcal $. A chart of $(U,\psi )$ of $\Mcal $ is called an $(r|s)$-submanifold chart of $\Mcal $ relative to $\Scal $ iff
\[ \psi ( U \cap \Scal ) = \psi (U) \cap ( \Krs \times \zz ) \]
where $ (0,0) \in {\mathbb{K}}^{(p-r|q-s)} $. We say that $\Scal $ is a $(r|s)$ submanifold of $\Mcal $ iff for each $x \in \Scal $ there exists a $(r|s)$-submanifold chart $(U,\psi )$ of $\Mcal $ relative to $\Scal $ such that $x \in U$. There is a subtle point to be made here and that is that the definition depends on a specific splitting $\Kpq=\Krs \times {\mathbb{K}^{(p-r|q-s)}}.$ In general many such splittings are possible. In our definition we choose one specific splitting and all submanifold charts are required to respect this particular splitting.
\end{definition}

\noindent
\begin{remark}
If $\Scal $ is a $(r|s)$-submanifold of $\Mcal $ let $\As $ denote the set of all pairs $( U \cap \Scal , \psiS )$ such that there exists an $(r|s)$-submanifold chart $(\psi ,U)$ of $\Mcal $ relative to $\Scal $ such that $\Scal \cap U \ne \emptyset $ and $\psiS : U \cap \Scal \rar \Krs $ is defined in terms of $\psi $ by requiring that $\psiS $ be the restriction of $\psi $ to $U \cap \Scal $ composed with the obvious projection of $\Krs \times \zz $ to $\Krs $ which discards the $\zz \in {\mathbb{K}}^{(p-r|q-s)}$. It is obvious and well-known that if $(U,\psi ) $ and $(V,\phi )$ are such charts with $U \cap V \cap \Scal \ne \emptyset $ then
\[ \phiS \circ \psiSI : \psiS (U \cap V \cap \Scal ) \rar \psiS (U \cap V \cap \Scal ) \]
is a $\Cinf $ mapping. Thus $\Scal $ inherits a $\Cinf $-manifold structure from $\BM $ which we denote $\BS $ when we wish to emphasize that it is a Banach manifold. Moreover $\phiS \circ \psiSI $ is essentially the restriction of $\phi \circ \psiI : \psi ( U \cap V ) \rar \phi (U \cap V )$ to $\psi (U \cap V ) \cap ( \Krs \times \zz )$ which maps this set\\
to $\phi (U \cap V ) \cap ( \Krs \times \zz )$ and consequently it is easy to see that $\phiS \circ \psiSI $ is a $\Ginf $- mapping. Indeed the inclusion mapping 
\[ i : \Krs \hrar \Krs \times \zz \hrar \Kpq \]
is a $\Ginf $-mapping as is also its restriction $i_{Q}$ to the open set $Q=\psiS (U \cap V \cap \Scal ) \subseteq \Krs$.
For $1\leq i \leq r $ and $1 \leq \au \leq s $
\[ {\phiS}^{i} \circ \psiSI = {\phi}^{i} \circ \psiI \circ i_{Q} \qquad \text{and} \qquad
   {\phiS}^{r+\au} \circ \psiSI = {\phi}^{r+\au} \circ \psiI \circ i_{Q}. \]
Consequently the components of $\phiS \circ \psiSI $ are $\Ginf $ maps and thus so is $\phiS \circ \psiSI $. This proves the next proposition. 
\end{remark}

\begin{prop} \label{P:fourPtwo}
If $\Scal $ is a $(r|s)$-submanifold of a $(p|q)$-supermanifold $\Mcal $ then $\Scal $ is a $(r|s)$-supermanifold. 
\end{prop}

\begin{corollary} \label{P:fourPthree}
If $\Scal $ is a $(r|s)$-submanifold of a $(p|q)$-supermanifold $\Mcal $ then the inclusion $i:\Scal \hrar \Mcal $ is a $\Ginf $-mapping. 
\end{corollary}

\noindent
\begin{proof}
Let $(U,\psi )$ be a $(r|s)$ submanifold chart of $\Mcal $ relative to $\Scal $. We must show that $\psi \circ i \circ \psiSI $ is a $\Ginf $-mapping. But $\psiSI = \psiI \circ i_{Q} $ where $ Q = \psiS (U \cap \Scal ) \subseteq \Krs $ and $i_{Q} $ is the 
inclusion $Q \hrar Q \times \zz \hrar \Kpq $. Thus $\psi \circ i \circ \psiSI = \psi \circ \psiI \circ i_{Q} = i_{Q} $ which is a $\Ginf $-mapping. 
\end{proof}

\begin{definition}
Let $\Mcal$ be a supermanifold of dimension $(p|q)$ with $\Scal \subseteq \Mcal$. A chart $(U,\psi ) \in \Am $ is called an initial submanifold chart relative to $\Scal $ centered at $x \in U$ iff
\begin{equation} 
\psi ( C_{x} (U \cap \Scal ) ) = \psi (U) \cap (\Krs \times \{ (0,0) \} )
\end{equation} 
relative to a specific splitting
\begin{equation} 
\Kpq = \Krs \times {\mathbb{K}}^{(p-r|q-s)}
\end{equation}  
and $C_{x} (U \cap \Scal )$ denotes the set of all $y \in U \cap \Scal $ such that there is a smooth curve in $\Mcal$ from $x$ to $y$ lying in $U \cap \Scal$. We say $\Scal $ is an {\it initial super submanifold } of $\Mcal$ of dimension $(r|s)$ iff for each $x\in \Scal $ there exists an initial submanifold chart relative to $\Scal $ centered at $x$ whose image is contained in $\Krs\subseteq \Kpq.$ See $\cite{KMS}$ for details regarding initial submanifolds of an ordinary manifold. 
\end{definition}
 
The authors are grateful to Ratiu for the last reference and for clarifying the status of these concepts for Banach Lie groups. He is, of course, not responsible for any misunderstanding of these ideas by the authors.

\begin{theorem} \label{E:initialsupermanifold}
Let $\Mcal $ be a supermanifold and $\Scal \subseteq \Mcal $ an initial super submanifold of $\Mcal $ of dimension $(r|s)$. Then there exists a unique $\Cinf $-manifold structure on $\Scal $ such that the injection $i:\BS \hookrightarrow \BM $ is an injective immersion. Moreover, $\Scal $ is in fact a supermanifold and $i$ is a $\Ginf$-mapping. 
\end{theorem} 

\noindent
\begin{proof}
Given that $\Scal $ is an initial super submanifold of $\Mcal $ it is clear that as a subset of $\BM $, $\BS $ is an initial submanifold of $\BM $. \\
It is known that  an initial submanifold of a Banach manifold, such as $\BM $, possesses a unique $\Cinf $-structure relative to which $i: \BS \hookrightarrow \BM $ is smooth. \\
Thus given an atlas $\Am $ of $\Mcal $ and $\Abm = \Am $ we have that the set of pairs 
\[(C_{x} (U \cap \Scal ),\psi |C_{x} (U \cap \Scal ))\]
such that  $x\in U, (U,\psi ) \in \Am, $ and $U \cap \Scal $ is nonempty is an atlas of $\Scal.$ Moreover $\Scal $ is a Banach manifold relative to this atlas and $i:\Scal \hrar \BM $ is smooth. To see that it is a supermanifold we must show that for two overlapping charts $(U,\psi )$, $(V,\phi )$ in $\Am $ which are used to define charts on $\Scal $ we have that
\begin{equation} 
\phiB \circ {\psiB}^{-1} : \psiB ( \UBx \cap \VBx ) \rar \phiB ( \UBx \cap \VBx )
\end{equation}
is of class $\Ginf $ where $\UBx = C_{x} (U \cap \Scal )$, $\VBx = C_{x} (V \cap \Scal )$ and $\psiB = \psi | \UBx $, $\phiB = \phi | \VBx $. Let $\phiJ $ denote the $J$-th component of $\phi $ and observe that for $u \in \psi (U \cap V) $
\begin{equation} \label{E:transitionmaps10}
{d}_{u}^{k}(\phiJ \circ \psiI )(V_1,V_2,\dots,V_k ) = \sumAApq {V}_{1}^{A_1}{V}_{2}^{A_2}\cdots {V}_{k}^{A_k}
\biggl( \frac{ {\Par}^k ( \phiJ \circ \psiI ) }{{\Par z}^{A_k} \cdots {\Par z}^{A_2}{\Par z}^{A_1}} \biggr)(u).
\end{equation}
for $V_1,V_2,\dots,V_k \in \Kpq$. Eq.(\ref{E:transitionmaps10}) holds by the definition of supermanifold which implies that the transition maps $\phiJ \circ \psiI$ are $\Ginf $. If we restrict to $u \in \psi ( \UBx \cap \VBx )$ and $V_1,V_2,\dots,V_k \in \Krs $ where we identify $\Krs $ with $\Krs \times \{ (0,0) \} \subseteq \Kpq,$ then 
\begin{equation} 
{d}_{u}^{k}(\phiBJ \circ \psiBI )(V_1,V_2,\dots,V_k ) = {d}_{u}^{k}(\phiJ \circ \psiI )(V_1,V_2,\dots,V_k )     
\end{equation}
Thus,
\begin{equation} 
{d}_{u}^{k}(\phiBJ \circ \psiBI )(V_1,V_2,\dots,V_k ) = \sumAApq {V}_{1}^{A_1}{V}_{2}^{A_2}\cdots {V}_{k}^{A_k}
\biggl( \frac{ {\Par}^k ( \phiJ \circ \psiI ) }{{\Par z}^{A_k} \cdots {\Par z}^{A_2}{\Par z}^{A_1}} \biggr)(u).
\end{equation}
Therefore, $\phiBJ \circ \psiBI $ is $\Ginf $ on $\psi (\UBx \cap \VBx ) $ by Theorem $\ref{E:Btheorem}$. We simply take $f = \psiBI $ and $\Ncal = \Kpq $ which is of course a trivial supermanifold. 

To see that $i:\Scal \hrar \Mcal $ is $\Ginf $ note that, using the same notation as above, $\psi \circ i \circ \psiBI $
is the inclusion of $\psiB (\UBx )$ into $\psi (U)$. To be more explicit, it is the inclusion 
\[ \psi (U) \cap (\Krs \times \{ (0,0) \} ) \hrar \psi (U) \subseteq \Kpq \]
which is clearly class $\Ginf $ because the inclusion 
\[ \Krs \overset{h}{\hrar} \Kpq \]
is $\Ginf $ since its components $h^I$ are.
\end{proof}


\begin{corollary} \label{cor:foliation}
Assume $\Mcal $ is a supermanifold of dimension $(p|q)$ and that $\Scal $ is a leaf of a foliation of the Banach manifold $\BM $ such that, for each $x\in T_x\Scal$ is a subspace of $T_x\Mcal$ of dimension $(r,s).$ Then $\Scal $ is an initial super submanifold of $\Mcal $ of dimension $(r|s)$ and consequently $\Scal $ is a supermanifold whose inclusion of $\Scal $ into $\Mcal $ is a $\Ginf $ mapping. 
\end{corollary}

\noindent
\begin{proof}
It is known that each leaf of a foliation of a Banach manifold $\BM $ is an initial submanifold of $\BM $ and consequently if $\Scal $ is such a leaf then it follows from the theorem that $\Scal $ is an initial super submanifold of $\Mcal $. The corollary follows. 
\end{proof}

\begin{prop} \label{P:fourPseven}
Assume that $\Mcal,\Ncal $ are supermanifolds, that $\Pcal $ is a supermanifold of $dim(r|s)$, that $\psi : \Mcal \rar \Ncal $ is a $\Ginf$ mapping, and that $i:\Pcal \rar \Ncal $ is a class $\Ginf $ injective immersion onto an initial submanifold $i(\Pcal)$ of $\Ncal$ of dimension $(r|s).$ If ${\psi}_{x_{o}}:\Mcal \rar \Pcal$ is the unique mapping such that $i \circ {\psi}_{x_{o}} = \psi,$ then it is of class $\Ginf$. 
\end{prop}

\noindent
\begin{proof}
First assume that $i( \Pcal ) $ is an initial submanifold of $\Ncal $ of dimension $(r|s)$ and that the inclusion $i:\Pcal \rar \Ncal $ is a class $\Ginf $ injective immersion. Notice that $B\Pcal$ is an initial submanifold of the Banach manifold $B\Ncal $ and that $\psi : \BM \rar B\Ncal$ is a $\Cinf $ mapping. It is known $\cite{KMS}$ that for Banach manifolds the unique mapping ${\psi}_{o}:\BM \rar B\Pcal $ such that $i \circ {\psi}_{o} = \psi$ is necessarily continuous and is in fact of class $\Cinf.$ 

To finish the proof, it suffices to show that each point $p \in \Pcal $ is in the domain $U$ of a chart $(U,y)$ of $\Pcal $ such that $y \circ {\psi}_{o}{|}_{{\psi}^{-1}_{o}(U)}$ is of class $\Ginf$ (observe that 
${\psi}^{-1}_{o}(U)$ is open in $\Mcal $). Let $p\in \Pcal$ and let $(V,x)$ be a chart of $\Ncal $ at $i(p)$. There exists ${j}_{1}<{j}_{2}<\dots <{j}_{t}$ such that $x^{{j}_{1}} \circ i, x^{{j}_{2}} \circ i, \dots ,x^{{j}_{t}} \circ i$ are components of a chart on a neighborhood ${U}_{p}$ of $U={i}^{-1}(V) \subseteq \Pcal.$ 
If $y=(x^{{j}_{1}} \circ i, x^{{j}_{2}} \circ i, \dots ,x^{{j}_{t}} \circ i)$ then for
$q \in {\psi}_{o}^{-1}(U_{p})$, $1 \leq k \leq t$,
\[ (y^{k} \circ {\psi}_{o})(q) = (x^{{j}_{k}} \circ i \circ {\psi}_{o})(q) = (x^{{j}_{k}} \circ \psi )(q) \]
and $y^{k} \circ {\psi}_{o} = x^{{j}_{k}} \circ \psi $ which is a class $\Ginf $ mapping. Since  $y \circ {\psi}_{o}$ is of class $\Ginf,$ it follows that  ${\psi}_{o}$ is a class $\Ginf$ mapping. 
\end{proof}

\section{Bundles and Supergroups}
\begin{definition}
A supermanifold $\Gcal $ which is also an abstract group is called a super Lie group if the group operations are $\Ginf $ with respect to the supermanifold structure on $\Gcal $. 
\end{definition}
\noindent
When $\Gcal$ is given the Banach manifold structure implicit in its definition the resulting Banach manifold is denoted by $\BG.$
\begin{definition}
A Banach manifold $\mathcal{B} $ which is also an abstract group is called a Banach Lie group if the group operations are $\Cinf $ with respect to the manifold structure on $\mathcal{B} $.
\end{definition}

\begin{remark}
Since $\Ginf $ functions are always class  $\Cinf $ functions, it follows that the Banach manifold $\BG$ corresponding to a super Lie group $\Gcal$  is necessarily a Banach Lie group.
\end{remark}

\noindent
Left invariant vector fields are defined just as in the classical case,

\begin{definition}
Let $\Gcal $ be a super Lie group with left translation map $l_{x}(g)=xg$. Then a vector field $X$ on $\Gcal $ is said to be left invariant if for $g,x\in \Gcal$ 
$$X(gx)=d_xl_g(X(x)).$$
For each  $v \in \TeG $  the vector field $X^v$ defined by 
\[ X(x) = d_{e}l_{x}(v) \]
for all $x \in \Gcal $ is left invariant and for every left invariant vector field $X$ there exists a $v\in T_e\Gcal$ such that $X=X^v.$ We denote the set of all  left invariant vector fields on $\Gcal $ by $\Lcal (\Gcal )$. Moreover $\Lcal(\Gcal)^0$ denotes the set of even left invariant vector fields while $\Lcal(\Gcal)^1$ denotes those which are odd.
\end{definition}
\noindent
The first assertion of the following Theorem is Theorem 3.4  in $\cite{RLie}.$

\begin{theorem} \label{E:superLieLIVF}
Let $\Gcal $ be an $(p|q)$-dimensional super Lie group, then $\Lcal (\Gcal )$ is an $(p|q)$-dimensional graded Lie left $\Lm $ module subject to the bracket operation $[ \ , \ ]:\Lcal (\Gcal ) \times \Lcal (\Gcal ) \rar \Lcal (\Gcal )$ defined by
\[ [X,Y] =XY - (-1)^{\eps (X) \eps (Y)}YX \]
for all $X,Y \in \Lcal (\Gcal )$. Moreover, there is a norm $|| \cdot ||$ on $\Lcal(\Gcal)$ such that 
it is a Banach space and

(1) $\Lcal(\Gcal)^0$ and $\Lcal(\Gcal)^1$ are closed subspaces of $\Lcal(\Gcal),$

(2) $\Lcal(\Gcal)$ is a Banach super Lie algebra in the sense that there exists $M>0$ such that \indent  $||[X,Y]|| \leq M||X|| \ ||Y||$ for all $X,Y\in \Lcal(\Gcal),$

(3) the Banach Lie algebra of the Banach Lie group $\BG$ is $\Lcal(\Gcal)^0.$
\end{theorem}

\noindent
\begin{proof}
The first assertion is proved in $\cite{RLie}$. To obtain a norm on $\Lcal(\Gcal)$ we first define a norm on $\gfrak=T_e\Gcal.$  Choose a chart $\psi=(u^1,u^2,\cdots ,u^{p+q})$ at the identity $e$ of $\Gcal.$ For $X\in T_e\Gcal,$ let 
$$X_{\psi}=(X^1_{\psi},X^2_{\psi},\cdots ,X^{p+q}_{\psi})\in \Lambda^{p+q}$$
where $X=\sum_AX_{\psi}^Ae_A$ and the basis $\{e_A\}$ of $\gfrak=T_e\Gcal$ is that defined by $e_A=\frac{\partial}{\partial u^A}.$ Now define $||X||=||(X^1_{\psi},X^2_{\psi},\cdots ,X^{p+q}_{\psi})||=\sum_A || X^A_{\psi}||$ which is the norm of $(X^1_{\psi},X^2_{\psi},\cdots ,X^{p+q}_{\psi})$ in $\Lambda^{p+q}.$ Clearly, $\gfrak$ is a Banach space with respect to this norm. It is equally clear that $\gz=\Lcal(\Gcal)^0$ and $\gfrak^1=\Lcal(\Gcal)^1$ are closed subspaces of $\gfrak.$ 

We show that the norm satisfies condition (2) of the Theorem. In this part of the proof we abandon the notation used  in the first paragraph choosing to represent elements of $\gfrak$ as the value $X_e$ of some  left invariant vector field $X\in \Lcal(\Gcal).$ Using this notation we define a norm on $\Lcal(\Gcal)$ by $||X||=||X_e|| $ where $||X_e||$ is the norm of $X_e$ as defined in the first paragraph.
Let $(\tilde e_A)_x=d_el_x(e_A), x\in\Gcal,$  denote the left invariant vector field defined by an element $e_A$ of the basis of $\gfrak.$ For $Z\in \Lcal(\Gcal)$ note that, because $d_el_x$ is even for $x\in \Gcal,$  $Z_x=d_el_x(Z_e)=\sum_Ad_el_x(Z^Ae_A)=\sum_A Z_A (\tilde {e}_{A})_x,$ for $Z_A\in \Lambda.$  
Define structure constants $f^C_{AB}\in \Lambda $ by $[\tilde e_A,\tilde e_B]=\sum_Cf^C_{AB}\tilde e_C$ and let $M>0$ be a number such that $||f^C_{AB}|| \leq M$ for all $A,B,C.$
We have for appropriate $\epsilon(A,B)\in \Zn_2,$
$$||[X,Y]||=||\sum_A\sum_B(-1)^{\epsilon(A,B)}X^AY^B[\tilde e_A,\tilde e_B]||
\leq \sum_{A,B,C}||X^A|| ||Y^B|| ||f^C_{AB}\tilde e_C|| $$   
$$\leq M(p+q)\sum_A||X^A||  \sum_B ||Y^B||=M(p+q) ||X|| ||Y||$$ 
and (2) follows. Part (3) follows from the fact that as Banach spaces $\Lcal(\Gcal)^0$ is isometric and isomorphic to $\gz= T_e^0\Gcal.$
\end{proof}                                                                                                             

\begin{remark} Notice that the norm defined on $\Lcal(\Gcal)$ above depends on the chart chosen at the identity $e$ and that,  relative to this norm,  $\Lcal(\Gcal)$ is isometric to the Banach space $\Kpq \oplus [({}^{1}\Lambda)^p\times ({}^{0}\Lambda)^q].$ Another chart produces a different norm on $\Lcal(\Gcal)$ but also provides an isometry from $\Lcal(\Gcal)$ onto $\Kpq \oplus [({}^{1}\Lambda)^p\times ({}^{0}\Lambda)^q].$ It follows that $\Lcal(\Gcal)$ relative to the first norm is isometric to $\Lcal(\Gcal)$ with the second norm, but the two spaces are not identical. Thus the topology on $\Lcal(\Gcal)$ is chart independent and so a subspace of $\Lcal(\Gcal)$ is closed relative to one norm iff it is relative to the other. This becomes important in our next theorem. We refer to a norm which is defined by some chart at the identity as an {\bf admissible norm.}
\end{remark}

\begin{definition}
Assume that $\gfrak$ is a super Lie algebra. We say that it is a Banach super Lie algebra if there is a norm on $\gfrak$ such that 

(1) $\gfrak$ is a Banach space relative to the norm such that both $\gz$ and $\gfrak^1$ are closed subspaces of $\gfrak,$ and

(2) there exists a number $M>0$ such that $|| [X,Y] || \leq M ||X|| ||Y|| $ for all $X,Y\in \gfrak.$ 
\end{definition}

We prepare to determine when a sub-super Lie algebra of $\Lcal (\Gcal )$ is $\Lcal (\Hcal )$ for some super Lie group $\Hcal.$ If $\hfrak$ is a sub-super Lie algebra of  $\Lcal(\Gcal),$ then we say that it is closed and split in  $\Lcal(\Gcal)$ iff it is closed with respect to some admissible norm on  $\Lcal(\Gcal)$ and there is a closed complementary subspace $\mfrak$ of $\hfrak$ in $\Lcal(\Gcal).$ More precisely, we require that  $\Lcal(\Gcal)=\hfrak \oplus \mfrak$ as graded normed linear spaces so that in particular $\hfrak^0$ is closed and split in $\Lcal(\Gcal)^0.$

\begin{remark} \label{E:RfivePeight}
Suppose $\Mcal, \Ncal$ are supermanifolds and that $\phi$ is a $\Ginf$ mapping from $\Mcal$ into $\Ncal.$ If $X$ is a vector field on $\Mcal$ and $Y$ is a vector field on $\Ncal,$ then we say $X$ is $\phi$-related to $Y$ if and only if $d_x\phi(X_x)=Y_{\phi(x)}$ for each $x\in \Mcal.$ For ordinary manifolds, $M,N$ it is well-known that if $X_1,X_2$ are vector fields on $M$ and $Y_1,Y_2$ are vector fields on $N $ such that $X_i$ is $\phi$-related to $Y_i$ for $i=1,2,$ then $[X_1,X_2]$ is $\phi$-related to $[Y_1,Y_2].$ This also holds in the present case for supermanifolds $\Mcal,\Ncal$ when $\phi$ is a $\Ginf$ mapping. The proof is identical to the classical proof and is left to the reader. This fact is needed in the proof of the next theorem.

\end{remark}

\begin{theorem} \label{E:subsuperLiegroup}
Let $\Gcal $ denote a type $(p|q)$ dimensional super Lie group and $\gfrak = \Lcal (\Gcal )$ its super Lie algebra of left invariant vector fields. Let $\hfrak \subseteq \gfrak $ be a $(r,s)$ dimensional sub-super Lie algebra of $\gfrak $ which is closed and split in $\Lcal(\Gcal).$ Then there is a type $(r|s)$ super Lie group $\Hcal $ which is a subgroup of $\Gcal $ such that $\Lcal (\Hcal ) = \hfrak $ and the inclusion $i: \Hcal \rar \Gcal $ is a $\Ginf $ injective immersion.
\end{theorem} 

\noindent
\begin{proof}
Let $\Gcal $ be a super Lie group of type $(p|q)$ and $\gfrak $ its Banach super Lie algebra of left invariant vector fields. Let $\hfrak \subseteq \gfrak $ be a sub-super Lie algebra of type $(r,s)$ which is closed and split. Then $\hz \subseteq \gz $ is a closed and split sub-Lie algebra of the Banach Lie algebra $\gz $. Moreover $\gz $ is the Lie algebra of the Banach Lie group $\BG $. Since $\hz $ is closed and split in $\gz $ it is known
(see $\cite{Lang}$) that there is a Banach Lie subgroup $H$ of $\BG $ with Lie algebra $\hz $. \\

Moreover $H$ can be obtained as the maximal integral submanifold through the identity of $\BG $ of the subbundle $E \rar \BG $ of the tangent bundle $T\BG  \rar \BG $ defined by $E_{x} = \delx (\hfrak_e^0 )$ for each $x \in \BG $ where $\hfrak_e=\{X_e | X\in \hfrak\}$ and $\gfrak_e=\{X_e | X\in \gfrak \}.$ Here $\hfrak_e^0 $ is identified as a closed split subspace of $\gfrak_e^0 $ which is identified with $T_{e}\BG. $ It is known that a leaf of a foliation is an initial submanifold (see the book by Kolar, Michor, and Slovak $\cite{KMS}$). Moreover it is known that the inclusion $i: H \hrar \BG $ is a smooth injective immersion. It follows from Corollary  $\ref{cor:foliation}$ that $H$ can be given a supermanifold structure and if we call $H$ with this structure $\Hcal $, then the Corollary also assures that the inclusion $i: \Hcal \hrar \Gcal $ is a $\Ginf $ mapping. Note that $E_x$ has dimension $(r,s)$ for each $x\in \Gcal,$ and $E_e=\hfrak^0_e.$ So $T_{e}H=\hfrak^0_e$ and charts take their values in the appropriate subspace $\Krs$ of $\Kpq.$ Since the charts of $\Hcal$ take their values in $\Krs$, $\Hcal$ has dimension $(r|s).$

Let $\mu : \Gcal \times \Gcal \rar \Gcal $ denote the group multiplication on $\Gcal $. It follows that $\mu \circ (i \times i): \Hcal  \times \Hcal \rar \Gcal $ is a $\Ginf $ mapping. 
Since $\Hcal $ is an initial submanifold of $\Gcal $ and $\mu (\Hcal \times \Hcal ) \subseteq \Hcal $ it follows from Proposition $\ref{P:fourPthree}$ that the mapping ${\mu}_{\Hcal} : \Hcal \times \Hcal \rar \Hcal $ such that $i \circ {\mu}_{\Hcal} = \mu \circ (i \times i )$ is a class $\Ginf $ mapping. A similar application of Proposition $\ref{P:fourPthree}$ shows that ${inv}_{\Hcal} (x) = x^{-1}$ is also a class $\Ginf $ mapping. Thus $\Hcal $ is a super Lie group and $i : \Hcal \rar \Gcal $ is a $\Ginf $-immersion. \\

Finally, since $\Hcal $ is a super Lie group, $l_{x}: \Hcal \rar \Hcal $ is a $\Ginf $-mapping for each $x \in \Hcal $ and $\dylx $ maps $\TyH $ into $\TxyH $ for all $x,y \in \Hcal $. In particular $\dylx $ also maps $\TyzH $ into $\TxyzH $ so that  for each $x\in \Hcal, \delx (\TezH ) = \TxzH $ and $\hfrak_e^0  = T_e\BH=T^0_e\Hcal.$  Thus $T^0_e\Hcal$ may be identified with $\hfrak^0.$ It is perhaps, not as obvious that $T^1_e\Hcal$ can be identified with $\hfrak^1.$

We show that $\hfrak$ is isomorphic to $\Lcal(\Hcal)$ as super Lie algebras in a succession of steps. To do this first observe that $\Lcal(\Gcal)$ can be identified with $T_e\Gcal$ by identifying $v\in T_e\Gcal$ with $X^v_{\Gcal}\in \Lcal(\Gcal)$ where $X^v_{\Gcal}(x)=d_el_x(v)$ for all $x\in \Gcal.$ Notice that since $\hfrak$ is a sub-super Lie algebra of $\Lcal(\Gcal), \hfrak$ is identified with $\hfrak_e \equiv \{X_e |  X\in \Lcal(\Gcal) \}$ ( notice the change in notation, $\hfrak_e$ here and below is a subset of $\Lcal(\Gcal)$ not $\Lcal(\Hcal)).$ Both $\hfrak_e$ and $T_e\Gcal$ are given a super Lie algebra structure by defining $[v,w]_{\Gcal}$ for $v,w\in T_e{\Gcal}$ by 
$$X^{[v,w]_{\Gcal}}_{\Gcal}=[X^v_{\Gcal},X^w_{\Gcal}].$$
Thus $\hfrak  \cong \hfrak_e$ which is a sub-super Lie algebra of $T_e\Gcal.$

We now show that $T_e\Hcal$ can also be identified as a sub-super Lie algebra of $T_e\Gcal$ to be followed later by a proof that $\hfrak_e=T_e\Hcal.$ To do this recall that $\iota: \Hcal\rar \Gcal$ is an immersed initial submanifold of $\Gcal$ and consequently that $d_e \iota: T_e \Hcal \rar T_e \Gcal$ is a right $\Lambda$-linear injection of $T_e\Hcal$ into $T_e\Gcal.$ For $v\in T_e\Hcal $ let $X^v_{\Hcal}$ denote the left invariant vector field on $\Hcal$ defined by $X^v_{\Hcal}(y)=d_el_y(v), y\in T_e\Hcal.$
For $v,w\in T_e\Hcal$ define $[v,w]_{\Hcal}$ by 
$$X^{[v,w]_{\Hcal}}_{\Hcal}=[X^v_{\Hcal},X^w_{\Hcal}].$$
Notice that for every $v\in T_e\Hcal,$ the vector field $X^v_{\Hcal}$ is $\iota$-related to $X^{d_e\iota(v)}_{\Gcal}.$
It follows from Remark $\ref{E:RfivePeight}$ that for $v,w\in T_e\Hcal ,$ $$ X^{[v,w]_{\Hcal}}_{\Hcal}=[X^v_{\Hcal},X^w_{\Hcal}]     \quad \quad and \quad \quad X^{[d_e\iota(v),d_e\iota(w)]_{\Gcal}}_{\Gcal}=[X^{d_e\iota(v)}_{\Gcal},X^{d_e\iota(w)}_{\Gcal}]$$ 
are $\iota$-related. 
Consequently  $d_e\iota[v,w]_{\Hcal}=[d_e\iota(v),d_e\iota(w)]_{\Gcal},$
and $(T_e\Hcal,[ , ]_{\Hcal})$ may be identified as a sub-super Lie algebra of $(T_e\Gcal,[ , ]_{\Gcal}).$

It remains only to show that $\hfrak_e$ and $T_e\Hcal$ are equal as subsets of $T_e\Gcal.$ To see this notice that a pure basis of $\hfrak_e$ can be extended to a pure basis of $T_e\Gcal.$ It follows that there exists a pure basis $\{e_A |1\leq A\leq p+q\}$ of $T_e\Gcal$ such that $\{e_A | A\in {\cal A} \},  {\cal A}=\{1,2,\cdots,r,p+1,p+2, \cdots,p+s\}$ is a pure basis of $\hfrak_e.$ Choose a chart $\psi:U\rar T^0_e\Gcal$ of $\Gcal$ at $e\in U.$ Then $\psi \circ \iota :\iota^{-1}(U)\rar \hfrak^0_e$ is a chart of $\Hcal$ at $e\in \iota^{-1}(U).$ If we define coordinate functions $(u^A)$ of $\psi$ by $\psi(x)=\sum_{A=1}^{p+q}u^A(x)e_A, x\in U,$ then we have coordinate functions defined on $\Hcal$ by $(\psi\circ \iota)(y)=\sum_{A\in {\cal A}}u^A(\iota(y))e_A ,y\in \iota^{-1)}(U).$ Thus $\{\frac{\partial}{\partial u^A}| A\in {\cal A}\}$ in $T_e\Hcal$ is identified with $\{e_A| A\in {\cal A}\}$ in $\hfrak_e$ and 
$$T_e\Hcal=\{ \sum_{A\in {\cal A}}\lambda^Ae_A \quad | \lambda^A\in \Lambda \}=\hfrak_e.$$
Consequently, we have that as super Lie algebras 
$$\hfrak=\hfrak_e=T_e\Hcal=\Lcal(\Hcal),$$
from which the theorem follows.
\end{proof}
\noindent
\begin{definition}
Let $\Gcal $ be a super Lie group and $ \gfrak $ its tangent module $\TeG $ at the identity $e$ of $\Gcal $. For each $v \in \gz $ we define a left invariant vector field $\Xv $ on $\BG$ by
\[ \Xv (x) = \delx (v) \in \TxzG = \TxBG  \]
for $x\in \BG.$
Let $\phiv : \Rn \times \BG \rar \BG $ denote the {\it flow of the vector field $\Xv $}  on $\BG.$ Thus 
\begin{equation} 
\ddt \phiv (t,x) = \Xv (\phiv (t,x) ) \qquad \text{where} \qquad
\phiv (0,x) = x.
\end{equation}
\end{definition}

\noindent
\begin{definition}
$exp$ is the mapping from $\gz $ into $\BG $ defined by $exp (v) \equiv \phiv (1,e)$. 
\end{definition}

\noindent
Note that $exp$ is $\Cinf$ mapping which is also a local diffeomorphism. Also, we can regard $exp$ as a mapping from $\gz $ into $\Gcal $ since as sets $\BG = \Gcal $. In fact it can be shown that $exp : \gz \rar \Gcal $ is a $\Ginf $-mapping. We now establish several lemmas towards that goal. \\

{\bf We fix the notation from this point up through the proof of Theorem 4.15. Let $\Gcal$ denote an arbitrary super Lie group and $\gfrak$ its tangent module $T_e\Gcal$ at the identity. Even vectors are denoted $\gz.$ We have a fixed pure basis $\{e_a\}$ of $\gfrak$ which can be taken to be the partials relative to a chart at $e.$}
\begin{definition}
The adjoint mapping defined on $\gfrak $ is  the mapping $ad: \gfrak \rar End(\gfrak )$ where, for $x,y\in \gfrak,$
\[ad(x)(y) = ad_{x}(y) = [x,y] \]
\end{definition}

\noindent
Observe that $ad_{\au x} = \au ad_{x}$ for all $\au \in \Lm $; the adjoint $ad$ on $\gfrak $ is right-$\Lm$-linear, thus $ad \in L^{-}(\gfrak,End(\gfrak))$. However, for a particular $x \in \gfrak $, we note that $ad_{x}(y \au) =ad_{x}(y) \au $ for all $\au \in \Lm $, thus $ad_{x} \in End^{+}(\gfrak )$.

\begin{lemma} \label{E:lemmafivethirteen}
Let $(End^{+}\gfrak )^0$ denote the linear space of all even left endomorphisms of $\gfrak.$ Once for all, identity these linear mappings with their matrices relative to our fixed basis of $\gfrak.$ For each matrix $M$ (representing such a linear mapping),  define,
\[ || \ M \ || = \sumijpq || \ \Mij \ || .\]
Assume that ${\{ a_k \}}^{\infty}_{k=0}$ are numbers in $\Kn $ such that $\sumkzf |a_k | \ || M ||^k $ converges for all $M\in (End^{+}\gfrak )^0$ such that $ || M || \leq R $ for some $R > 0 $. Let $B_{R}(0)$ be the {\bf open} ball at zero in $(End^{+}\gfrak )^0,$ then $f:B_{R}(0) \rar (End^{+}\gfrak )^0$ defined by $f(M) = \sumkzf a_k M^k $ is of class $\Ginf $.
\end{lemma}

\begin{remark}
Having chosen a basis $\{ \ei , \ea \} $ of $\gfrak$  the {\bf even} endomorphisms of $\gfrak $ are identified with matrices with a (p,q) block-form,
\begin{equation} 
M \in (End^+\gfrak)^0 \ \implies \ M =
\begin{array}{l}
{\begin{pmatrix}
A & B  \\
C & D  \\
\end{pmatrix}} 
\end{array}
\end{equation}
where $A_{mn},D_{\au,\beta} \in \zLm $ and $B_{m \beta},C_{\au,n} \in \oLm.$ Notice that as a vector space over $\Kn$ $(End^+\gfrak)^0$ may be identified with $\Kn^{(p^2+q^2+2pq)}.$ 
\end{remark}
\noindent
We now prove the Lemma.

\begin{proof}
Note that for $||M|| < R$,
\[ f(M+H) = f(M) + \sumkof a_k \sumizk (M^{k-i-1}HM^{i} ) + O(H^2). \]
Thus,
\[ \dMfH = \sumkof a_k \sumizk (M^{k-i-1}HM^{i} ). \]
The components of this matrix are
\begin{equation} 
d_{M}f_{bc}(H) = \sumkof a_k \sumizk \summn (M^{k-i-1})_{bm}(H)_{mn}(M^{i})_{nc}  = \summn H_{mn} {\Lm }^{mn}_{bc}(M) 
\end{equation} 
where, for some $\epsilon_{bmn}\in \Zn_2$
\[{\Lm }^{mn}_{bc}(M) = \sumkof a_{k} \sumizk (-1)^{\epsilon_{bmn}}(M^{k-i-1})_{bm}(M^{i})_{nc}. \]
Thus $\frac{\Par f_{bc}}{\Par z_{mn}}$ exists and is equal to ${\Lm }^{mn}_{bc}$; moreover the components  $f_{bc} $ of $f$ are of class $G^1$ on $B_{R}(0)$. Thus $f$ is of class $G^1$ on $B_{R}(0)$. Observe that
\[ d_{M} \bigl( \frac{\Par f_{bc}}{\Par z_{mn}} \bigr) (H) = {\sum}_{r,s} H_{rs}{\Lm }^{mnrs}_{bc}(M) \]
where for some $\epsilon_{brs},\epsilon_{bmnrs}\in \Zn_2,$ 
\[ {\Lm }^{mnrs}_{bc}(M) = \sumkof a_{k} \sumizk {\sum}_{j=0}^{k-i-1} (-1)^{\epsilon_{brs}}({M}^{k-i-1-j-1})_{br}
                            (M^{j})_{sm} (M^{i})_{nc} \] 
\[ \qquad \qquad \qquad  \ \ \ + \sumkof a_{k} \sumizk (-1)^{\epsilon_{bmnrs}}({M}^{k-i-1})_{bm} {\sum}_{l=0}^{i} ({M}^{i-l-1})_{nr}({M}^{l})_{sc}.\]
Thus $\frac{\Par f_{bc}}{\Par z_{mn}}$ is of class $G^1$ and consequently $f$ is of class $G^2$ on $B_{R}(0)$.
An inductive argument with calculations similar to those above  show that all the partials of $f_{bc}$ exist and are power series in the components of powers of $M$ which converge on $B_{R}(0)$. Moreover their Frechet derivatives are linear in the components of $H$. Thus $f$ is of class $\Ginf $. We leave the details to the reader.
\end{proof} 
                    
\begin{corollary}
Let $\gfrak $ be any super Lie algebra such as the one defined above and let $\gz $ be its even elements. Define a mapping $f$ from $\gz $ into $(End^+\gfrak)^0 $ by
\[ X \overset{f}{\longmapsto} {\int}^{1}_{0} e^{-s \ ad_{X}} ds. \]
Then f is of class $\Ginf $.
\end{corollary}

\begin{proof}
First note that if $X \in \gz$ and $ad_{X} (Y) = [X,Y]$ then since $X$ is even and \\
$\eps ([X,Y]) = \eps (X) + \eps (Y) = \eps (Y)$ we find that $ad_{X} $ is an even left endomorphism of $\gfrak.$ The composite of even endomorphisms is even, thus the series
\[ e^{-s \ ad_{X} } = \sumkzf \frac{1}{k!}(-s)^{k} \ (ad_{X})^{k} \]
is an even left endomorphism of $\gfrak.$ This series is absolutely and uniformly convergent on every ball about zero relative to the matrix norm defined in the lemma. It follows from the lemma that the mapping from $(End^+\gfrak)^0 $ to itself defined by
\[ M \longmapsto e^{-s \ M} \]
is a $\Ginf $ mapping. 

To finish the proof we must show that $ad:X\rightarrow ad_X$ is a class $\Ginf$ mapping. The mapping $ad: \gz \rar (End^+\gfrak)^0 $ is linear over $\Kn $ as is clear from $ad_{X}(Y) = [X,Y] $ and the definition of the Lie bracket.
Hence the best linear approximation to the adjoint mapping is itself; $d_{X}(ad) = ad $. Thus the mapping $X \mapsto d_{X}(ad)$ is constant, its higher derivatives are zero. To see that $ad$ is class $\Ginf $ we have only to show that it is of class $G^{1},$ so we must show that for $X\in \gz, H \in \gz, d_{X} (ad^c_a)(H) $ is linear in the components of $H$ where the $ad^c_a:\gz \rar \Lambda$ are the component mappings of $ad$  defined by representing $ad_X$  as a matrix .\\  Since $ad_X$ is left linear its matrix is defined by $ad_X(e_a)=[X,e_a]=\sum X^b[e_b,e_a]= \sum X^bf^c_{ba}e_c$ so that $ad^c_a(X)= \sum X^bf^c_{ba}.$
Now observe that $d_Xad^c_a(H)=ad^c_a(H)= \sum H^bf^c_{ba}$ which is linear in the components of $H.$ It follows from Proposition $\ref{P:threePfourteen}$
that $ad$ is a class $\Ginf$ mapping,  hence $X \mapsto e^{-s \ ad_{X}}$ is the composite of $\Ginf $ maps and is consequently $\Ginf $ for each $s \in \Rn$. Finally integrate to obtain the desired result. 
\end{proof}                                                   

Notice that the proof  that $ad: \gfrak \rar End(\gfrak )$  is a class $\Ginf$ mapping is completely analogous to this proof since it is also linear over $\Kn $ and posseses the required properties with respect to the module operations over $\Lambda.$  Moreover the mapping $ad$ regarded as a mapping  from $\gz$ to $(End^+\gz)$ is also a class $\Ginf$ mapping. Its components $ad^c_a$ are obtained as before even though the basis is not a basis of $\gz.$ \\




\noindent
\begin{lemma} \label{E:lemmafivefifteen}
$exp:\gz \rar \Gcal $ is a class $G^1$ mapping. 
\end{lemma}

\noindent
\begin{proof}
We need to compute the Frechet derivative of $exp$ at $X \in \gz $. Since $\BG $ is a Banach Lie group we have the following formula for the Frechet derivative (see $\cite{DK}$),
\begin{equation} 
d_{X} (exp) (H) = \delexpX \biggl( {\int}^{1}_{0} e^{-s \ ad_{X}} (H) ds \biggr).
\end{equation}
Define a function $ F: \gz \rar (End^+\gfrak)^0 $ by 
\begin{equation} 
F(X)(H) =  {\int}^{1}_{0} e^{-s \ ad_{X}} (H) ds .
\end{equation}
It follows from Corollary 5.15 that $F$ is a class $\Ginf$ mapping.
Notice that $d_{X} (exp) (H) = \delexpX (F(X)(H))$ even though,  in this formula, not only is $H$ restricted to $\gz,$ but it is also the case that  $F(X)(H)\in \gz.$
We have insisted, however,  that $F(X)$ be defined on all of $\gfrak$ since we need the identity $F(X)(H)=H^iF(X)(e_i)+H^{\tilde\alpha}F(X)(e_{\tilde\alpha})$ which requires that $F(X)$ be defined on odd elements of $\gfrak.$ On the other hand this very formula shows  that the mapping from $\gz$ to $(End^+ \gz )^0$ defined by $X\rar F(X)|_{\gz}$ is also a class $\Ginf$ mapping. We will occasionally abuse notation by failing to distinguish between the two mappings.
Let $\mu :\GG \rar \Gcal $ be the class $\Ginf$  group multiplication of  the supergroup $\Gcal.$ We have that, \\
\begin{equation} 
\begin{array}{lll}
d_{X} (exp) (H) &=& \delexpX \bigl( F(X)(H) \bigr) \\
                &=& d_{e} \bigl[ \mu (exp(X), \cdot ) \bigr] ( F(X)(H) ) \\
		&=& (d_{2}\mu )(exp(X),e)(F(X)(H)).
\end{array}
\end{equation}
Where $d_{2}\mu $ denotes the Frechet derivative with respect to the second slot of $\mu $. If $H = \sumip H^{i}e_{i} + \sumaq \Hta \ea $ with respect to the pure basis $\{ e_{i},\ea \} $ of $\gfrak $ then
\begin{equation} \notag
\begin{array}{lll}
d_{X} (exp) (H) &=& (d_{2}\mu )(exp(X),e)(F(X)(\sumip H^{i}e_{i} + \sumaq \Hta \ea)) \\
                &=& \sumip H^i (d_{2}\mu )(exp(X),e)(F(X)(e_{i}))+
		   \sumaq \Hta (d_{2}\mu )(exp(X),e)(F(X)(\ea)) \\
		&=& \sumip H^i d_{X} (exp) (e_{i}) + \sumaq \Hta d_{X} (exp)(\ea ).
\end{array}
\end{equation}
To pull the ``scalars" out of $d_{2}\mu $ in the above we used the following observation. Since $\mu : \GG \rar \Gcal $ is $\Ginf $ so is the mapping with one argument fixed, that is $({\mu}_{a})(x) \equiv \mu(a,x) $ is $\Ginf $. Therefore $d_e\mu_a$ is a mapping from the full tangent module $T_e\Gcal=\gfrak$ into $T_a\Gcal$ such that
\[ (d_e{\mu}_{a})(H) = \sumip H^i (d_e{\mu}_{a})(e_{i}) + \sumaq \Hta (d_e{\mu}_{a})(\ea ). \]
We have shown that $d_{X} (exp)$ is linear over the components $H^{i},\Hta $ and hence that $exp$ is superdifferentiable at $X$  for each $X\in \gz$. It follows that $exp$ is of class $G^1$.
\end{proof} 

\noindent
\begin{lemma} \label{E:lemmafivesixteen}
$exp:\gz \rar \Gcal $ is a class $G^2$ mapping.
\end{lemma}

\noindent
\begin{proof}
Let $H,K \in \gz $ and let $\lambda \mapsto \Xl $ be a curve in $\gz $ whose value at $\lambda=0$ is $X_{0}$ and whose derivative at zero is given by  $ \ddl (\Xl ){|}_{\lambda = 0} = K $. \\
For each $\lambda$ we have
\[ d_{\Xl}(exp)(H) = (d_{2}\mu ) (exp(\Xl ), e ) ( F( \Xl ) (H) ) \]
from which it follows that
\begin{equation} \notag
\begin{array}{lll}
{d}^{2}_{X_0} (exp) (K,H) & \equiv & \ddl \biggl[ d_{\Xl}(exp)(H) \biggr] \\ && \\
                        &=& \ddl \biggl[ (d_{2}\mu )(exp(\Xl), e )(F(\Xl )(H)) \biggr] \\ && \\
		        &=& d_{1}(d_{2}\mu )(exp(X_{0},e ) \bigl( \ddl (exp(\Xl ))( F(X_{0})(H) \bigr) \\ 
			& & \ \ + \ (d_{2}\mu )(exp(X_{0},e ) \bigl(\ddl (F(\Xl ))(H) \bigr) \\ && \\
			&=& d_{1}(d_{2}\mu )(exp(X_{0},e ) \bigl( d_{X_{0}}(exp)(K)( F(X_{0})(H)) \bigr) \\
			& & \ \ + \ (d_{2}\mu )(exp(X_{0},e)((d_{X_{0}}F)(K),H) \\ && \\
			&=& d_{1}(d_{2}\mu )(exp(X_{0},e ) \bigl( (d_{2}\mu )(exp(X_{0},e ))
			                                        (F(X_{0})(K) (F(X_{0})(H)) \bigr) \\
			& & \ \ + \ (d_{2}\mu )(exp(X_{0},e ) \bigl( {d}_{X_{0}}F(K)(H) \bigr) .
\end{array}
\end{equation}
Since $F$ and $\mu$ are of class $\Ginf $ we find that we can expand ${d}^{2}_{X_0} (exp) (K,H) $ linearly in the components of $K$ and $H$ as is required by Theorem $\ref{E:Btheorem}$ in order that $exp$ be a class $G^2$ mapping. Thus $exp$ is of class $G^2$.
\end{proof}


\begin{theorem} \label{E:expisGinf}
$exp: \gz \rar \Gcal $ is a class $\Ginf $ mapping. 
\end{theorem}

\noindent
\begin{proof}
The proof is by induction on k.
Inductively, assume that for each k there exists a $\Ginf $ mapping  $\lk$ from $ \gz \times \Gcal$ into the space of all $\Lambda$-multi-linear mappings from $\gk $ into $\TG  $ such that
\[ {d}_{X}^{k}(exp) ( v_1,v_2,\dots ,v_k ) = \lk (X, exp(X))( v_1,v_2,\dots ,v_k ). \]
Note first that there exists such $\lk $ for $k=1,2 $ by the previous two lemmas. In particular,
\[ d_{X}(exp)(v) = {\lambda}^{1}(X,exp(X))(v) \]
where
\[ {\lambda}^{1}(X,a)(v) = (d_{2}\mu)(a,e)(F(X)(v)).\]
Likewise,
\[ {d}^{2}_{X}(exp)(v_1,v_2) = {\lambda}^{2}(X,exp(X))(v_1,v_2) \]
where
\[{\lambda}^{2}(X,a)(v_1,v_2 ) =
 d_{1}(d_{2}\mu )(a,e)\bigl( (d_{2}\mu)(a,e)(F(X)(v_1 )(F(X)(v_2 )))
                                   + (d_{2}\mu )(a,e)( \bigl( {d}_{X}F)(v_1 )(v_2 ) \bigr). \]  
Now assume the existence of $\lk $ and prove ${\lambda}^{k+1}$ exists. If $\lambda \rightarrow X_{\lambda}$ is a curve in $\gz $ whose value at zero is $X_{0}$ and whose derivative at zero is $v_1,$ then
\begin{equation} \notag
\begin{array}{lll}
{d}^{k+1}_{X_{0}} (exp) ( v_1,v_2,\dots ,v_{k+1} ) &=
                            & d_{X_0}\biggl[{d}^{k} (exp) ( v_2,v_3,\dots ,v_{k+1} )\biggr](v_1)\\&\\
                            
             &=& d_{X_0}\biggl[{d}^{k} (exp) ( v_2,v_3,\dots ,v_{k+1})\biggr](\ddl \Xl |_{\lambda=0})\\&\\
                            & = & \ddl \biggl[ {d}^{k}_{\Xl}(exp)
                        ( v_2,v_3,\dots ,v_{k+1} ) {\biggr]}_{\lambda = 0 } \\ && \\
                        &=& \ddl \biggl[ \lk(\Xl , exp(\Xl) )( v_2,v_3,\dots ,v_{k+1} ) \biggr]_{\lambda=0}\\ && \\
		        &=& [(d_{1}\lk)(X_{0},exp(X_{0}))(v_{1} )]( v_2,\dots ,v_{k+1} ) \\
			& & \ \ +[(d_{2} \lk ) (X_{0},exp(X_{0})) \bigl( \ddl (exp \Xl ) {|}_{\lambda = 
			0}\bigr)]( v_2,\dots ,v_{k+1} ) \\ && \\
		        &=& [(d_{1}\lk)(X_{0},exp(X_{0}))(v_{1} )]( v_2,\dots ,v_{k+1} ) \\
			& & \ \ + [(d_{2} \lk ) (X_{0},exp(X_{0})) \bigl( {\lambda}^{1}(X_{0},exp(X_{0}))
			        (v_{1} )]( v_2,\dots ,v_{k+1} ).
\end{array}
\end{equation}


\noindent
Thus,
\[ {d}_{X_0}^{k+1}(exp) ( v_1,v_2,\dots ,v_{k+1} ) = {\lambda}^{k+1} (X, exp(X),e)( v_1,v_2,\dots ,v_{k+1} ). \]
where
\begin{equation} \notag
\begin{array}{lll}
{\lambda}^{k+1}(X,a)( v_1,v_2,\dots ,v_{k+1} ) 
     &=& [(d_{1}\lk)(X,a)(v_{1} )]( v_2,\dots ,v_{k+1} ) \\
     & & \ \ + [(d_{2} \lk ) (X,a) \bigl( {\lambda}^{1}(X,a)
			        (v_{1} ) \bigr)] ( v_2,\dots ,v_{k+1} ). \\
\end{array}
\end{equation}	
By the induction hypothesis it follows that ${\lambda}^{k+1}$ is of class $\Ginf$ and ${d}^{k+1}_{X_{0}} (exp) ( v_1,v_2,\dots ,v_{k+1} )$ is multi-linear over $\gz$ for each $X_0\in \gz.$ It follows from Theorem $\ref{E:Btheorem}$ that $exp$ is a class $\Ginf $ mapping from $\gz $ to 
$\Gcal.$
\end{proof}			
		

\noindent
\begin{theorem} \label{E:subgroupisLiegroup} 
Let $\Gcal $ be a $(p|q)$-super Lie group and $\Scal $ a subgroup which is also an initial $(r|s)$-submanifold of $\Gcal $. Then $\Scal $ is a $(r|s)$-super Lie group.
\end{theorem}

\noindent
\begin{proof}
Let $i:\Scal \hrar \Gcal $ denote the inclusion mapping and $\muS,\muG $ the group "multiplications" on $\Scal $ and $\Gcal $ respectively, then by Theorem 4.5 ${\mu}_{\Gcal} \circ (i \times i)$ is the composite of $\Ginf $ mappings and so is of class $\Ginf $. Since $\Scal $ is an initial submanifold, it follows from Proposition 4.7 $\ref{P:fourPthree}$ that the unique mapping ${\mu}_{\Scal} : \Scal \times \Scal \rar \Scal $ such that $i \circ {\mu}_{\Scal} = {\mu}_{\Gcal} \circ (i \times i)$ is of class $\Ginf $. A similar argument shows that $inv_{\Scal }$ is a class $\Ginf $ mapping. The Theorem follows.
\end{proof}

\noindent
\begin{definition}
If $\Gcal $ is a $(p|q)$ super Lie group and $\Scal \subseteq \Gcal $ is a subgroup which is also a $(r|s)$-submanifold of $\Gcal $ then we say $\Scal $ is a {\it sub-super Lie group } of $\Gcal $.
\end{definition}


\noindent
\begin{remark}
If $\Scal $ is a closed sub-super Lie group of a super Lie group of $\Gcal $ then $\BS $ is a closed sub-Lie group of $\BG $ as  Banach Lie groups. Moreover the coset space $\BG / \BS $ is known to be a Banach manifold and $\BG \rar \BG / \BS $ is a principal fiber bundle with structure group $\BS $.
\end{remark}

\begin{theorem} \label{E:quotient}
If $\Gcal $ is a $(p|q)$ super Lie group and $\Scal $ is a closed $(r|s)$ sub-super Lie group of $\Gcal $ then $\Gcal / \Scal $ is a $(p-r|q-s)$ supermanifold. Moreover $\Gcal \rar \Gcal / \Scal $ is a $\Ginf $-mapping and is a principal fiber bundle with structure group the super Lie group $\Scal $. All local trivializing maps are $\Ginf $-maps. 
\end{theorem}

\noindent
\begin{proof}
One only needs to check that the mappings which define the bundle structure of $\BG \rar \BG / \BS $ are in fact $\Ginf $-maps so there is little to prove. We sketch the main features of the proof for the convenience of the reader but in fact the argument is borrowed from Br$\ddot{o}$cker and Dieck $\cite{BD}$\\
First notice that since $i:\Scal \hrar \Gcal $ is $\Ginf $ the mapping $d_{e}i: \TezS \hrar \TezG $ is injective. Choose a pure basis $\{ e_{i} , \ea \}$ , $1 \leq i \leq r$ and $1 \leq \au \leq s$ of $\TeS $, and extend it to a pure basis
$\{ e_{i} , \ea \}$ , $1 \leq i \leq p$ and $1 \leq \au \leq q$ of $\TeG $. Thus,
\[ \TezS \cong \Krs \hrar \Krs \times {\mathbb{K}}^{(p-r|q-s)} \cong \TezG \]
and one may factor $\TezG = \TezS \times \Mcale $ as Banach spaces where
\[v \in \Mcale \ \iff \ v = \mathop{{\sum}}_{j=r+1}^{p}v^{j}e_{j}+\mathop{{\sum}}_{\au=s+1}^{q}{\tilde{v}}^{\au}\ea  \]
where $v^{j},{\tilde{v}}^{\au} \in \zLm $. The Banach structure is given by the norm on $\TezG $ which is defined by,
\[ || \ \sumip H^{i} e_{i} + \sumaq \Hta \ea \ || = \sumip |H^{i}{|}_{\Lm} + \sumaq |\Hta {|}_{\Lm} \]
for $H^i \in \zLm $ and $\Hta \in \oLm $ and $| \cdot {|}_{\Lm} $ is the norm on the Banach algebra of supernumbers $\Lm $. The definition for the norm on subspaces of $\TezG $ is obvious. \\

Now define $\Mcalee = \{ X \in \Mcale \ | \ || X || < \eps \} $ for $\eps > 0$ and let $\Dcale = exp( \Mcale )$. Recall that $exp: \TezG \rar \Gcal $ is both a local $\Cinf $ diffeomorphism and a $\Ginf $-mapping. Consider $\mu: \Dcale \times \Scal \rar \Gcal $ defined by $\mu (g,s) = gs$ in $\Gcal $. We claim that for $\eps $ small enough $\mu $ is an embedding. To see this first note that $(d\mu)_{(e,e)}|(\TezDe \times \{ 0 \} )$ and $(d\mu)_{(e,e)}|(\{ 0 \} \times \TezS )$ are identity maps on $\TezDe $ and $\TezS $ respectively. So $(d\mu)_{(e,e)}(v,w)=v+w $ and if $(d\mu)_{(e,e)}(v,w)=0 $ then $v = -w \in \TezDe \cap \TezS = \{ 0 \} $ and $ ker(d\mu)_{(e,e)} = \{ (0,0) \} $. By the inverse function theorem for Banach manifolds there exists an open set $U$ about $e$ in $\Scal $ and $\eps > 0 $ small enough \\
so that $\mu :\Dcale \times U \rar \Dcale U $ is a $\Cinf $ diffeomorphism. It is also a $\Ginf $-mapping since the group operation on $\Gcal $ is a $\Ginf $-mapping and since the inclusions $\Dcale \times U \hrar \Dcale \times \Scal \hrar \Gcal \times \Scal \hrar \Gcal \times \Gcal $ are all $\Ginf $-mappings. Note that for $s \in \Scal $ the right multiplication map $\Rs : \Gcal \rar \Gcal $ defined by $\Rs (x) = xs $ is a $\Ginf $-mapping and so is
\[ \mu | (\Dcale \times (Us)) = 
\Rs \circ [ \mu | (\Dcale \times \Scal ) ] \circ [ id_{\Dcale} \times R_{s^{-1}} ]. \]  \\
Moreover $\mu |(\Dcale \times (Us)) $ is a $\Cinf $ diffeomorphism from $\Dcale \times (Us) $ onto $\Dcale Us $ for each $s \in \Scal $ and $\mu | (\Dcale \times \Scal ) $ is a local $\Cinf $-diffeomorphism and a $\Ginf $-mapping. We claim that for small enough $\eps $, $\mu | (\Dcale \times \Scal ) $ is injective. Indeed if one chooses $V \subseteq \Gcal $ open about $e$ such that $(V^{-1}V)\cap \Scal \subseteq U $ then for each $\epsp < \eps $ , $\epsp >0 $ such that $\Dcalep \subseteq V $ one can show that $\mu | (\Dcalep \times \Scal ) $ is injective. \\


Thus we have the existence of $\eps > 0 $ such that $\mu: \Dcale \times \Scal \rar \Dcale \Scal $ is an embedding.\\


We now show how to obtain a $\Ginf $ structure on the coset space $\Gcal / \Scal $. Let $\eta : \Gcal \rar \GmS $ denote the mapping which sends $x \in \Gcal $ to the coset $\eta (x) \in \GmS $. For $g \in \Gcal $ let $U_{g} = g\Dcale \Scal $ and notice that $U_{g} = \mu (\Dcale \times \Scal ) $ is open in $\Gcal $. Since $U_{g} $ is the union of cosets $\eta (U_{g}) $ is open in the quotient topology on $\GmS $. Let $\psigI $ denote the inverse of a chart where $\psigI : \Dcale \rar \eta ( U_{g} \Scal ) $ is defined by
\[ \Dcale \lrar \Dcale \times \{ e \} \rar \Dcale \times \Scal
    \overset{\mu}{\lrar} \Dcale \Scal 
    \overset{l_{g}}{\lrar} g\Dcale \Scal = U_{g} 
    \overset{\eta}{\lrar} \eta ( U_{g} ). \]
For $g,h \in \Gcal $ such that the relevant maps are well defined,
\begin{equation} \notag
\begin{array}{lll}
(\psih \circ \psigI )(x) &=& \psih (\psigI (x) ) \\
                         &=& \psih (\eta ( l_g (\mu (x,e))))\\
                         &=& \psih ( \eta ( l_h (l_{h^{-1}} l_g ) ( \mu (x,e)))) \\
			 &=& \psih ( \eta (l_h ( \mu (h^{-1}gx,e)))) \\
			 &=& \psih (\psihI (h^{-1}gx,e)) \\
			 &=& l_{h^{-1}g}(x).

\end{array}
\end{equation}	
Thus $\psih \circ \psigI $ is a $\Ginf $-mapping and consequently $\{ (\eta (U_{g} ), \psig ) \ | \ g \in \Gcal \}$ is a $\Ginf $ structure on $ \GmS $.\\

We now produce a $\Ginf $ local trivialization of $\Gcal $ as  bundle over $\GmS $ \\


For $g \in \Gcal $ let $\phigI : \eta ( U_{g} ) \times \Scal \rar U_{g} \subseteq \Gcal $ be the inverse of our proposed trivialization mapping where $\phigI $ is defined by
\[ \eta (U_{g}) \times \Scal 
   \overset{\psig \times id}{\lrar} \Dcale \times \Scal
   \overset{\mu}{\lrar} \Dcale \Scal
   \overset{l_{g}}{\lrar} g \Dcale \Scal = U_{g} \]
meaning,
\[ (x,s) \mapsto (\psig (x),s ) \mapsto \mu (\psig (x),s ) \mapsto l_{g}(\mu (\psig (x),s )). \]
For appropriate $g,h \in \Gcal $
\begin{equation} \notag
\begin{array}{lll}
(\phih \circ \phigI )(x,s) &=& \phih (l_{g} (\mu (\psig (x),s))) \\
                           &=& \phih (l_{h} ((h^{-1}g)\mu (\psig (x),s))) \\
                           &=& \phih (l_{h} ( (h^{-1}g) \psig (x) s )) \\
			   &=& \phih (l_{h} ( (\psih \circ \psigI)(\psig (x)s))) \\
			   &=& \phih (l_{h} ( \psih (x) s ) ) \\
			   &=& \phih (l_{h} ( \mu (\psih (x) , s ))) \\
			   &=& \phih (\phihI (x,s)) \\
			   &=& (x,s).
\end{array}
\end{equation}
Thus two "adjacent" local trivializing maps agree and one has a principal bundle structure on $\Gcal \rar \GmS $.
\end{proof}


\section{A Super Version of Lie's Third Theorem}

Our next result requires us to show that if one has a supersmooth ($G^{\infty}$) vector field on the even part of a super Lie algebra and if this vector field depends supersmoothly on a parameter then the solution depends supersmoothly on both the parameter and the initial condition. \newline

Consider then a Banach super Lie algebra $\gfrak$ and a function $F:\gz \times \gz \rightarrow \gz$ which we interpret as a parameterized vector field on $\gz.$ What does it mean to say $F$ is a $G^{\infty}$ function? We choose a basis of $\gfrak$ and identify $\gz$ with $\Kpq$ via the obvious globally defined chart. We actually choose two copies of the same chart but denote the components of the first by $(u_1,u_2,\cdots, u^{p+q})$ and its copy by
$(v^1,v^2,\cdots ,v^{p+q}).$ So coordinates on $\gz\times \gz$ will be denoted by $(u^1,u^2,\cdots,u^{p+q},v^1,v^2,\cdots ,v^{p+q})$  although strictly speaking these should be reordered so that all even coordinates come first in the $2(p+q)$-tuple and the odd coordinates last so that the chart has its values in $\Kn^{2p|2q}.$ Throughout this section, $E$ will denote the Banach space $\gz \times \gz$ with the norm defined below. Thus $E$ is a $G^{\infty}$ manifold with a single global chart.
Now $F$ is a $G^{\infty}$ function iff all its component functions are.

In that which follows we will assume $F$ is of class $G^{\infty}$ in which case it is necessarily of class $C^{\infty}$ on the Banach space $E.$ Additionally, by Theorem $\ref{E:Btheorem}$ there will exist continuous functions
$$b^u_{A_1A_2\cdots A_k}, \quad \quad  b^v_{B_1B_2\cdots B_l}, \quad \quad  b^{uv}_{A_1A_2\cdots A_kB_1B_2\cdots B_l}$$
such that 
$$(d^k_1F)_{(x,y)}(X_1,X_2,\cdots ,X_k)=\sum X_1^{A_1}X_2^{A_2}\cdots X_k^{A_k}b^u_{A_1A_2\cdots A_k}(x,y),$$
$$(d^l_2F)_{(x,y)}(Y_1,Y_2,\cdots ,Y_l)=\sum Y_1^{B_1}Y_2^{B_2}\cdots Y_l^{B_l}b^v_{B_1B_2\cdots B_l}(x,y),$$
$$(d^k_1d^l_2F)_{(x,y)}(X_1,X_2,\cdots ,X_k,Y_1,Y_2,\cdots ,Y_l)=\sum X_1^{A_1}\cdots X_k^{A_k}Y_1^{B_1}\cdots Y_l^{B_l}b^{uv}_{A_1\cdots A_kB_1\cdots B_l}.$$

Here $d^r_iF$ denotes the $i$-th partial Frechet derivative iterated $r$ times for $i=1,2$ while $(d^k_1d^l_2F$ denotes $l$ iterations of the second partial Frechet derivative followed by $k$ iterations of the first partial Frechet derivative.  Not all cases are exhibited above; one should consider iterated partial derivatives obtained via any permutation of first and second partial Frechet derivatives and similar formula would exist for each reordering. All such conditions characterize when a class $C^{\infty}$ function $F$ is of class $G^{\infty}$ by Theorem $\ref{E:Btheorem}$.  

If  $\gfrak$ denotes such a Banach super Lie algebra let $E=\gz\times \gz$ denote the Banach space with norm defined by $||(X,Y)||_E=max \{||X||_{\gz}, ||Y||_{\gz}\}$ for $(X,Y)\in E.$ We write $B_r(0)$ to denote $\{X\in \gz | \quad  ||X||_{\gz}<r \}$ and $B^E_r(0)$ for $\{(X,Y)\in E | \quad  ||(X,Y)||<r \}.$ We also drop the subscripts on both $||\cdot||_E$ and $|| \cdot ||_{\gz}$ below since it should be obvious from the context which norm is intended. \newline

\begin{lemma} \label{E:lemmasixPone}
Let $F:\gz \times \gz \rightarrow \gz$ be a class $G^{\infty}$ mapping such that  $F(0,0)=0,$ and such that for some positive number $M$ and each positive number $r,$ $||(d_2F)_u|| \leq rMe^{rM}$ for all $u\in B^E_r(0).$ Then, for some $r>0,$ there exists a unique mapping $f:[0,1]\times B^E_r(0)\rightarrow \gz$ such that \newline
(1) for each $t\in [0,1]$ the mapping from $E$ into $\gz$ defined by $u\rightarrow f(t,u)$ is a class $G^{\infty}$ mapping and \newline
(2) $\frac {df}{dt}(t,X,Y)=F(X,f(t,X,Y))$ and $f(0,X,Y)=Y$ for all $(X,Y)\in B^E_r(0).$ \newline
\end{lemma}

\begin{proof}
First we show that there exists a $\Cinf$ mapping $f$ which satisfies condition $(2)$ of the lemma.
For this purpose consider the mapping $\tilde F:E\rightarrow E$ defined by $\tilde F(X,Y)=(0,F(X,Y)).$ Then $\tilde F$ is a smooth vector field on $E$ such that $\tilde F(0,0)=0$ and by Corollary 4.1.25 of 
$\cite{AMR}$ there exists $r>0$ such that whenever $u\in E$ and $||u||<r$ there exists an integral curve of $\tilde F$ through
$u$ which is defined on $[-1,1].$ Since there exists a flow box of $\tilde F$ at $(0,0)$ on $E$ it follows that, for some $r>0,$ there exists a smooth function $\tilde f:[-1,1]\times B^E_r(0)\rightarrow E$ such that for $(t,u)\in [-1,1] \times B^E_r(0)$
$$\frac{d\tilde f}{dt}(t,u)=\tilde F(\tilde f(t,u))  \quad \quad  \tilde f(0,u)=u.$$
Since, for $u=(X,Y)\in B^E_r(0),$ $\tilde F(\tilde f(t,u))=(0,F(\tilde f(t,u)))$ and $\frac {d\tilde f}{dt}=(\frac {d\tilde f_1}{dt}, \frac {d\tilde f_2}{dt})$ it follows that $\frac {d\tilde f_1}{dt}=0$ and $\frac {d\tilde f_2}{dt}=F(\tilde f(t,u))=F(X,\frac {d\tilde f_2}{dt}(t,u)). $ Consequently, $f\equiv  \tilde f_2$ is a smooth mapping from $[0,1]\times B^E_r(0)$ into $\gz$ such that 
$$\frac {df}{dt}(t,X,Y)=F(X,f(t,X,Y))  \quad  \ and \   \quad f(0,X,Y)=Y$$
for all $(X,Y)\in B^E_r(0).$ \newline

We must now show that (1) of the Lemma holds. To do this we require an explicit formula which shows how the derivatives of the function $(X,Y)\rightarrow f(t,X,Y)$ depend on $X$ and $Y$ for each fixed $t\in [0,1].$\newline

Let $\Fcal $ denote the Banach space of all continous maps $g$ from $I=[0,1]$ into $\gz $ equipped with the sup-norm:
\[ || \ g \ || = {lub}_{t\in I} |  g(t)  |.  \]

It is our intent to show that the mapping $h:B^E_r(0)\rightarrow \Fcal$ defined by $h(u)(t)=f(t,u)$ for $u\in B^E_r(0),t\in [0,1]$ is of class $G^{\infty},$ that is, we will show that the mapping from $B^E_r(0)$ to $\gz$ defined by $u\rightarrow h(u)(t)$ is of class $G^{\infty}$ for each $t\in[0,1].$ It will then follow that the solution of our differential equation is a $G^{\infty}$ function of $(X,Y)$ where $X\in \gz$ is a parameter and $Y$ is an initial condition of the differential equation. To avoid excessive language we simply say $h$ is of class $G^{\infty}$ in this situation.

Notice that $\gfrak\times \gfrak$ is a Banach super vector space such that $(\gfrak \times \gfrak)^0=
\gz \times \gz=E$ and $(\gfrak \times \gfrak)^1=\gfrak^1 \times \gfrak^1.$ A function such  as $h:B^E_r(0)\rightarrow \Fcal$ is of class $G^{\infty}$ iff the function $h_t: E \rightarrow \gz$  defined by $h_t(w)=h(w)(t)$ is of class 
$\Ginf$ for each $t$ and by Proposition $\ref{P:threePfourteen}$  this is true iff it is of class $C^{\infty}$ and the derivatives $d_w^kh_t^B$ of the components $h^B_t$  are multi-linear.
Here $d_w^kh_t^B$ is a mapping from $(T_wE)^k=E^k$ into $\Lambda.$ Since this is a condition on the components $h_t^B$ of $h_t$ we may write 
$d_w^kh_t=d_w^kh_t^Be_B$ and think of it as a $\gz$-valued function. Indeed, $h_t=h_t^Be_B$ where
$\{e_B\} $ is a basis of $\gfrak$ (not $\gz$) and so $h^1_t,h^2_t,\cdots h_t^p$ are even functions while
$h_t^{p+1},h_t^{p+2},\cdots,h_t^{p+q}$ are odd. Thus $d_w^kh^B_t$ maps into $\Lambda^0$ for $1\leq B\leq p$ and maps into $\Lambda^1$ for $p+1\leq B\leq p+q$ from which it follows that $d_w^kh_t=d_w^kh_t^Be_B$ is $\gz$-valued.

 Observe that multi-linearity implies  that for each $t\in [0,1],$ there exist continuous functions $\gamma^{uv}_{tC_1C_2\cdots C_l}:B^E_r(0)\rightarrow \gz$ such that $$d^l_wh_t(Y_1,Y_2,\cdots,Y_l)=Y_1^{C_1}Y_2^{C_2}\cdots Y_l^{C_l}\gamma^{uv}_{tC_1C_2\cdots C_l}(w)$$
 for $Y_1,Y_2,\cdots ,Y_l \in E.$
We do not need to write explicit formulas for the functions $\gamma^{uv}_{tC_1C_2\cdots C_l}$ we need only know that if $Y_1,Y_2,\cdots ,Y_l$ are arbitrary elements of $E$ and are written in terms of their components relative to the basis $\frac{\partial}{\partial u^A},\frac{\partial}{\partial v^B}$ then 
$d^l_wh_t(Y_1,Y_2,\cdots,Y_l)$ is linear in the components of each $Y_i.$ The order in which the components $Y_1^{C_1}Y_2^{C_2}\cdots Y_l^{C_l}$ occur will not be important since they can be permuted up to signs and the definition of 
$\gamma^{uv}_{tC_1C_2\cdots C_l}$ can be adjusted in a manner to agree with Theorem $\ref{E:Btheorem}$.

Define $K: E \times \Fcal \rar \Fcal $ by 
$$K((X,Y),g)(t) = Y+{\int}^{t}_0F(X,g(s))ds$$ 
for $(X,Y)\in E.$  Notice that $K((X,Y),g)=g$ iff
\[ \frac{dg}{dt} = F(X,g(t)) \qquad \text{ and } \qquad g(0)=K((X,Y),g)(0)=Y. \]

If $f$ is the smooth solution of the vector field $F$ obtained above and $h:B^E_r(0)\rightarrow \Fcal $ is defined by $h(u)(t)=f(t,u)$ for $t\in [0,1], u\in B^E_r(0),$ then  
$h$ is smooth (since solutions depend smoothly on parameters and initial conditions) and
\[ K(u,h(u))=h(u) \qquad \qquad \forall \ u \in B^E_{\rcal}(0) \subset E \]
Thus if $H(u,f) \equiv f-K(u,f), $ then
\[ H(u,h(u)) \equiv 0 \]
and for $\lambda \rar u_{\lambda} $ a curve through $u$ in $B^E_r(0)$ and $\delta = \frac{d}{d\lambda} ( u_{\lambda} ){|}_{\lambda = 0}$
\[ H(u_{\lambda} , h(u_{\lambda} )) = 0\]
we have
\[ (d_{1}H)_{(u,h(u))}(\delta ) + (d_{2}H)_{(u,H(u))}((dh)_{u}(\delta) ) = 0.\]
If we can show that $\rcal >0$ can be chosen small enough so that $(d_{2}H)_{(u,h(u))}:\Fcal \rar \Fcal $ has an inverse for all $(u,h(u)),$ then it will follow that 
\[ (d_{2}H)((dh)_{u}(\delta)) = -(d_{1}H)(\delta) \] and that
\begin{equation} \label{E:star}
\boxed{ (dh)_{u}(\delta) = -{(d_{2}H)}^{-1}_{(u,h(u))}((d_{1}H)_{(u,h(u))}(\delta ). } 
\end{equation}
This explicit formula for $dh$ will enable us to show that $h$ is of class $G^1$ and eventually that it is of class $G^{\infty}.$ In order to obtain the required $r>0$ first notice that $d_2H$ can be written in terms of $d_2K$ which, in turn, can be written in terms of $d_2F.$
Indeed, if $\lambda \rar f_{\lambda}$ is a curve in $\Fcal$ through $f\in \Fcal,$ then
\begin{equation}
\begin{array}{ll} \notag
(d_{2}H)_{(u,f)} \bigl( \ddl ( \fl ) \liszero \bigr) 
                &= \ddl (H (u, \fl)) \liszero \\
                &= \ddl (\fl - K(u,\fl )) \liszero \\
		&= \ddl (\fl ) \liszero - (d_{2}K)_{(u,f)} (\ddl (\fl ) \liszero )
\end{array}
\end{equation}
and denoting $\delf \equiv \ddl (\fl ) \liszero $, we have
\begin{equation} \label{E:aster}
\boxed{ (d_{2}H)_{(u,f)}(\delf ) = \delf - (d_{2}K)_{(u,f)}(\delf ) } 
\end{equation}
It follows that $(d_2H)_{(u,f)}=I_{\Fcal} -(d_2K)_{(u,f)}$ as operators and so if the
operator norm of $(d_2K)_{(u,f)}$ is smaller than 1, then $(d_2H)_{(u,f)}$ will be invertible.
But we also know that $K((X,Y), \fl )(t) = Y + {\int}_{0}^{t} F (X, \fl (s) ) ds$ so that
\begin{equation}
\begin{array}{ll} \notag
(d_{2}K)_{((X,Y),f)}(\delta_f)(t) &= \ddl (K((X,Y),\fl )(t))  \\
                    &= {\int}_{0}^{t} \ddl (F (X,\fl (s)) ds \\
		    &= {\int}_{0}^{t} (d_2F)_{(X,f(s))} ( \delf (s)) ds. \\
\end{array}
\end{equation}
Note that, $$||(d_2K)_{(u,f)}(\delta_f)|| \leq \int_0^1||(d_2F)_{(X,f(s))}||\quad || \delf (s)|| ds \leq rMe^{rM}||\delta_f ||$$
and $||(d_2K)_{(u,f)}|| \leq rMe^{rM} <1$ for appropriately chosen $r>0.$ 
Now let $w_{\lambda}=(\Xl,\Yl)\in B^E_r(0)$ be a curve through $w=(X,Y)$ and $\delta=\frac{d}{d\lambda}(w_{\lambda})|_{\lambda=0}=(\delta_1,\delta_2)\in E=\gz \times \gz,$
then
\[ H( w_{\lambda} , f )(t) = f(t) - K(w_{\lambda} ,f )(t) = f(t) -\Yl - {\int}_{0}^{t}F(\Xl,f(s))ds \]
and $ (d_{1}H)_{(w,f)}(\delta )(t) = -\delta_2-\int_0^t(d_1F)_{(X,f(s))}(\delta_1)ds. $

Let $\dhat(t)=(d_1H)_{(w,h(w))}(\delta)(t),$ then 
$$\dhat(t)=-\delta_2^B\frac{\partial}{\partial v^B}-\delta_1^A\int_0^t(d_1F)_{(X,h(w)(s))}(\frac{\partial}{\partial u^B})ds.$$

It follows that eqn.[\ref{E:star}] becomes
\begin{equation}
\begin{array}{ll}
(dh)_{w} (\delta ) 
    &= -(d_{2}H)^{-1}(d_{1}H)_{(w,h(w))}(\delta )) \\
    &= (d_{2}H)^{-1} ( \dhat ) \\
    &= [ I_{\Fcal} - (d_{2}K)_{(w,h(w))}]^{-1} (\dhat ) \qquad \text{ by eqn.[$\ref{E:aster}$]} \\
    &= [ I_{\Fcal}+(d_{2}K)_{(w,h(w))}+(d_{2}K)_{(w,h(w))} \circ (d_{2}K)_{(w,h(w))}+\cdots ] (\dhat) \\
    &= \dhat +(d_{2}K)_{(w,h(w))}(\dhat)+(d_{2}K)_{(w,h(w))}(d_{2}K_{(w,h(w))}(\dhat ))+\cdots 
                                                      + (d_{2}K)_{(w,h(w))}^{l}(\dhat)+\cdots.\\
\end{array}
\end{equation}

We now show that $h$ is of class $G^1.$  For $\dhat$ as defined above,

$(d_2K)_{(w,h(w))}(\dhat)(t)=$
$$\int_0^t(d_2F)_{(w,h(w)(s))}(-\delta_2^B\frac{\partial}{\partial v^B}-\delta_1^A\int_0^s(d_1F)_{(X,h(w)(r))}(\frac{\partial}{\partial u^B})dr)ds=\delta^B_2\gamma^v_B(w)(t)+\delta^A_1\gamma^u_A(w)(t)$$
where
$$\gamma^u_A(w)(t)=-\int_0^t(d_2F)_{(w,h(w)(s))}(\int_0^s(d_1F)_{(X,h(r))}(\frac{\partial}{\partial u^B})dr)ds$$
and
$$\gamma^v_B(w)(t)=
-\int_0^t(d_2F)_{(w,h(w)(s))}(\frac{\partial}{\partial v^B})ds.$$
Write 
$(d_2K)_{(w,h(w))}(\dhat )(t)=\sum_C\delta^C\gamma_C(w)(t)$ where the components $\delta^C$ include all the components of both $\delta_1$ and $\delta_2$ and the $\gamma_C$ include both types of indexed functions $\gamma^u_A$ and $\gamma^v_B.$
We have
$[(d_2K)_{(w,h(w))}]^l(\hat{\delta} )$$$=
[(d_2K)_{(w,h(w))}]^{l-1}(\sum_B \delta^C\gamma_C(w))=
\sum_C\delta^C[(d_2K)_{(w,h(w))}]^{l-1}(\gamma_C(w))=\sum_C\delta^C\hat b^{l}_C(w)$$
for some set of continuous functions $\hat b^{l}_C$ from $B^E_r(0)$ to $\Fcal.$
It follows that if  $w\in B^E_r(0)$ and $\delta=(\delta_1,\delta_2)\in E=\gz\times \gz$  then,
$$d_wh_t(\delta)=(dh)_w(\delta)(t)=\sum_C\delta^C\sum_{l=0}^{\infty}\hat b^{lv}_B(w)(t)$$
and by Theorem $\ref{E:Btheorem}$, $h$ is of class $G^1.$ 

To show that $h$ is of class $G^2$ one selects a curve $\lambda \rightarrow w_{\lambda}$ in $B^E_r(0)$ with value $w$ at $\lambda=0$ and with $\delta_1=\frac{d}{d\lambda}(w_{\lambda})|_{\lambda=0}\in E.$ Then, for fixed $\delta_2\in E,$ 
$d^2_wh(\delta_1,\delta_2)=\frac{d}{d\lambda}(d_{w_{\lambda}}h)(\delta_2)$ can be computed in terms of $d_1(d_2K)_{(w,h(w))}$ and $d_2(d_2K)_{(w,h(w))}.$ Indeed, $d^2_wh(\delta_1,\delta_2)$ is a sum of terms each of which is multi-linear in both variables $\delta_1,\delta_2.$ 
Explicitly, by Equation (51), if we define $\dhat_1,\dhat_2$ as above, then
$$d^2_wh( \delta_1, \delta_2)=
 \sum_{l=1}^{\infty}\{\sum_{i=0}^l [(d_2K)_{(w,h(w))}]^i\frac{d}{d\lambda}((d_2K)_{(w_{\lambda},h(w_{\lambda}))})([(d_2K)_{(w,h(w))}]^{l-i-1}(\dhat_2))|_{\lambda=0}$$ where
  
\noindent $\frac{d}{d\lambda}((d_2K)_{(w_{\lambda},h(w_{\lambda}))})([(d_2K)_{(w,h(w))}]^{l-i-1}(\dhat_2))|_{\lambda=0}$
$$= (d_1d_2K)_{(w,h(w))}(\delta_1,[(d_2K)_{(w,h(w))}]^{l-i-1}(\dhat_2))+
 (d_2d_2K)_{(w,h(w))}((dh)_w(\delta_1),[(d_2K)_{(w,h(w))}]^{l-i-1}(\dhat_2)).$$
 
 In order to show that $d^2_wh$ is 2-multi-linear it is useful to first show that for arbitrary \newline $\phi_1,\phi_2\in E,$ $(d_1d_2K)(\phi_1,\hat\phi_2)(t)$ is bilinear in the components of $\phi_1,\phi_2.$
 Here, for arbitrary $\phi \in E,$ $\hat \phi(t)=-\phi_2^B\frac{\partial}{\partial v^B}-\phi_1^A\int_0^t(d_1F)_{(X,h(w)(s))}(\frac{\partial}{\partial u^B})ds.$
 
Recall that  
 $(d_{2}K)_{((X,Y),f)}(\delta_f)(t) = \int_{0}^{t} (d_2F)_{(X,f(s))} ( \delf (s)) ds,$
and note that for $\phi_1=(\phi_X,\phi_Y)\in T_{(X,Y)}E=E$ and $\phi_2\in E$ with $\hat \phi_2$ as defined above 
 $$(d_1d_2K)(\phi_1,\hat \phi_2)(t)=\int_0^t (d_1d_2F)_{(X,f(s))}(\phi_X,\hat \phi_2(s))ds.$$
 Since $(d_1d_2F)_{(X,f(s))} (\phi_X,\hat \phi_2(s))$ is linear in the components of $\phi_1,\phi_2$  we see that 
 $(d_1d_2K)(\phi_1,\hat \phi_2)(t)$ is  linear in the components of $\phi_1$ and $\phi_2.$
Thus $(d_1d_2K)$ is 2-multi-linear in both input variables. Since this holds for arbitrary $\hat \phi_2$ and since for any $\delta_2$ we can define  $\phi_2$ by taking it to be $[(d_2K)_{(w,h(w))}]^{l-i-1}(\dhat_2))$ and since the latter is linear in its components  we see that the composite
$$(d_1d_2K)_{(w,h(w))}(\delta_1,[(d_2K)_{(w,h(w))}]^{l-i-1}(\dhat_2))$$
is linear in the components of $\delta_2.$ It is obvious that it is also linear in the components of $\delta_1$ and so is multi-linear in the components of $\delta_1,\delta_2.$

A similar argument shows that $(d_2d_2K)_{(w,h(w))}((dh)_w(\delta_1),[(d_2K)_{(w,h(w))}]^{l-i-1}(\dhat_2))$
is also multi-linear in the components of both $\delta_1$ and $\delta_2.$ It follows that $d^2_wh( \delta_1, \delta_2)$ is as well. Consequently, $h$ is of class $G^2$ as asserted above.

To see that $h$ is of class $G^{\infty}$ we must examine higher derivatives of both $F$ and $K.$
Since $F$ is of class $G^{\infty}, (d_2^lF)_w:\gz \times \gz \rar \gz$ is multi-linear over $\gz$ for each positive integer $l$ and $w\in B^E_r(0).$  Moreover, we have for $w=(X,Y)\in B^E_r(0)$ and $Y_1,Y_2,\cdots, Y_l \in \Fcal $ 
$$(d^l_2K)_{(w,h(w))}(Y_1,Y_2,\cdots ,Y_l)(t)=\int_0^t(d^l_2F)_{(X,h(w)(s))}(Y_1(s),Y_2(s),\cdots ,Y_l(s))ds.$$
Consider the special case when each $Y_i:I \rar \gz$ is either constant or satisfies the condition $Y_i(t)=\dhat_i(t)=-\delta_2^{iB}\frac{\partial}{\partial v^B}-\delta_1^{iA} \int_0^t(d_1F)_{(X,h(w)(s))}(\frac{\partial}{\partial u^B})ds$ for supernumbers $\delta^{iA}_1,\delta^{iB}_2.$ In this case, it follows that $(d^l_2K)_{(w,h(w))}(Y_1,Y_2,\cdots ,Y_l)(t)$ is linear in the components of $Y_1,Y_2,\cdots ,Y_l$ since  
$(d^l_2F)_{(X,h(w)(s))}(Y_1(s),Y_2(s),\cdots ,Y_l(s))$ is (recall that $F$ is of class $\Ginf$ and so is multi-linear over $\gz$). Similar results hold  in the ``mixed" case.  For example, if $w\in  B^E_r(0)$ and $X_1,X_2,\cdots, X_k,Y_1,Y_2,\cdots Y_l\in  \Fcal,$ then 

$(d^k_1d^l_2K)_{(w,h(w))}(X_1,X_2,\cdots ,X_k,Y_1,Y_2,\cdots ,Y_l)(t)$
$$=\int_0^t(d_1^kd_2^lF)_{(X,h(w)(s))}(X_1(s),X_2(s),\cdots ,X_k(s),Y_1(s),Y_2(s),\cdots ,Y_l(s))ds$$
and the latter is linear in the components of $X_i(s), Y_j(s)$ whenever each of them is either constant or assumes the special form of $\dhat(s)$ for some $\delta\in E.$

To show that $h$ is of class $\Ginf$ one proceeds inductively. Assume that $d^k_w{ h_t} :E^k \rar \gz$ is multi-linear over $\gz$ for each $t\in I,w\in B^E_r(0)$ and that $d^k_wh:E^k\rar \Fcal$ is a linear combination of terms each of which is a composite of the functions $d^r_wh,$ $r<k,$ and iterates and composites of partial Frechet derivatives of $K$ (see, for example,  the formula for $d^2h$ above). Notice that if $w \in B^E_r(0), t\in I,$ then for $\delta_1,\delta_2,\cdots, \delta_{k+1}\in E$
$$d^{k+1}_wh_t(\delta_1,\delta_2,\cdots, \delta_{k+1})=d^{k+1}_wh(\delta_1,\delta_2,\cdots, \delta_{k+1})(t)=d_w[d^{k}_wh(\delta_2,\cdots, \delta_{k+1})](\delta_1)(t)$$
and
$$d_w[d^{k}_wh(\delta_2,\cdots, \delta_{k+1})](\delta_1)(t) =\frac{d}{d\lambda}(d^k_{w_{\lambda}}h)(\delta_2,\cdots,\delta_{k+1})|_{\lambda=0}(t)$$ 
where $\lambda \rar w_{\lambda}$ is a curve through $w\in B^E_r(0)$ such that $\delta_1=\frac{d}{d\lambda}(w_{\lambda})|_{\lambda=0}.$
Now the inductive assumption guarantees that $d_{w_{\lambda}}h$ is multi-linear over $\gz$ and that it is sum of terms each of which is a composite of functions $d_w^rh, r<k,$ and iterates of various partial derivatives of $K$ evaluated at the point $(w_{\lambda}, h(w_{\lambda}).$ Derivatives of such terms with respect to $\lambda$ generically increase the number of partial Frechet derivatives of $K$ and will increase the number of derivatives of $h$ by at most one. Thus $d^{k+1}_wh$ has the same form as that of $d^k_wh$ with at most one extra derivative of $h.$ It follows from the discussion in the paragraph preceding this one and from calculations analogous to those  showing that $d^2_wh_t$ is bi-linear that $d^{k+1}_wh_t:E^{k+1} \rar \gz$ is multi-linear over $\gz.$ This inductive argument assures that $d^l_wh_t: E^l \rar \gz$ is multi-linear over $\gz$ for every positive integer $l$ and every $w\in B^E_r(0)$ and so by Proposition $\ref{P:threePfourteen}$ it follows that
 $h_t: B^E_r(0)\rar \gz$ is of class $\Ginf$ for each $t\in I.$ The lemma follows.
\end{proof}

\begin{corollary} \label{cor:sixPtwo}
Let $F:\gz \times \gz \rightarrow \gz $ be defined by 
$$F(X,Z)=X+\sum_{k=1}^{\infty}\frac{B_k}{k!}ad_Z^k(X)$$
where $B_0,B_1,B_2, \cdots$ are the Bernoulli numbers.
Then there exists a positive number r and  a function $W:[0,1]\times B^E_r(0)\rightarrow \gz$ such that \newline
(1) for each $t\in [0,1]$ the mapping from $B^E_r(0)$ into $\gz$ defined by $u\rightarrow W(t,u)$ is a class $G^{\infty}$ mapping and \newline
(2) $\frac {dW}{dt}(t,X,Y)=F(X,W(t,X,Y))$ and $W(0,X,Y)=Y$ for all $(X,Y)\in B^E_r(0).$ \newline
\end{corollary}

\begin{proof}
Given $F$ as defined above, observe that  it follows from Lemma 5.13 and the second half of the proof of Corollary 5.15  that $F$ is of class $G^{\infty}$ since 
\[ F(X,Z) = \sum_{k=1}^{\infty} F_k(X,Z)\]
where \[ F_{k}(X,Z) = {ad}_{Z}^{k}(X) \]
is $\Lm$-linear in $X$ and is the diagonal of a $\Lm $-multilinear mapping in $Z$ up to  signs. \\
(Here $(z_{1},z_{2},\dots,z_{k}) \mapsto [z_{1},[z_{2},[\cdots,[z_{k},X],\cdots ]$ is multilinear up to signs in the $z_{i}$ over $\Lm $ for each $i=1,2,\dots k$ and ${ad}_{Z}^{k}(X)$ is the diagonal of this map in the $z$ variables.) Also, notice that for $X,Z\in B_r(0),$
$$||(d_2F)_{(X,Z)}(H)||\leq \sum_{k=1}^{\infty}k\frac{|B_k|}{k!}M^k ||Z||^{k-1} ||X|| ||H||
\leq rM\sum_{k=1}^{\infty}\frac{1}{(k-1)!}(rM)^{k-1}||H||= rMe^{rM}||H||.$$
Since $F(0,0)=0$ it follows from the Lemma that there exists a function $W:[0,1]\times B_r(0)\times  B_r(0)\rightarrow \gz$ which satisfies (1) and (2) of the Lemma. The Corollary follows. 
\end{proof}

\subsection{A Super Lie Theorem}
\begin{theorem}
Assume that $\gfrak $ is a Banach super Lie algebra such that \\
(1) \ $\gz $ is enlargable with Lie group the Banach Lie group $G $, and \\
(2) for all $ g \in G $, $Ad_{g} : \gz \rar \gz $ is $\zLm $-linear. \\
Then there exists a $\Ginf $-atlas on $G$ such that the corresponding supermanifold $\Gcal $ is a super Lie group with respect to the group operations on $G$. Moreover  the even factor $\Lcal(\Gcal)^0$ of the super Lie algebra of left-invariant vector fields on $\Gcal $ is Lie algebra isomorphic  to $\gfrak^0.$ In general, it does not follow that $\gfrak$ is isomorphic to $\Lcal(\Gcal)$ as super Lie algebras. In fact, $\gfrak$ can be given two different super Lie structures such that $\Lcal(\Gcal)^0$ is Lie algebra isomorphic to $\gz.$
\end{theorem}

\noindent
\begin{proof}
Using our Lemma and Corollary, the proof follows that of Duistermaat and Kolk $\cite{DK}$. Let 
\[ \gez = \{ X \in \gz \ | \ \frac{e^{ad_{X}}-I}{ad_{X}} = \sum_{k=0}^{\infty} \frac{1}{(k+1)!}{ad}^{k}_{X} \ \ 
                             \text{has an inverse} \ \} \]
If $\eta :\gez \rar \gz $ is defined by $\eta (X)=\frac{e^{ad_{X}}-I}{ad_{X}} $ then the inverse of $\eta (X)$
is given by $\zeta (X)$ where $\zeta $ is the mapping from $\gez $ to $\gz $ defined by
\[ \zeta (X) = \frac{ad_{X}}{e^{ad_{X}}-I}= \sum_{k=0}^{\infty} \frac{B_{k}}{k!}{ad}_{X}^{k} \]
where $B_{0},B_{1},B_{2},\dots $ are the Bernoulli numbers (see $\cite{DK}$). We know that $X \mapsto ad_{X}$ is a $\Ginf $ mapping from $\gz $ to $End(\gz ) $, moreover we also know that the mappings from $End(\gz ) $ to $End(\gz ) $ defined by 
\[ A \mapsto \sum_{k=0}^{\infty} \frac{1}{(k+1)!}{A}^{k} \qquad \text{and} \qquad 
   A \mapsto \sum_{k=0}^{\infty} \frac{B_{k}}{k!}{A}^{k} \]
are $\Ginf $-mappings (by Lemma 5.13 and the proof of the second half of Corollary 5.15) \\
Define a mapping $F:\gez \times \gez \rar \gz $ by
\[ F(X,Z) = \zeta (Z)(X) = \sum_{k=0}^{\infty} \frac{B_{k}}{k!}{ad}_{Z}^{k}(X). \]
It follows from the last Corollary that there exists a function $W:[0,1]\times B_r(0)\times B_r(0) \rightarrow \gz$ such that \newline
(1) for each $t\in [0,1]$ the mapping from $B_r(0)\times B_r(0)$ into $\gz$ defined by $(X,Y)\rightarrow W(t,X,Y)$ is a class $G^{\infty}$ mapping and \newline
(2) $\frac {dW}{dt}(t,X,Y)=F(X,W(t,X,Y))$ and $W(0,X,Y)=Y$ for all $(X,Y)\in B^E_r(0).$ \newline
Thus if we define $\mu : \gz \times B_{r}(0) \rar \gz $ by
\[ \mu (X,Y) = W(1,X,Y), \] 
then $\mu $ is a class $\Ginf $-mapping. It follows from the argument of the proof of Theorem 1.6.1 of Duistermaat and Kolk $\cite{DK}$ that
\[ exp ( \mu (X,Y) ) = exp ( X )exp ( Y ) \]
for all $X,Y \in B_{r}(0) \subset \gez$. Now we know $exp : \gz \rar G$ is a $\Cinf $-diffeomorphism on a small ball about $0 \in \gz $ (here $G$ is the Banach Lie group having $\gz $ as its Lie algebra). For each $x \in G$ define ${\kappa}^{x}(y) = log ( l_{x^{-1}}(y))$ where $log=exp^{-1}$. Then ${\kappa}^{x}$ is a local $\Cinf $ diffeomorphism. Duistermaat and Kolk show that for $x,y$ such that ${\kappa}^{y} \circ ({\kappa}^{x})^{-1}$ is defined it follows that
\[ ({\kappa}^{y} \circ ({\kappa}^{x})^{-1})(X) = Y \qquad \iff \qquad Y = \mu (\mu (Y_{o},-X_{o}),X) \]
for a choice of $X_{o},Y_{o}$ in $dom({\kappa}^{x}) \cap dom({\kappa}^{y})$. Thus
\[ ({\kappa}^{y} \circ ({\kappa}^{x})^{-1} )(X) = \mu ( \mu (Y_{o},-X_{o}),X) \]
and consequently, ${\kappa}^{y} \circ ({\kappa}^{x})^{-1}$ is a $\Ginf $ mapping. It follows that the family of maps $\{ {\kappa}^{x} \}$ is a $\Ginf $-atlas on $G$ and we denote the resulting supermanifold by $\Gcal $. Following Duistermaat and Kolk once more, let $m: \Gcal \times \Gcal \rar \Gcal $ be defined by $m(x,y)=xy^{-1}$ where $m$ is just the group operations on G. We show $m$ is of class $\Ginf $. We have
\begin{equation}
\begin{array}{ll} \notag
\bigl( {\kappa}^{xy^{-1}} \circ m \circ [ ({\kappa}^{x})^{-1} \times ({\kappa}^{y})^{-1} ] \bigr) (X,Y) 
       &= {\kappa}^{xy^{-1}} \bigl( m ( ({\kappa}^{x})^{-1}(X), ({\kappa}^{y})^{-1}(Y) ) \bigr) \\
       &= {\kappa}^{xy^{-1}} \bigl( ({\kappa}^{x})^{-1}(X)({\kappa}^{y})^{-1}(Y))^{-1} \bigr) \\
       &= ad_{y} ( \mu (X, -Y))
\end{array}
\end{equation}
since
\begin{equation}
\begin{array}{ll} \notag
x exp(X) (y exp (Y))^{-1} &= x exp(X)exp(-Y)y^{-1} \\
                          &= xy^{-1}y exp ( \mu (X, -Y) )y^{-1} \\
			  &= xy^{-1} exp (ad_{y} ( \mu (X,-Y))).
\end{array}
\end{equation}
Since $ad_{y}: \gz \rar \gz $ is a $\Ginf $-mapping for all $y \in G$ we see that
\[ {\kappa}^{xy^{-1}} \circ m \circ [ ({\kappa}^{x})^{-1} \times ({\kappa}^{y})^{-1} ] \]
is a $\Ginf $ mapping since it is just the map
\[ (X,Y) \mapsto ad_{y}(\mu (X, -Y)). \]
Thus $\Gcal $ is a super Lie group. Finally notice that $\Lcal(\Gcal)$ can be identified as a super Lie algebra with $T_e\Gcal$
and consequently $\Lcal(\Gcal)^0 =T_e^0\Gcal=T_e\BG= \gz$ as Lie algebras.

Notice, however,  that without further restrictions $\Lcal(\Gcal)$ can not generally be shown to be isomorphic to $\gfrak$ as super Lie algebras. To see this note that we can define a trivial super Lie algebra structure on $\gfrak$ which is generically distinct from the given one in such a way that the Lie structure on $\gz$ is not affected. Indeed, keep the module structures on $\gfrak$ intact so as to leave our construction of $\Gcal$ unchanged. On the other hand distort the super Lie structure by requiring that the Lie structure of $\gz$ remain unchanged, but require that $[\gz,\gfrak^1]=0$ and $[\gfrak^1,\gfrak^1]=0.$ 
Thus generically, one has two super Lie structures on $\gfrak$ which give rise to the same super Lie group $\Gcal.$
\end{proof}

\begin{remark} If $\Gcal$ and $\Hcal$ are super Lie groups and $\phi:\Gcal \rar \Hcal $ is a class $\Ginf$ homomorphism then the mapping $d_e\phi$ is a homomorphism from the super Lie algebra $T_e\Gcal$ to the super Lie algebra $T_e\Hcal $ ( using their obvious identifications with the super Lie algebras of left invariant vector fields). Moreover the diagram

\begin{center}
\setlength{\unitlength}{0.675cm}
\begin{picture}(6,6)
\put(0.85,0.25){\makebox{$\Gcal$}}
\put(0.75,3.75){\makebox{$\gfrak$}}
\put(4.25,3.75){\makebox{$\hfrak$}}
\put(4.25,0.25){\makebox{$\Hcal$}}

\put(0,2.2){\makebox{${exp}$}}
\put(2.3,4.3){\makebox{$d_{e}\phi$}}
\put(2.5,-0.1){\makebox{$\phi$}}
\put(4.7,2.2){\makebox{$exp$}}

\put(1.5,4){\vector(1,0){2.5}}
\put(1,3.5){\vector(0,-1){2.5}}
\put(4.5,3.5){\vector(0,-1){2.5}}
\put(1.5,0.5){\vector(1,0){2.5}}
\end{picture}
\end{center}

is commutative. The proof of this result is almost identical to the proof in the usual Lie group case and is  left to the reader. The point is that since $d_e\phi$ is $\Lambda$ left-linear it is also a class $\Ginf$ mapping so the entire diagram is in the $\Ginf$ category. In particular notice that if $\phi$ is an injective inmmersion then this shows that the exponential mapping on $\Hcal$ is simply the restriction of the exponential mapping on $\Gcal$ to $\phi(\Hcal).$ This fact makes it possible to make contact with the physicist usual technique for identifying the super Lie groups of matrices of given super Lie algebras of matrices.

\end{remark}

\section{Formal Supergroups }
The physics literature sometimes defines supergroups in terms of a formal multiplication. The rules for muliplying group elements are obtained from the Campbell-Hausdorff formula (see $\cite{BB}$). Apparently this is related to the approach taken by Berezin and Leites $\cite{Leites}$, Kac $\cite{KAC}$, and Kostant $\cite{Kostant}$ which are analogus to the formal groups in ordinary Lie theory (see $\cite{Serre}$ for example).
The formal approach assumes a certain algebraic structure as the starting point. We, in contrast, have shown that the exponential function is a $\Ginf$ mapping and can use our general results to prove that the relevant algebraic structure is correct. 

Let $\Vcal=\Vcal^0\oplus \Vcal^1$ denote a graded left $\Lambda$-module which is finitely and freely generated over $\Lambda.$ Once for all, select a fixed pure basis of $\Vcal$ of type $(p,q)$ and recall that left $\Lambda$ endomorphisms of $\Vcal^0$ may be represented by matrices.
\begin{equation}
  M =
\begin{array}{l}
{\begin{pmatrix}
A & B  \\
C & D  \\
\end{pmatrix}} 
\end{array}
\end{equation}
where $A,B,C,D$ are respectively, $p\times p, p\times q, q\times p, q\times q$ matrices over $\Lambda$ which respect the grading. Also recall that $M$ is even iff both $A$ and $D$ have only even entries while $B$ and $C$ have only odd entries. Similarly $M$ is odd iff both $A$ and $D$ have only odd entries while $B$ and $C$ have only even entries.

Let $\Wcal=\Vcal^0$ and notice that $\Wcal$ is isomorphic as a vector space to $\Kpq.$ We denote the set of all matrices $M$ defined above by $gl(\Wcal)$ and observe that it is a super Lie algebra with respect to the bracket
$$[M,N]=MN-(-1)^{\epsilon(M)\epsilon(N)}NM.$$
Moreover it is a Banach super Lie algebra relative to the norm $|| M ||=\sum_{i,j=1}^{p+q}|| M_{ij}||_{\Lambda}.$ Clearly the subspace of even elements $gl^0(\Wcal)$ is a Banach Lie algebra whose Lie bracket is induced by the multiplication of the associative Banach algebra structure on $gl^0(\Wcal).$ It is well-known that the group of units $Gl^0(\Wcal)$ of this associative Banach algebra is open in the associative algebra. Moreover it is a Banach Lie group whose Lie algebra is precisely the Lie algebra structure on $gl^0(\Wcal)$ induced by the associative structure (see Neeb [N]). 

According to our general construction we may define a $G^{\infty}$ atlas of charts on $Gl^0(\Wcal)$ having values in the even part of the super Lie algebra $gl(\Wcal).$ We denote the resulting super manifold by $Gl^s(\Wcal).$

\begin{prop}
The super Lie algebra of left invariant vector fields $\Lcal(Gl^s(\Wcal))$ of $Gl^s(\Wcal)$ is isomorphic to the super Lie algebra $gl(\Wcal).$
\end{prop}

\begin{proof} We omit most of the details as they closely follow the usual proof that the Lie algebra of $Gl(n,\Rn)$ is $gl(n,\Rn).$ One shows that if $B$ in $ gl(\Wcal)$ is identified with the vector tangent to $Gl^s(\Wcal)$ at the identity $e,$ then the left invariant vector field on $Gl^s(\Wcal)$ is given by 
$$X^B(A)=d_el_A(B)=\sum_{i,j}(AB)_{ij}\frac {\partial}{\partial x_{ij}}|_{A}$$
for $A\in Gl^s(\Wcal).$  Here $x_{ij}$ is a chart on $Gl^s(\Wcal)$ where $x_{ij}(M)$ denotes the $ij$-component of the matrix $M;$ thus $x_{ij}$ is a $\Lambda$-valued function on $Gl^s(\Wcal).$ Notice that $x_{ij}$ is even for $1 \leq i,j \leq p$ and for $p+1 \leq i,j \leq p+q$ but otherwise is odd. Using the fact that $X^B=\sum_{i,j,k}x_{ik}B_{kj}\frac {\partial}{\partial x_{ij}}$
one can now show that $X^{[M,N]}=[X^M,X^N]$ as in the usual case. The tedious details require careful,  but straightforward,  considerations of parities; they are left to the reader.

\end{proof}

It should now be clear that the various supergroups of matrices which occur in the physics literature are indeed super Lie groups relative to the Rogers definition of a supermanifold  when the supernumbers are infinitely generated. The usual physics treatments of the subject begin with a super Lie algebra of matrices and then define the corresponding super Lie group by a formula which is tacitly assumed to be a supermanifold. The Campbell-Hausdorff formula is then used to link the super Lie algebra to its super Lie group. We have developed the machinery necessary to understand the super manifold structure of the underlying super Lie group and have shown that the exponential mapping is indeed a $G^{\infty}$ mapping. These results then justify the physicist's intuition and also show how the super Lie group structure in the matrix case derives from more general principles.

\begin{definition} If $\gLie$ is a graded Lie algebra over $\Cn,$ then its Grassmann shell is the super Lie
algebra $\gLiehat=\Lambda\otimes \gLie$ defined by $[\lambda X,\mu Y]=\lambda \mu(-1)^{\epsilon(\mu)\epsilon(X)}[X,Y]_{\gLie}$ for $\lambda,\mu\in \Lambda$ and $X,Y\in \gLie.$ More generally, one says that a super Lie algebra $\gfrak$ is a conventional Berezin superalgebra of dimension $(p,q)$ if and only if it possesses a pure basis  for which the structure constants have no soul.
\end{definition}
 
 Notice that if we  choose a basis $\{ \Em , \Ea \} $, $m=1,2,\dots p $ and $\au = 1,2, \dots q $ of a graded Lie algebra $\gLie $ over $\Cn$ where we define $\Em $ to be even and $\Ea $ to be odd, then it is also a pure basis of the Grassmann shell of $\gLiehat$ and  the corresponding structure constants relative to this basis are complex numbers. Thus the Grassmann shell of a graded Lie algebra over $\Cn$ is a special type of conventional Berezin superalgebra.
 
 \begin{theorem} Let $\gLie=\gLie^0\oplus \gLie^1$ denote a $(p,q)$ graded Lie algebra over $\Cn.$ Then there exists a super Lie group $\Hcal$ whose super Lie algebra of left invariant vector fields is isomorphic to the Grassmann shell $\gLiehat$ of $\gLie.$
 \end{theorem}
 \begin{proof} First apply Ado's theorem for the case of graded Lie algebras over $\Cn$ (see $\cite{KAC}$ page 79). This theorem assures us that there exists an even injective homomorphism $ \phi:\gLie \hrar gl(r|s,\Cn) $. We choose $gl(r|s,\Lm ) $ to be the set of all left endomorphisms on a $(r,s)$ supervector space $\Vcal$ and identify these endomorphisms with their corresponding $(r+s)\times (r+s) $ $\Lm $-valued matrices. Now identify $\gLie$ with its image in $gl(r|s,\Lambda)$ and choose a basis $\{ \Em,\Ea \} ,1\leq m \leq p, 1\leq \alpha \leq q$ of $\gLie.$ Finally, extend this $(p,q)$ basis to a basis $\{ \Em,\Ea \}, \quad m=1,2,\dots (r+s)^2, \quad \au = 1,2, \dots (r+s)^2 $ of  $gl(r|s,\Cn ) $. The Grassmann shell of $gl(r|s,\Cn)$ is 
\[ gl(r|s,\Lm ) \equiv \{ \summor {\xi}^m \Em + \sumaos {\xi}^{\au} \Ea \ | \ {\xi}^{m},{\xi}^{\au} \in \Lm \}. \]
Likewise the Grassmann shell of $\gLie $ , denoted $\gLiehat $, is constructed by replacing complex scalars by Grassmann scalars. Notice that the injective homomorphism naturally extends to the Grassmann shell, thus we injectively embed the Grassmann shell of the graded Lie algebra into matrices having Grassmann supernumbers as entries:
\[ \phi : \gLiehat \hrar gl(r|s,\Lm ). \]
Now we know that $gl(r|s,\Lm)=gl(\Vcal^0)$ is the Lie super algebra of the super Lie group $G^s(\Vcal^0)$ and that $\gLiehat$ is a sub-super Lie algebra of $gl(\Vcal^0).$ It follows from Theorem $\ref{E:subsuperLiegroup}$ that there is a super Lie group $\Hcal$ which is a sub-super Lie group of $Gl^s(\Vcal^0)$ having $\gLiehat$ as its super Lie algebra of left invariant vector fields.  Thus we have the commutative diagram.
 
\[
 \begin{CD}
  \gLie @>  >> gl(r|s,\Cn) \\
  @VVV @VVV \\
  \gLiehat @> >> gl(r|s,\Lm ) 
  \end{CD}
\]

\end{proof}

\end{document}